\documentclass[aps, prl, reprint, twocolumn, longbibliography]{revtex4-1} 

\usepackage{graphicx,graphics,color} 
\usepackage{amsmath, amssymb,amsfonts} 
\usepackage{bm,times,xspace} 
\usepackage[colorlinks,citecolor=blue,linkcolor=red]{hyperref} 
\usepackage{braket}
\usepackage{lineno}  %appending line number

\begin{document}
% \linenumbers

\title{Thermodynamic Geometry of Nonequilibrium Fluctuations in Cyclically Driven Transport}

\author{Zi Wang}
\author{Jie Ren}
\email{Corresponding Email: Xonics@tongji.edu.cn}
\affiliation{%
Center for Phononics and Thermal Energy Science, China-EU Joint Lab on Nanophononics, Shanghai Key Laboratory of Special Artificial Microstructure Materials and Technology, School of Physics Science and Engineering, Tongji University, Shanghai 200092, China
}%
\date{\today}

\begin{abstract}
Nonequilibrium thermal machines under cyclic driving generally outperform steady-state counterparts. However, there is still lack of coherent understanding of versatile transport and fluctuation features under time modulations.
Here, we formulate a theoretical framework of thermodynamic geometry in terms of full counting statistics of nonequilibrium driven transports. We find that, besides the conventional dynamic and adiabatic geometric curvature contributions, the generating function is divided into an additional nonadiabatic contribution, manifested as the metric term of full counting statistics. 
This nonadiabatic metric generalizes recent results of thermodynamic geometry in near-equilibrium entropy production to far-from-equilibrium fluctuations of general currents. Furthermore, the framework proves  geometric thermodynamic uncertainty relations of near-adiabatic thermal devices, constraining fluctuations in terms of statistical metric quantities.
We exemplify the theory in experimentally accessible driving-induced quantum chiral transport and Brownian heat pump. 
\end{abstract}

\maketitle

{\it Introduction.}--In the past decade, significant advances have been achieved in both experiments and theories that allow for direct manipulations of thermodynamics of small setups~\cite{blickle2012realization,rossnagel2014nanoscale,martinez2016brownian,martinez2016engineered,josefsson2018quantum,talkner2020colloquium,wang2022inelastic,saha2021maximizing,holubec2021fluctuations}. These systems are subject to large fluctuations that are detrimental to their stable output. Recently, it has been shown that cyclically driven thermal devices can be tuned to be more stable and perform better than their steady state counterparts~\cite{barato2016cost,holubec2018cycling,miller2021thermodynamic}, igniting a surge of interest into the stochastic thermodynamics of this regime.

The concept of geometry provides deep insights into the nonequilibrium cyclic driving. Its manifestation in transport was originally introduced in the Thouless pump, relating the quantization of pumped charge with the overall integral of the underlying non-trivial Berry curvature~\cite{thouless1983quantization, berry1984quantal}. This idea also generalizes to open systems~\cite{brouwer1998scattering,sinitsyn2007universal,rahav2008directed}. In thermal devices, the geometric-phase-like contribution provides a way of directing heat flow~\cite{ren2010berry,ren2012geometric,yuge2012geometrical,chen2013dynamic,wang2017unifying,watanabe2017geometric,nie2020berry,wang2022geometric,wang2022diffusive,monsel2022geometric} and constructing heat engines~\cite{giri2017geometric,bhandari2020geometric,hino2021geometrical,hayakawa2021geometrical}. These geometric results, ranging from quantum Markovian systems to classical diffusive dynamics, are mainly restricted to the adiabatic slow driving protocols. By utilizing controls,  nonadiabatic pump effect can be eliminated at the expense of extra dissipation~\cite{takahashi2020nonadiabatic,funo2020shortcuts}. The leading order nonadiabatic dissipation in the finite but small driving frequency regime~\cite{andresen1984thermodynamics} is captured by the concept of thermodynamic metric~\cite{salamon1983thermodynamic,crooks2007measuring,feng2009far,sivak2012thermodynamic}. Yet, in the arbitrarily fast regime, the average entropy production assumes another geometric interpretation that is lower bounded by the Wasserstein distance~\cite{aurell2012refined,nakazato2021geometrical,van2021geometrical,van2023thermodynamic}, providing insights into the optimal control of the dissipation during finite-time processes~\cite{aurell2011optimal,muratore2013heat,muratore2014nanomechanical}. The above thermodynamic metric structures, defined on the probability manifold, make the derivation of efficiency-power trade-off~\cite{guarnieri2019thermodynamics,brandner2020thermodynamic,hayakawa2021geometrical,eglinton2022geometric} and the optimal protocol design~\cite{miller2019work,abiuso2020optimal,miller2020geometry,abiuso2020geometric,proesmans2020finite, chen2021extrapolating, alonso2021geometric, frim2022geometric, frim2022optimal, li2022geodesic,scandi2022minimally,abiuso2022thermodynamics,dago2022dynamics} straightforward. 

However, previous nonadiabatic results based on the metric structure are merely restricted to the analysis of average work or entropy production without temperature bias. This leads to little understanding of the generic transport behaviors in open systems with multiple reservoirs and strong nonequilibrium bias, let alone the transport fluctuations thereof. Therefore, important questions arise naturally: How to analyze general currents and fluctuations in nonadiabatic cyclic thermal devices? Can the nonadiabatic driven transport be characterized by a nonequilibrium thermodynamic metric structure? If so, what are the general constraints on transport fluctuations caused thereby?  

In this Letter, we solve the problems by formulating a geometric scheme of the generating function of currents, representing the nonadiabatic effects on each order of current moments as a metric term of full counting statistics. Based on this statistical metric structure of nonequilibrium transport, we derive geometric thermodynamic uncertainty relations (Geometic TURs) to constrain the current fluctuations under the near-adiabatic driving in terms of statistical metric quantities. Originally, TUR was proposed~\cite{barato2015thermodynamic} and proved theoretically~\cite{gingrich2016dissipation,gingrich2017inferring} and experimentally~\cite{pal2020experimental,yang2020phonon} within the steady states of classical Markovian dynamics, which bounds the precision of fluctuating current $Q$ in terms of the entropy production. TUR was subsequently generalized to the finite-time regime~\cite{pietzonka2017finite,liu2020thermodynamic,saryal2021bounds}, quantum systems endowed with coherence effects~\cite{hasegawa2020quantum,hasegawa2021thermodynamic,van2022thermodynamics}, setups with broken time-reversal symmetry~\cite{macieszczak2018unified,proesmans2019hysteretic}, and even systems with feedback controls~\cite{liu2020thermodynamic,potts2019thermodynamic}. Also, the well established fluctuation theorems~\cite{esposito2009nonequilibrium,campisi2011colloquium} prove the fluctuation theorem uncertainty relations~\cite{hasegawa2019fluctuation}. For reviews on TUR, see Ref.~\cite{horowitz2020thermodynamic}; for subsequent applications to the thermodynamic inference, see Refs.~\cite{seifert2019stochastic,cao2022effective,cao2022improved}. Importantly, Koyuk, Seifert and Pietzonka have derived a set of modified TURs, applicable to driven systems, by taking into account of the dependence of currents on the driving frequency~\cite{koyuk2018generalization,koyuk2019operationally,koyuk2020thermodynamic}.

%%%%%%%%%%%%%%%%%%%%%%%%%%%%%%%%%%%%%%%%%%%%%%
\begin{figure}[tp]
\centering
\includegraphics[width=\linewidth]{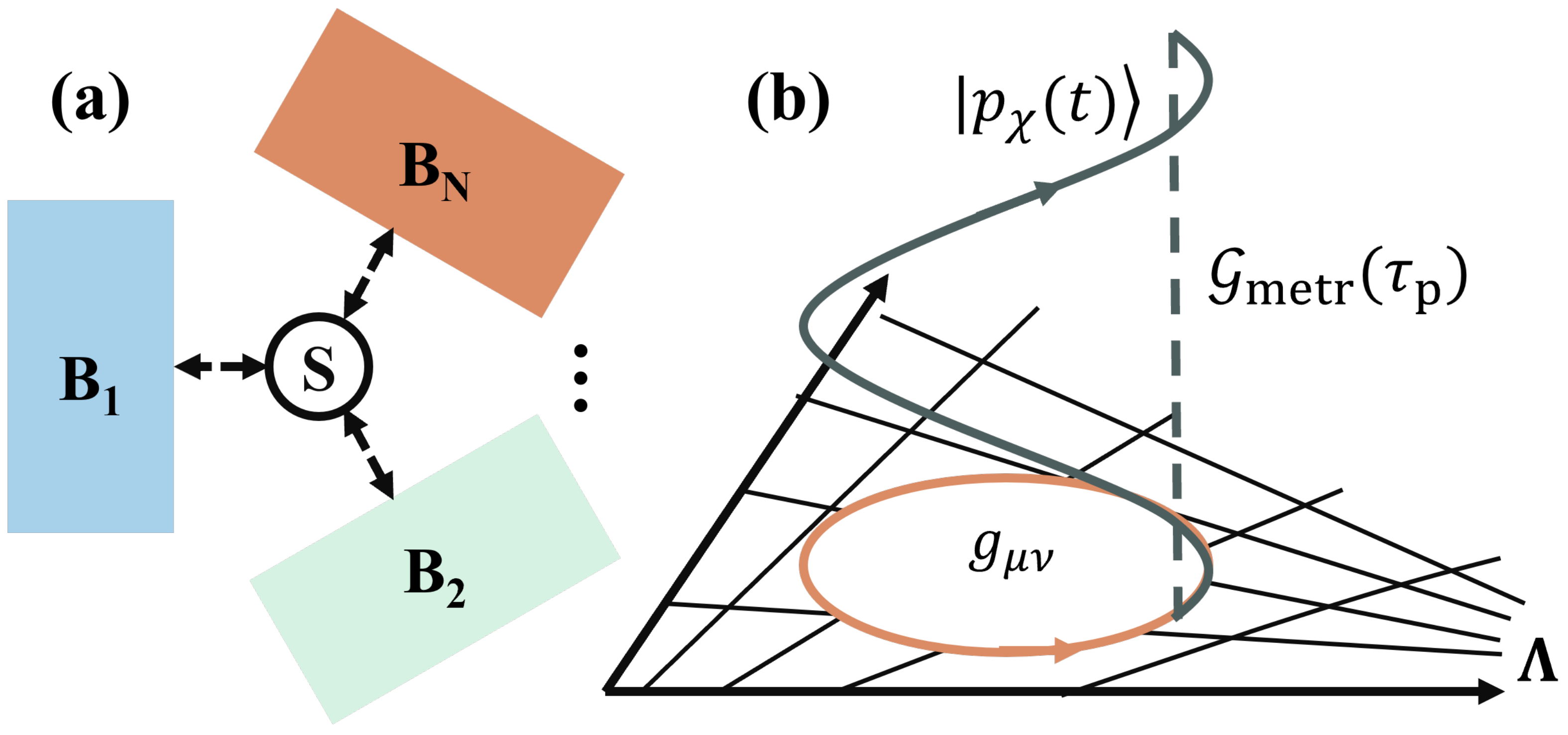}
\caption{{\bf The metric geometry in cyclically driven transport.} (a) The nonequilibrium cyclically driven system S coupled to multiple reservoirs $B_\nu$ ($\nu = 1, 2, ..., N$). (b) The metric structure of the cumulant generating function (CGF) ${\mathcal G}_{\rm metr}(\tau_{\rm p}) = \int_0^{\tau_{\rm p}} dt g_{\mu \nu} \dot{\Lambda}_\mu \dot{\Lambda}_\nu$ in the curved parameter space $\boldsymbol{\Lambda}$. The dashed line represents a metric CGF contribution ${\mathcal G}_{\rm metr}(t)$ to the twisted distribution $\ket{p_\chi(t)}$ and describes the fluctuations in the nonadiabatic regime with an arbitrary driving speed, with $\chi$ being an auxiliary counting parameter for the interested current. }
\label{fig0}
\end{figure}
%%%%%%%%%%%%%%%%%%%%%%%%%%%%%%%%%%%%%%%%%%%%%%%%%

Our results in this Letter advance the understanding of nonadiabatic geometric effects in parametrically driven thermal devices, which can be far from equilibrium. This allows for a study of fluctuating devices with various thermal functionalities regarding both average performance and fluctuation strength, paving the way towards designing precise thermal devices under non-equilibrium reservoirs and non-adiabatic cyclic modulations.

{\it Setups.}--We consider a cyclically driven open system coupled to multiple reservoirs $B_\nu$, which is schematically shown in Fig.~\ref{fig0}(a). The protocol parametrized as ${\boldsymbol \Lambda}(t+\tau_{\rm p})={\boldsymbol \Lambda}(t)$ forms a closed curve $\partial \Omega$ in the parameter space $\boldsymbol{\Lambda}$, with $\tau_{\rm p}$ being the driving period. 
Without loss of generality, we here take the discrete state case as an example. Similar discussions also apply for continuous cases. As such, the system distribution function is $\ket{p(t)} := (p_1, ...,p_N)^{\rm T}$, with $p_i(1 \leq i \leq N)$ describing the probability of occupying state $i$. The transition rate along $j \to i$ induced by the $\nu$-th reservoir is $k_{ij}^\nu$ ($i\neq j$), which can be time dependent under the protocol ${\boldsymbol \Lambda}(t)$. The master equation is thus written as $\partial_t \ket{p(t)} = \hat{L}(t) \ket{p(t)}$ with $L_{ij} = \sum_\nu (k_{ij}^\nu - \delta_{ij} \sum_{l \neq i}k_{li}^\nu)$ conserving the probability during transitions by $\sum_{i} L_{ij} = 0$.

To each transition path $k_{ij}^\nu$, we associate an increment of the accumulated current $\Delta Q = d_{ij}^\nu$~\cite{horowitz2020thermodynamic}. 
Stochastic exchanges between the system and reservoirs, like the current of particle number, heat, or work, are described by the antisymmetric tensor $d_{ij}^\nu = -d_{ji}^\nu$. While, the symmetric $d_{ij}^\nu=d_{ji}^\nu$ corresponds to time-reversal invariant quantities like dynamic activity ($d_{ij}^\nu=1$)~\cite{maes2020frenesy}. The evolution of the full counting statistics of accumulated currents can be considered by the twisted operator $\hat{L}_{\chi}$ with the counting field $\chi$:  
\begin{equation}
\label{eq:master eq}
\partial_t \ket{p_\chi(t)} = \hat{L}_\chi(t) \ket{p_\chi(t)},
\end{equation}
where the matrix elements are $L_{\chi, ij} = \sum_{\nu} k_{ij}^\nu e^{\chi d_{ij}^\nu}$ for $i \neq j$ and $L_{\chi, ii} = L_{ii}$. By defining the cumulant generating function (CGF) ${\mathcal G} := \ln {\mathcal Z} = \ln \braket{1|p_\chi(t)}$, the $n$-th cumulant of stochastic accumulated current $Q$ at time $t$ is obtained by taking the $n$-order derivative of CGF with respect to $\chi$, as $\braket{Q^n}_{\rm c} = \partial_\chi^n {\mathcal G}|_{\chi=0}$. Here, $\bra{1}$ is a vector with all elements being $1$ and ${\mathcal Z}$ is the moment generating function encoding each order of moments by $\braket{Q^n} = \partial_\chi^n {\mathcal Z}|_{\chi = 0}$. The non-Hermitian $\hat{L}_\chi$ can be decomposed into $\hat{L}_\chi = \sum_{n=0}^N E_n \ket{r_n}\bra{l_n}$, where the left and right eiegnvectors are biorthogonal $\braket{l_m|r_n} = \delta_{m,n}$ and $n=0$ corresponds to the unique steady state (we assume the ground state of $\hat{L}_\chi$ is nondegenerate). 
%Note that $\bra{l_0}_{\chi = 0} = \bra{1}$ , since $\sum_i L_{ij}=0$. 
For details of the dynamics of the twisted master equation, see Sec.~\uppercase\expandafter{\romannumeral1} of~\cite{supp}.

{\it Thermodynamic Geometry of Full Counting Statistics.}--Here, we sketch the derivation scheme of our most general geometric formulation. For derivation details, see Sec.~\uppercase\expandafter{\romannumeral2} of~\cite{supp}. Supposing that after several driving cycles, the system enters its cyclic state, satisfying the Floquet theorem 
\begin{equation}
\label{eq:floquet}
\ket{p_\chi(t)} = e^{{\mathcal G}(t)} \ket{\phi(t)}= e^{{\mathcal G}_{\rm dyn}(t) + {\mathcal G}_{\rm geo}(t)} \ket{\phi(t)}, 
\end{equation}
where $\ket{\phi(t+\tau_{\rm p})} = \ket{\phi(t)}$ is a cyclic state and $\ket{p_\chi(t)}$ only accumulates a CGF ${\mathcal G}(\tau_{\rm p}) = {\mathcal G}_{\rm dyn}(\tau_{\rm p}) + {\mathcal G}_{\rm geo}(\tau_{\rm p})$ during one driving period. It shows clearly that in addition to the dynamic-phase-like steady states contribution  ${\mathcal G}_{\rm dyn} (\tau_{\rm p}) := \int_0^{\tau_{\rm p}} dt E_0(t)$, there is a general geometric contribution 
\begin{equation}
\label{eq:general geometric}
{\mathcal G}_{\rm geo}(\tau_{\rm p}) = -\oint_{\partial \Omega} d\Lambda_\mu \braket{l_0|\partial_\mu \phi(t)}, 
\end{equation}
where we define $\partial_\mu := \partial_{\Lambda_\mu}$ for short. ${\mathcal G}_{\rm geo}$ is formally analogous to the Aharonov-Anandan phase in driven quantum systems~\cite{aharonov1987phase}, containing both the adiabatic and nonadiabatic effects. $\mathcal{G}_{\rm dyn}$ is simply an average over instantaneous steady states, while $\mathcal{G}_{\rm dyn}$ has no static analogues. Specifically, one can decompose the state $\ket{\phi}$ into the adiabatic and nonadiabatic components, which are respectively the instantaneous steady state $\ket{r_0}$ and the transverse states perpendicular to $\ket{r_0}$, as: 
\begin{equation}
\label{eq:dyson}
\ket{\phi(t)} =\ket{r_0(t)} + \hat{G} \ket{\partial_t \phi(t)},
\end{equation}
where the operator $\hat{G} := (\hat{L}_\chi - E_0)^+ (\hat 1-\ket{\phi}\bra{l_0})$, with the pseudo-inverse $(\hat{L}_\chi - E_0)^+=\sum_{n \neq 0} \frac{1}{E_n-E_0}\ket{r_n} \bra{l_n}$. The first term is the adiabatic trajectory and the second term signifies the non-adiabatic excitations. 

By substituting Eq.~(\ref{eq:dyson}) into Eq.~(\ref{eq:general geometric}), we find that the geometric CGF is generally divided into two parts: ${\mathcal G}_{\rm geo}={\mathcal G}_{\rm curv} + {\mathcal G}_{\rm metr}$. The first part is the adiabatic Berry-curvature-like CGF ${\mathcal G}_{\rm curv} = \oint_{\partial \Omega} d \Lambda_\mu A_\mu = \int_\Omega dS_{\mu \nu} F_{\mu \nu}$~\cite{ren2010berry,nie2020berry,wang2022diffusive,giri2017geometric,bhandari2020geometric,hino2021geometrical} with the geometric connection $A_\mu = -\braket{l_0|\partial_\mu r_0}$ and the antisymmetric curvature $F_{\mu \nu} = \braket{\partial_\nu l_0|\partial_\mu r_0} - \braket{\partial_\mu l_0|\partial_\nu r_0}$, governing the current statistics in the adiabatic regime. This regime is fertile in constructing precise and efficient adiabatic thermal machines~\cite{holubec2021fluctuations,barato2016cost,holubec2018cycling,miller2021thermodynamic}.

Of our prime interest is actually the second part, the nonadiabatic metric component: 
\begin{eqnarray}
\label{eq:nonadiabatic cgf}
{\mathcal G}_{\rm metr}(\tau_{\rm p}) &=& \int_0^{\tau_{\rm p}} g_{\mu \nu} \dot{\Lambda}_\mu \dot{\Lambda}_\nu dt, 
\\
\text{with} \;\; g_{\mu \nu} &:=& \frac{1}{2} \left [ \braket{\partial_\mu l_0|\hat{G}|\partial_\nu \phi} + \braket{\partial_\nu l_0|\hat{G}|\partial_\mu \phi} \right ], 
\nonumber
\end{eqnarray}
from which the full nonadiabatic effect on each order of fluctuation cumulants can be derived $\braket{Q_{\rm metr}^n}_{\rm c} := \partial_\chi^n {\mathcal G}_{\rm metr}|_{\chi = 0} = \int_0^{\tau_{\rm p}} g_{\mu \nu}^{Q^n} \dot{\Lambda}_\mu \dot{\Lambda}_\nu dt$, with the corresponding metric for cumulants being $g_{\mu \nu}^{Q^n} := \partial_\chi^n g_{\mu \nu}|_{\chi =0}$. This metric structure in CGF is illustrated in Fig.~\ref{fig0}(b). In contrast to the time-antisymmetric ${\mathcal G}_{\rm curv}$ that reverses upon time reversal, the time-symmetric metric tensor (also symmetric in the sense of $g_{\mu \nu} = g_{\nu \mu}$) indicates that the nonadiabatic component ${\mathcal G}_{\rm metr}$ provides a time-reversal invariant contribution of each current and the corresponding fluctuations. We note that although Eq.~(\ref{eq:nonadiabatic cgf}) is merely a formal solution, our following concrete results follow from it. 

The statistical metric Eq.~(\ref{eq:nonadiabatic cgf}) describes the full non-adiabatic effect on arbitrary transport fluctuations. In the near-adiabatic regime, the state $\ket{\phi(t)}$ reduces to $\ket{r_0(t)}$ and the metric simplifies to the leading order of nonadiabaticity, as
\begin{equation}
\label{eq:near_eq_cgf}
\mathfrak{g}_{\mu \nu} = \sum_{n \neq 0} \frac{\braket{\partial_\mu l_0|r_n}\braket{l_n|\partial_\nu r_0}+(\mu \leftrightarrow  \nu)}{2(E_n - E_0)}, 
\end{equation}
which describes the near-adiabatic currents and fluctuations. Here, $(\mu \leftrightarrow  \nu)$ means interchanging indices. Previous works on the thermodynamic geometry can be derived by restricting to this near-equilibrium regime and consider only the average entropy production~\cite{crooks2007measuring,sivak2012thermodynamic,brandner2020thermodynamic} or its fluctuation~\cite{miller2020geometry} in setups with a single reservoir.

It is worth noting that, in the geometry of optimal transport, the cost of changing between distributions, i.e., the average entropy production, is determined by other metrics on the probability manifold~\cite{van2021geometrical,nakazato2021geometrical,van2023thermodynamic}, both for the overdamped~\cite{aurell2011optimal,aurell2012refined}, underdamped Brownian~\cite{muratore2014nanomechanical} and discrete master equation case~\cite{muratore2013heat,van2023thermodynamic}. The minimization of average entropy production naturally reduces to finding the geodesic between initial and final distributions, whose length is bounded from below by the Wasserstein distance~\cite{aurell2012refined,van2023thermodynamic}, leading to the optimal Landauer erasure~\cite{aurell2012refined,proesmans2020finite,dago2022dynamics}. Distinct from the above regime, the metric Eq.~(\ref{eq:nonadiabatic cgf}) here works in the geometry of parameter space, which is valid for any currents and fluctuations of interest under cyclic parametric driving. In what follows, we will discuss implications of the statistical metric of CGF on average currents and fluctuations, separately.

{\it Metric Structure and Average Currents.}--Here, we consider consequences of the CGF metric on the average current. Details of calculation are summarized in Sec.~\uppercase\expandafter{\romannumeral3} of~\cite{supp}. We show that the average current during one period is $\braket{Q}:=\partial_\chi{\mathcal G}(\tau_{\rm p})|_{\chi=0} = \int_0^{\tau_{\rm p}} dt \braket{1|\hat{J}(t)|p(t)}$, with the current operator being $\hat{J}:= \partial_\chi \hat{L}_\chi|_{\chi=0}$ and the cyclic distribution being $\ket{p(t)} = \ket{p_\chi(t)}|_{\chi=0}$. Particularly, we derive the nonadiabatic metric structure for current $Q$, as:
\begin{eqnarray}
\label{eq:avg nonadiabatic}
\braket{Q_{\rm metr}} &:=&\left.\partial_\chi {\mathcal G}_{\rm metr}(\tau_{\rm p})\right|_{\chi=0}= \int_0^{\tau_{\rm p}} g_{\mu \nu}^{Q} \dot{\Lambda}_\mu \dot{\Lambda}_\nu dt, 
\\
\text{with}\;\; g_{\mu \nu}^Q &=& \frac{1}{2} \left [\bra{1}\hat{J} \hat{L}^+ \partial_\mu (\hat{L}^+ \ket{\partial_\nu p})+(\mu \leftrightarrow \nu) \right],  \nonumber
\end{eqnarray}
where $g_{\mu \nu}^Q = \partial_\chi g_{\mu \nu}|_{\chi =0}$ is a symmetric metric with respect to the average accumulated current $\braket{Q}$ and $\hat{L}^+$ is the pseudo-inverse of $\hat{L}_{\chi}|_{\chi=0}$. It is worth noting that Eq.~(\ref{eq:avg nonadiabatic}) works for arbitrary nonadiabatic driving speed. If one replaces $\ket{p}$ by the instantaneous steady states $\ket{\pi}$ in Eq.~(\ref{eq:avg nonadiabatic}), one will enter in the near-adiabatic regime and obtain the corresponding metric $\mathfrak{g}_{\mu \nu}^Q = \left [\bra{1}\hat{J} \hat{L}^+ \partial_\mu (\hat{L}^+ \ket{\partial_\nu \pi})+(\mu \leftrightarrow \nu) \right]/2$, which describes the leading order finite-time effect.

Here, we discuss the application of Eq.~(\ref{eq:avg nonadiabatic}) to thermodynamic optimization. The metric of total entropy production can be expressed as $\tilde{g}_{tt} = g_{tt}^\Sigma +\partial_t(\sigma_p - \sigma_\pi)$, where $g_{tt}^\Sigma$ describes the reservoir entropy production due to heat currents and $\partial_t(\sigma_p - \sigma_\pi)$ is the system entropy production rate, with $\sigma_p = -\sum_n p_n \ln p_n$. Note that over one whole period, $\braket{\Sigma} := \int_0^{\tau_{\rm p}} dt g_{tt}^\Sigma=\int_0^{\tau_{\rm p}} dt \tilde{g}_{tt}$ since $\ket{p}$ and $\ket{\pi}$ is cyclic in time. The positivity of total entropy production guarantees $\tilde{g}_{tt}>0$, which allows us to obtain a thermodynamic speed limit $\tau_{\rm p} \geq L^2/\braket{\Sigma}$, bounding the system evolution speed with entropy production and non-equilibrium thermodynamic length $L = \int_0^{\tau_{\rm p}} dt \sqrt{\tilde{g}_{tt}}$. The equality is obtained when the entropy production rate is constant and this endows us an entropy minimization principle $\partial_t \tilde{g}_{tt}=0$.

We note that the pseudo-Riemannian-metric $g_{\mu \nu}^Q$ is not promised to be positive-definite. Nevertheless, this "non-positive definite" sacrifice allows us to generalize the previous thermodynamic geometry framework to non-equilibrium transports for generic currents in finite-time driving regimes. For example, in the near-adiabatic regime, we can obtain the vector field in the parameter space along which the non-adiabatic pump current vanishes $\mathfrak{g}_{\mu \nu}^Q \dot{\Lambda}^\mu \dot{\Lambda}^\nu=0$. This provides us a geometric view point of the non-adiabatic control over the time-dependent pump effect. For details of this optimization principle, see Sec.~\uppercase\expandafter{\romannumeral4} of~\cite{supp}. The concise average current expression Eq.~(\ref{eq:avg nonadiabatic}) is simply a consequence of our general result Eq.~(\ref{eq:nonadiabatic cgf}), which generally encodes the statistical information of each order fluctuation cumulants $\braket{Q_{\rm metr}^n}_{\rm c}$. 

%%%%%%%%%%%%%%%%%%%%%%%%%%%%%%%%%%%%%%%%%%%%%%
\begin{figure}[tp]
\centering
\includegraphics[width=\linewidth]{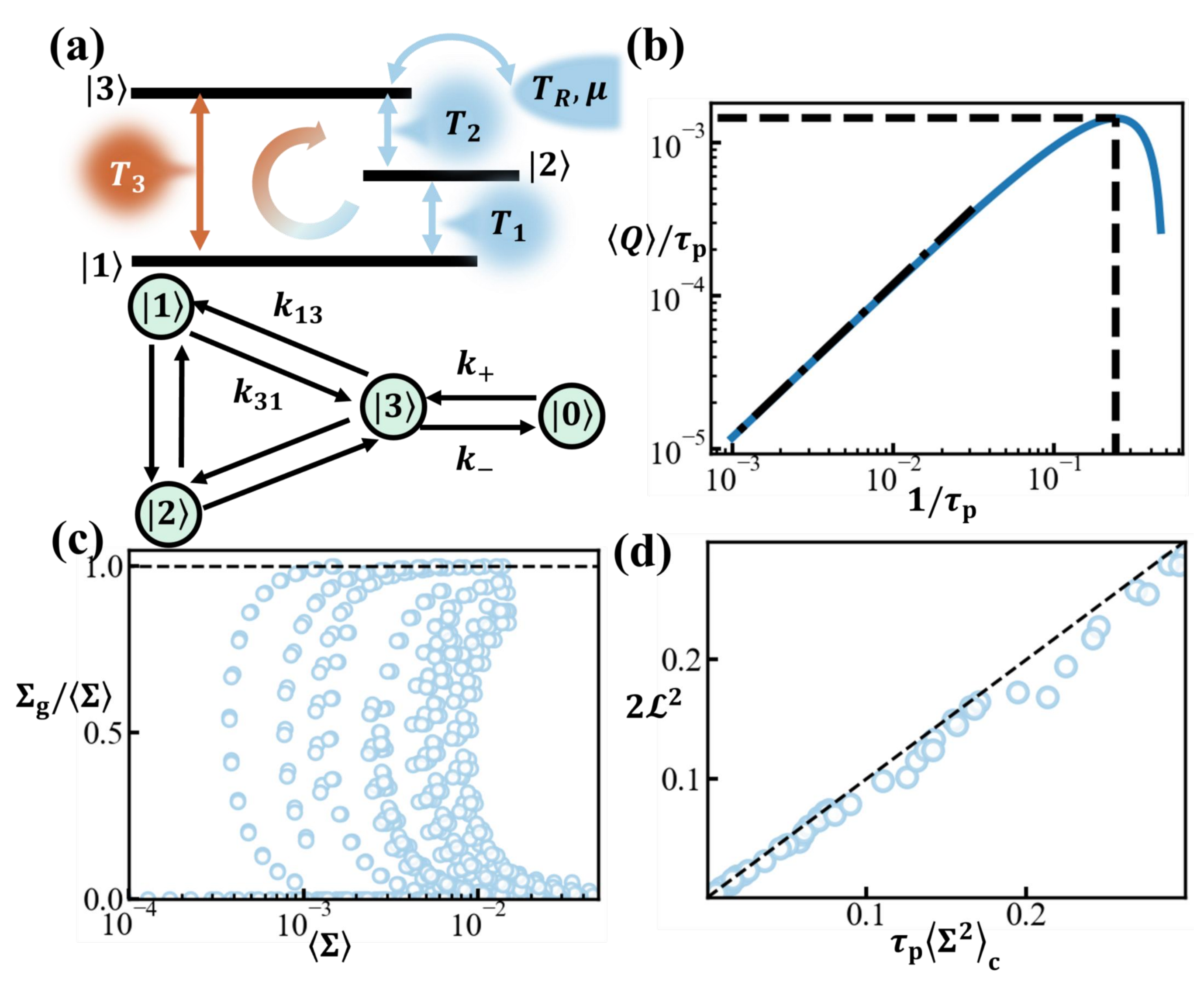}
\caption{{\bf The nonequilibrium quantum tricycle model with energy levels $\epsilon_n$ of quantum dots being driven.} (a) The system setup and its transition graph. Three quantum dots with tunable energy levels are mediated by three thermal photonic/phononic reservoirs. The level $\ket{3}$ is in addition coupled to an electron reservoir. (b) The nonadiabatic average heat flux versus the inverse period ( $\phi=2\pi/3$ and $T_1=T_2=T_3=T_R$). The dot-dash line is for the adiabatic component and the dash line is for the optimal period. (c) The geometric TUR ($\Sigma_{\rm g} := 2(\braket{Q_{\rm dyn}}-\braket{Q_{\rm metr}})^2/\braket{Q^2}_c$) is verified. (d) The geometric bound on the fluctuation of entropy production $\braket{\Sigma^2}_{\rm c} \tau_{\rm p} \geq 2 {\mathcal L}^2$ is verified. } 
\label{fig_qd}
\end{figure}

%%%%%%%%%%%%%%%%%%%%%%%%%%%%%%%%%%%%%%%%%%%%%%%%%

{\it Geometric TURs and Fluctuations.}--Here, by restricting to the near-adiabatic regime, we show that the fluctuations encoded by Eq.~(\ref{eq:near_eq_cgf}) are constrained by a kind of Geometric TURs, wherein the two geometric terms originated from ${\mathcal G}_{\rm geo}={\mathcal G}_{\rm curv} + {\mathcal G}_{\rm metr}$ play a central role. Based on the fluctuation-response inequality~\cite{dechant2020fluctuation}, which is a nonlinear generalization of the Cramer-Rao bound, we obtain the Geometric TURs (see Sec.~\uppercase\expandafter{\romannumeral5} of~\cite{supp} for details), as:
\begin{equation}
\label{eq:gtur}
\braket{\Sigma} \geq 2\frac{(\braket{Q_{\rm dyn}}-\braket{Q_{\rm metr}})^2}{\braket{Q^2}_{\rm c}} := \Sigma_{\rm g}, 
\end{equation}
where $\Sigma$ is the entropy production during one driving period, $Q_{\rm dyn}$ and $Q_{\rm metr}$ are respectively the dynamic and nonadiabatic metric components of an arbitrary time-antisymmetric current (that can be particle number, heat, or work). Both the variance $\braket{Q^2}_{\rm c}$ and entropy production $\braket{\Sigma}$ contain contributions of the dynamic, adiabatic curvature and nonadiabatic metric origins. Eq.~(\ref{eq:gtur}), consistent with Ref.~\cite{koyuk2019operationally}, generalizes the adiabatic limit results in a thermoelectric heat engine~\cite{lu2022geometric}. It clearly unveils the role played by the near-adiabatic metric structure and paves the way towards geometric inference and optimization.

Now let us show some direct consequences of the Geometric TURs on the near-adiabatic but finite-time processes.  
If the reservoirs are instantaneously isothermal with each other, the dynamic components vanish in the sense of mean values $\braket{Q_{\rm dyn}}=\braket{\Sigma_{\rm dyn}}=0$, but not necessary for the fluctuation $\braket{Q^2_{\rm dyn}}_{\rm c}$ of an arbitrary current. Meanwhile, $\braket{\Sigma_{\rm curv}}=0$ due to the vanishing quasistatic entropy production. By rewriting Eq.~(\ref{eq:gtur}) as 
$\braket{Q^2}_{\rm c}\geq 2 (\braket{Q_{\rm dyn}}-\braket{Q_{\rm metr}})^2/(\braket{\Sigma_{\rm dyn}}+\braket{\Sigma_{\rm curv}}+\braket{\Sigma_{\rm metr}})$, we can obtain a geometric bound for the current fluctuation
\begin{equation}
\label{eq:tur no dyn}
\braket{Q^2}_{\rm c} \geq 2\frac{\braket{Q_{\rm metr}}^2}{\braket{\Sigma_{\rm metr}}}, 
\end{equation}
which becomes tighter for faster drivings. By taking the entropy production as the current ($Q:=\Sigma$) and considering the positive-definiteness of $g_{\mu \nu}^\Sigma$, we can bound the fluctuation of entropy production $\Sigma$ by the thermodynamic length ${\mathcal L}$, as:
\begin{equation}
\label{eq:entropy variance bound}
\braket{\Sigma^2}_{\rm c} \geq \frac{2 {\mathcal L}^2}{\tau_{\rm p}}, 
\end{equation}
where ${\mathcal L} := \oint_{\partial\Omega} \sqrt{\mathfrak{g}_{\mu \nu}^\Sigma d \Lambda_\mu d \Lambda_\nu}$ is a geometric quantity independent of the parametrization of protocol. Here, we use both Eq.~(\ref{eq:tur no dyn}) and the Cauchy-Schwarz inequality~\cite{crooks2007measuring,brandner2020thermodynamic}: $\braket{\Sigma^2}_{\rm c} \geq 2\braket{\Sigma_{\rm metr}} = 2\int_0^{\tau_{\rm p}} \mathfrak{g}_{\mu \nu}^{\Sigma} \dot{\Lambda}_\mu \dot{\Lambda}_\nu dt \geq 2{\mathcal L}^2/\tau_{\rm p}$. This result can be understood as a kind of fluctuation-dissipation inequality. The geometric bound Eq.~(\ref{eq:entropy variance bound}) connects the entropy production fluctuation in near-equilibrium finite-time processes to previously defined thermodynamic length~\cite{salamon1983thermodynamic,crooks2007measuring}, providing a basis for inferring the statistical distribution of entropy production in cyclically driven processes. 

In the following, we will validate the metric structure Eq.~(\ref{eq:near_eq_cgf}) and the Geometric TURs [Eq.~(\ref{eq:gtur}) and~(\ref{eq:entropy variance bound})] using two examples.

{\it Discrete Master Equation System.}--Our first model is the nonequilibrium quantum tricycle generating the chiral current by the cyclic driving, illustrated in Fig.~\ref{fig_qd}(a), which is inspired by the classical stochastic pump model~\cite{astumian2002brownian,rahav2008directed} and steady-state continuous thermal devices~\cite{kosloff2014quantum}. The system Hamiltonian $\hat{H} = \hat{H}_{S} + \hat{H}_{R} + \hat{H}_{SR}+\hat{H}_{B}+\hat{H}_{SB}$ is composed of the three quantum dot levels $\hat{H}_S = \sum_{n = 1}^3 \epsilon_n \hat{c}_n^\dagger \hat{c}_n$, the electron reservoirs $\hat{H}_R = \sum_{k} \epsilon_{k} \hat{d}_{k}^\dagger \hat{d}_{k}$, the tunneling term $\hat{H}_{SR} = \sum_{k} t_{k} (\hat{d}_{k}^\dagger \hat{c}_3 + \hat{c}_3^\dagger \hat{d}_{k})$, the Bosonic thermal reservoir $\hat{H}_{B} = \sum_{\nu=1;k}^{\nu=3} \epsilon_{\nu, k} \hat{a}_{\nu,k}^\dagger \hat{a}_{\nu,k}$, and the system-reservoir coupling term $\hat{H}_{SB} = \sum_{\nu=1;k}^{\nu=3} r_{\nu,k} (\hat{a}_{\nu,k} +\hat{a}_{\nu,k}^\dagger)(\hat{c}_{\nu}^\dagger \hat{c}_{\nu+1} + \hat{c}_{\nu+1}^\dagger \hat{c}_{\nu})$. Here, $\nu = 4$ denotes the same site as $\nu=1$. $\hat{H}_{SB}$ mediates the transitions between quantum dots $\ket{i}$ and $\ket{i+1}$ ($1 \leq i \leq 3$) and $\hat{H}_{SR}$ enables electrons to hop into (out of) the system through the transition $\ket{0} \to \ket{3}$ ($\ket{3} \to \ket{0}$). We restrict ourselves to the Coulomb blockade and the weak coupling regime. For the twisted master equation and driving protocols, see Sec.~\uppercase\expandafter{\romannumeral6}A in~\cite{supp}. 

By driving the energy level of quantum dots $\epsilon_n$ out of phase, e.g. $\epsilon_n(t) = \epsilon_n^0 + \delta \sin[2\pi t/ \tau_{\rm p} + (n-1) \phi]$ with $\delta$ being the driving amplitude, we realize the driving induced chiral current even in the absence of biases. As shown in Fig.~\ref{fig_qd}(b), $\braket{Q}/\tau_{\rm p}$ is decreased by the nonadiabatic effect and reaches its maximum $-\braket{Q_{\rm curv}}^2/(4\tau_{\rm p} \braket{Q_{\rm metr}})$ at the optimal period $-2\tau_{\rm p} \braket{Q_{\rm metr}}/\braket{Q_{\rm curv}}$ as denoted by the dashed lines. In contrast to the nonzero pumping, by merely driving the well depth (energy level) of the classical analog satisfying the Arrhenius law of transition rate, the chiral current is prohibited by the no-pumping theorem~\cite{rahav2008directed,chernyak2008pumping}. As shown in Fig.~\ref{fig_qd}(c), the average entropy production can be bounded and inferred by the chiral current fluctuations, satisfying Eq.~(\ref{eq:gtur}). Also, as shown in Fig.~\ref{fig_qd}(d), the fluctuation of the entropy production itself is bounded from left by the thermodynamic length, validating the geometric bound Eq.~(\ref{eq:entropy variance bound}).

{\it Continuous Brownian System.}--Here, we show that Eq.~(\ref{eq:entropy variance bound}) can be saturated in a Brownian heat pump engine. We consider two linearly coupled harmonic oscillators between two reservoirs of temperature $T_i$~\cite{ren2012geometric}. The Langevin dynamics is $\Gamma \dot{\bf x} =  K {\bf x} + {\boldsymbol \xi}(t)$, where ${\bf x} = (x_1, x_2)^{\rm T}$ is the oscillators' position and ${\boldsymbol \xi} = (\xi_1, \xi_2)^{\rm T}$ is a vector of independent Gaussian white noise satisfying $\braket{\xi_i}=0, \braket{\xi_i(t_1) \xi_j(t_2)} = 2\gamma_i T_i \delta_{ij}\delta(t_1-t_2)$. The viscosity and stiffness matrices are
$\Gamma = \binom{\gamma_1, 0}{0, \gamma_2}, K = k\binom{-1,1}{1,-1}$.  

By analytical calculation, when $\boldsymbol{\Lambda} = (k, \gamma_1)^{\rm T}$ is driven, the metric for the average entropy production $\mathfrak{g}_{\mu \nu}^\Sigma$ and entropy variance $\mathfrak{g}_{\mu \nu}^{\Sigma^2}$ in the isothermal case ($T_1=T_2$) satisfies $\mathfrak{g}_{\mu \nu}^{\Sigma^2} = 2\mathfrak{g}_{\mu \nu}^{\Sigma}$. Our bound Eq.~(\ref{eq:entropy variance bound}) is saturable by reparametrizing the protocol in terms of the thermodynamic length, i.e., the time spent around a parameter point being $dt = (\tau_{\rm p}/{\mathcal L}) \sqrt{\mathfrak{g}_{\mu \nu}^{\Sigma} d\Lambda_\mu d\Lambda_\nu}$~\cite{crooks2007measuring,brandner2020thermodynamic}. For details, see the Sec.~\uppercase\expandafter{\romannumeral6}B of~\cite{supp}.

{\it Summary.}--
We have proposed a general framework of thermodynamic geometry in terms of full counting statistics for analyzing the transport fluctuations in nonequilibrium driven systems. Our theory can study the fluctuation properties of arbitrary currents among multiple reservoirs under finite-time modulations. As an illustration, we have proved and validated the geometric TURs, relating the current fluctuations and entropy production in near-adiabatically driven systems. We have verified the results in a quantum chiral transport and Brownian heat pump, both analytically and numerically. This geometry framework can be readily adopted to study the effect of quantum phenomena (like quantum coherence~\cite{brandner2017universal,ptaszynski2018coherence,camati2019coherence}, squeezing~\cite{schaller2014open,wang2021nonequilibrium}) on the performance and TUR of heat engines in the finite-time regime. Also, deriving optimal protocols with minimal fluctuations under cyclic parametric driving with arbitrary speed is an important future direction. 

%%%%%%%%%%%%%%%%%%%%%%%%%%%%%%%%%%%%%%%%%%%%%%%%%%%%%

\begin{acknowledgments}

%%%%%%%%%%%%%%%%%%%%%%%

We acknowledge the support from the National Natural Science Foundation of China (No. 11935010), the Natural Science Foundation of Shanghai (Grant Nos. 23ZR1481200 and 23XD1423800), and the Opening Project of Shanghai Key Laboratory of Special Artificial Microstructure Materials and Technology.
\end{acknowledgments}

%%%%%%%%%%%%%%%%%%%%%%%%%%%%

\bibliography{ref.bib}

\begin{thebibliography}{98}%
\makeatletter
\providecommand \@ifxundefined [1]{%
 \@ifx{#1\undefined}
}%
\providecommand \@ifnum [1]{%
 \ifnum #1\expandafter \@firstoftwo
 \else \expandafter \@secondoftwo
 \fi
}%
\providecommand \@ifx [1]{%
 \ifx #1\expandafter \@firstoftwo
 \else \expandafter \@secondoftwo
 \fi
}%
\providecommand \natexlab [1]{#1}%
\providecommand \enquote  [1]{``#1''}%
\providecommand \bibnamefont  [1]{#1}%
\providecommand \bibfnamefont [1]{#1}%
\providecommand \citenamefont [1]{#1}%
\providecommand \href@noop [0]{\@secondoftwo}%
\providecommand \href [0]{\begingroup \@sanitize@url \@href}%
\providecommand \@href[1]{\@@startlink{#1}\@@href}%
\providecommand \@@href[1]{\endgroup#1\@@endlink}%
\providecommand \@sanitize@url [0]{\catcode `\\12\catcode `\$12\catcode
  `\&12\catcode `\#12\catcode `\^12\catcode `\_12\catcode `\%12\relax}%
\providecommand \@@startlink[1]{}%
\providecommand \@@endlink[0]{}%
\providecommand \url  [0]{\begingroup\@sanitize@url \@url }%
\providecommand \@url [1]{\endgroup\@href {#1}{\urlprefix }}%
\providecommand \urlprefix  [0]{URL }%
\providecommand \Eprint [0]{\href }%
\providecommand \doibase [0]{http://dx.doi.org/}%
\providecommand \selectlanguage [0]{\@gobble}%
\providecommand \bibinfo  [0]{\@secondoftwo}%
\providecommand \bibfield  [0]{\@secondoftwo}%
\providecommand \translation [1]{[#1]}%
\providecommand \BibitemOpen [0]{}%
\providecommand \bibitemStop [0]{}%
\providecommand \bibitemNoStop [0]{.\EOS\space}%
\providecommand \EOS [0]{\spacefactor3000\relax}%
\providecommand \BibitemShut  [1]{\csname bibitem#1\endcsname}%
\let\auto@bib@innerbib\@empty
%</preamble>
\bibitem [{\citenamefont {Blickle}\ and\ \citenamefont
  {Bechinger}(2012)}]{blickle2012realization}%
  \BibitemOpen
  \bibfield  {author} {\bibinfo {author} {\bibfnamefont {Valentin}\
  \bibnamefont {Blickle}}\ and\ \bibinfo {author} {\bibfnamefont {Clemens}\
  \bibnamefont {Bechinger}},\ }\bibfield  {title} {\enquote {\bibinfo {title}
  {Realization of a micrometre-sized stochastic heat engine},}\ }\href
  {\doibase 10.1038/nphys2163} {\bibfield  {journal} {\bibinfo  {journal} {Nat.
  Phys.}\ }\textbf {\bibinfo {volume} {8}},\ \bibinfo {pages} {143--146}
  (\bibinfo {year} {2012})}\BibitemShut {NoStop}%
\bibitem [{\citenamefont {Ro\ss{}nagel}\ \emph {et~al.}(2014)\citenamefont
  {Ro\ss{}nagel}, \citenamefont {Abah}, \citenamefont {Schmidt-Kaler},
  \citenamefont {Singer},\ and\ \citenamefont {Lutz}}]{rossnagel2014nanoscale}%
  \BibitemOpen
  \bibfield  {author} {\bibinfo {author} {\bibfnamefont {J.}~\bibnamefont
  {Ro\ss{}nagel}}, \bibinfo {author} {\bibfnamefont {O.}~\bibnamefont {Abah}},
  \bibinfo {author} {\bibfnamefont {F.}~\bibnamefont {Schmidt-Kaler}}, \bibinfo
  {author} {\bibfnamefont {K.}~\bibnamefont {Singer}}, \ and\ \bibinfo {author}
  {\bibfnamefont {E.}~\bibnamefont {Lutz}},\ }\bibfield  {title} {\enquote
  {\bibinfo {title} {Nanoscale heat engine beyond the carnot limit},}\ }\href
  {\doibase 10.1103/PhysRevLett.112.030602} {\bibfield  {journal} {\bibinfo
  {journal} {Phys. Rev. Lett.}\ }\textbf {\bibinfo {volume} {112}},\ \bibinfo
  {pages} {030602} (\bibinfo {year} {2014})}\BibitemShut {NoStop}%
\bibitem [{\citenamefont {Mart{\'\i}nez}\ \emph
  {et~al.}(2016{\natexlab{a}})\citenamefont {Mart{\'\i}nez}, \citenamefont
  {Rold{\'a}n}, \citenamefont {Dinis}, \citenamefont {Petrov}, \citenamefont
  {Parrondo},\ and\ \citenamefont {Rica}}]{martinez2016brownian}%
  \BibitemOpen
  \bibfield  {author} {\bibinfo {author} {\bibfnamefont {Ignacio~A}\
  \bibnamefont {Mart{\'\i}nez}}, \bibinfo {author} {\bibfnamefont {{\'E}dgar}\
  \bibnamefont {Rold{\'a}n}}, \bibinfo {author} {\bibfnamefont {Luis}\
  \bibnamefont {Dinis}}, \bibinfo {author} {\bibfnamefont {Dmitri}\
  \bibnamefont {Petrov}}, \bibinfo {author} {\bibfnamefont {Juan~MR}\
  \bibnamefont {Parrondo}}, \ and\ \bibinfo {author} {\bibfnamefont
  {Ra{\'u}l~A}\ \bibnamefont {Rica}},\ }\bibfield  {title} {\enquote {\bibinfo
  {title} {Brownian carnot engine},}\ }\href {\doibase 10.1038/nphys3518}
  {\bibfield  {journal} {\bibinfo  {journal} {Nat. Phys.}\ }\textbf {\bibinfo
  {volume} {12}},\ \bibinfo {pages} {67--70} (\bibinfo {year}
  {2016}{\natexlab{a}})}\BibitemShut {NoStop}%
\bibitem [{\citenamefont {Mart{\'\i}nez}\ \emph
  {et~al.}(2016{\natexlab{b}})\citenamefont {Mart{\'\i}nez}, \citenamefont
  {Petrosyan}, \citenamefont {Gu{\'e}ry-Odelin}, \citenamefont {Trizac},\ and\
  \citenamefont {Ciliberto}}]{martinez2016engineered}%
  \BibitemOpen
  \bibfield  {author} {\bibinfo {author} {\bibfnamefont {Ignacio~A}\
  \bibnamefont {Mart{\'\i}nez}}, \bibinfo {author} {\bibfnamefont {Artyom}\
  \bibnamefont {Petrosyan}}, \bibinfo {author} {\bibfnamefont {David}\
  \bibnamefont {Gu{\'e}ry-Odelin}}, \bibinfo {author} {\bibfnamefont
  {Emmanuel}\ \bibnamefont {Trizac}}, \ and\ \bibinfo {author} {\bibfnamefont
  {Sergio}\ \bibnamefont {Ciliberto}},\ }\bibfield  {title} {\enquote {\bibinfo
  {title} {Engineered swift equilibration of a brownian particle},}\ }\href
  {\doibase 10.1038/nphys3758} {\bibfield  {journal} {\bibinfo  {journal} {Nat.
  Phys.}\ }\textbf {\bibinfo {volume} {12}},\ \bibinfo {pages} {843--846}
  (\bibinfo {year} {2016}{\natexlab{b}})}\BibitemShut {NoStop}%
\bibitem [{\citenamefont {Josefsson}\ \emph {et~al.}(2018)\citenamefont
  {Josefsson}, \citenamefont {Svilans}, \citenamefont {Burke}, \citenamefont
  {Hoffmann}, \citenamefont {Fahlvik}, \citenamefont {Thelander}, \citenamefont
  {Leijnse},\ and\ \citenamefont {Linke}}]{josefsson2018quantum}%
  \BibitemOpen
  \bibfield  {author} {\bibinfo {author} {\bibfnamefont {Martin}\ \bibnamefont
  {Josefsson}}, \bibinfo {author} {\bibfnamefont {Artis}\ \bibnamefont
  {Svilans}}, \bibinfo {author} {\bibfnamefont {Adam~M}\ \bibnamefont {Burke}},
  \bibinfo {author} {\bibfnamefont {Eric~A}\ \bibnamefont {Hoffmann}}, \bibinfo
  {author} {\bibfnamefont {Sofia}\ \bibnamefont {Fahlvik}}, \bibinfo {author}
  {\bibfnamefont {Claes}\ \bibnamefont {Thelander}}, \bibinfo {author}
  {\bibfnamefont {Martin}\ \bibnamefont {Leijnse}}, \ and\ \bibinfo {author}
  {\bibfnamefont {Heiner}\ \bibnamefont {Linke}},\ }\bibfield  {title}
  {\enquote {\bibinfo {title} {A quantum-dot heat engine operating close to the
  thermodynamic efficiency limits},}\ }\href {\doibase
  10.1038/s41565-018-0200-5} {\bibfield  {journal} {\bibinfo  {journal} {Nat.
  Nanotechnol.}\ }\textbf {\bibinfo {volume} {13}},\ \bibinfo {pages}
  {920--924} (\bibinfo {year} {2018})}\BibitemShut {NoStop}%
\bibitem [{\citenamefont {Talkner}\ and\ \citenamefont
  {H\"anggi}(2020)}]{talkner2020colloquium}%
  \BibitemOpen
  \bibfield  {author} {\bibinfo {author} {\bibfnamefont {Peter}\ \bibnamefont
  {Talkner}}\ and\ \bibinfo {author} {\bibfnamefont {Peter}\ \bibnamefont
  {H\"anggi}},\ }\bibfield  {title} {\enquote {\bibinfo {title} {Colloquium:
  Statistical mechanics and thermodynamics at strong coupling: Quantum and
  classical},}\ }\href {\doibase 10.1103/RevModPhys.92.041002} {\bibfield
  {journal} {\bibinfo  {journal} {Rev. Mod. Phys.}\ }\textbf {\bibinfo {volume}
  {92}},\ \bibinfo {pages} {041002} (\bibinfo {year} {2020})}\BibitemShut
  {NoStop}%
\bibitem [{\citenamefont {Wang}\ \emph
  {et~al.}(2022{\natexlab{a}})\citenamefont {Wang}, \citenamefont {Wang},
  \citenamefont {Lu},\ and\ \citenamefont {Jiang}}]{wang2022inelastic}%
  \BibitemOpen
  \bibfield  {author} {\bibinfo {author} {\bibfnamefont {Rongqian}\
  \bibnamefont {Wang}}, \bibinfo {author} {\bibfnamefont {Chen}\ \bibnamefont
  {Wang}}, \bibinfo {author} {\bibfnamefont {Jincheng}\ \bibnamefont {Lu}}, \
  and\ \bibinfo {author} {\bibfnamefont {Jian-Hua}\ \bibnamefont {Jiang}},\
  }\bibfield  {title} {\enquote {\bibinfo {title} {Inelastic thermoelectric
  transport and fluctuations in mesoscopic systems},}\ }\href {\doibase
  10.1080/23746149.2022.2082317} {\bibfield  {journal} {\bibinfo  {journal}
  {Adv. Phys.-X}\ }\textbf {\bibinfo {volume} {7}},\ \bibinfo {pages} {2082317}
  (\bibinfo {year} {2022}{\natexlab{a}})}\BibitemShut {NoStop}%
\bibitem [{\citenamefont {Saha}\ \emph {et~al.}(2021)\citenamefont {Saha},
  \citenamefont {Lucero}, \citenamefont {Ehrich}, \citenamefont {Sivak},\ and\
  \citenamefont {Bechhoefer}}]{saha2021maximizing}%
  \BibitemOpen
  \bibfield  {author} {\bibinfo {author} {\bibfnamefont {Tushar~K}\
  \bibnamefont {Saha}}, \bibinfo {author} {\bibfnamefont {Joseph~NE}\
  \bibnamefont {Lucero}}, \bibinfo {author} {\bibfnamefont {Jannik}\
  \bibnamefont {Ehrich}}, \bibinfo {author} {\bibfnamefont {David~A}\
  \bibnamefont {Sivak}}, \ and\ \bibinfo {author} {\bibfnamefont {John}\
  \bibnamefont {Bechhoefer}},\ }\bibfield  {title} {\enquote {\bibinfo {title}
  {Maximizing power and velocity of an information engine},}\ }\href
  {https://www.pnas.org/content/118/20/e2023356118.short} {\bibfield  {journal}
  {\bibinfo  {journal} {Proc. Natl. Acad. Sci. U.S.A}\ }\textbf {\bibinfo
  {volume} {118}},\ \bibinfo {pages} {e2023356118} (\bibinfo {year}
  {2021})}\BibitemShut {NoStop}%
\bibitem [{\citenamefont {Holubec}\ and\ \citenamefont
  {Ryabov}(2021)}]{holubec2021fluctuations}%
  \BibitemOpen
  \bibfield  {author} {\bibinfo {author} {\bibfnamefont {Viktor}\ \bibnamefont
  {Holubec}}\ and\ \bibinfo {author} {\bibfnamefont {Artem}\ \bibnamefont
  {Ryabov}},\ }\bibfield  {title} {\enquote {\bibinfo {title} {Fluctuations in
  heat engines},}\ }\href {\doibase 10.1088/1751-8121/ac3aac} {\bibfield
  {journal} {\bibinfo  {journal} {J. Phys. A Math. Theor.}\ }\textbf {\bibinfo
  {volume} {55}},\ \bibinfo {pages} {013001} (\bibinfo {year}
  {2021})}\BibitemShut {NoStop}%
\bibitem [{\citenamefont {Barato}\ and\ \citenamefont
  {Seifert}(2016)}]{barato2016cost}%
  \BibitemOpen
  \bibfield  {author} {\bibinfo {author} {\bibfnamefont {Andre~C.}\
  \bibnamefont {Barato}}\ and\ \bibinfo {author} {\bibfnamefont {Udo}\
  \bibnamefont {Seifert}},\ }\bibfield  {title} {\enquote {\bibinfo {title}
  {Cost and precision of brownian clocks},}\ }\href {\doibase
  10.1103/PhysRevX.6.041053} {\bibfield  {journal} {\bibinfo  {journal} {Phys.
  Rev. X}\ }\textbf {\bibinfo {volume} {6}},\ \bibinfo {pages} {041053}
  (\bibinfo {year} {2016})}\BibitemShut {NoStop}%
\bibitem [{\citenamefont {Holubec}\ and\ \citenamefont
  {Ryabov}(2018)}]{holubec2018cycling}%
  \BibitemOpen
  \bibfield  {author} {\bibinfo {author} {\bibfnamefont {Viktor}\ \bibnamefont
  {Holubec}}\ and\ \bibinfo {author} {\bibfnamefont {Artem}\ \bibnamefont
  {Ryabov}},\ }\bibfield  {title} {\enquote {\bibinfo {title} {Cycling tames
  power fluctuations near optimum efficiency},}\ }\href {\doibase
  10.1103/PhysRevLett.121.120601} {\bibfield  {journal} {\bibinfo  {journal}
  {Phys. Rev. Lett.}\ }\textbf {\bibinfo {volume} {121}},\ \bibinfo {pages}
  {120601} (\bibinfo {year} {2018})}\BibitemShut {NoStop}%
\bibitem [{\citenamefont {Miller}\ \emph {et~al.}(2021)\citenamefont {Miller},
  \citenamefont {Mohammady}, \citenamefont {Perarnau-Llobet},\ and\
  \citenamefont {Guarnieri}}]{miller2021thermodynamic}%
  \BibitemOpen
  \bibfield  {author} {\bibinfo {author} {\bibfnamefont {Harry J.~D.}\
  \bibnamefont {Miller}}, \bibinfo {author} {\bibfnamefont {M.~Hamed}\
  \bibnamefont {Mohammady}}, \bibinfo {author} {\bibfnamefont {Mart\'{\i}}\
  \bibnamefont {Perarnau-Llobet}}, \ and\ \bibinfo {author} {\bibfnamefont
  {Giacomo}\ \bibnamefont {Guarnieri}},\ }\bibfield  {title} {\enquote
  {\bibinfo {title} {Thermodynamic uncertainty relation in slowly driven
  quantum heat engines},}\ }\href {\doibase 10.1103/PhysRevLett.126.210603}
  {\bibfield  {journal} {\bibinfo  {journal} {Phys. Rev. Lett.}\ }\textbf
  {\bibinfo {volume} {126}},\ \bibinfo {pages} {210603} (\bibinfo {year}
  {2021})}\BibitemShut {NoStop}%
\bibitem [{\citenamefont {Thouless}(1983)}]{thouless1983quantization}%
  \BibitemOpen
  \bibfield  {author} {\bibinfo {author} {\bibfnamefont {D.~J.}\ \bibnamefont
  {Thouless}},\ }\bibfield  {title} {\enquote {\bibinfo {title} {Quantization
  of particle transport},}\ }\href {\doibase 10.1103/PhysRevB.27.6083}
  {\bibfield  {journal} {\bibinfo  {journal} {Phys. Rev. B}\ }\textbf {\bibinfo
  {volume} {27}},\ \bibinfo {pages} {6083--6087} (\bibinfo {year}
  {1983})}\BibitemShut {NoStop}%
\bibitem [{\citenamefont {Berry}(1984)}]{berry1984quantal}%
  \BibitemOpen
  \bibfield  {author} {\bibinfo {author} {\bibfnamefont {Michael~Victor}\
  \bibnamefont {Berry}},\ }\bibfield  {title} {\enquote {\bibinfo {title}
  {Quantal phase factors accompanying adiabatic changes},}\ }\href {\doibase
  10.1098/rspa.1984.0023} {\bibfield  {journal} {\bibinfo  {journal} {Proc. R.
  Soc. A}\ }\textbf {\bibinfo {volume} {392}},\ \bibinfo {pages} {45--57}
  (\bibinfo {year} {1984})}\BibitemShut {NoStop}%
\bibitem [{\citenamefont {Brouwer}(1998)}]{brouwer1998scattering}%
  \BibitemOpen
  \bibfield  {author} {\bibinfo {author} {\bibfnamefont {P.~W.}\ \bibnamefont
  {Brouwer}},\ }\bibfield  {title} {\enquote {\bibinfo {title} {Scattering
  approach to parametric pumping},}\ }\href {\doibase
  10.1103/PhysRevB.58.R10135} {\bibfield  {journal} {\bibinfo  {journal} {Phys.
  Rev. B}\ }\textbf {\bibinfo {volume} {58}},\ \bibinfo {pages}
  {R10135--R10138} (\bibinfo {year} {1998})}\BibitemShut {NoStop}%
\bibitem [{\citenamefont {Sinitsyn}\ and\ \citenamefont
  {Nemenman}(2007)}]{sinitsyn2007universal}%
  \BibitemOpen
  \bibfield  {author} {\bibinfo {author} {\bibfnamefont {N.~A.}\ \bibnamefont
  {Sinitsyn}}\ and\ \bibinfo {author} {\bibfnamefont {Ilya}\ \bibnamefont
  {Nemenman}},\ }\bibfield  {title} {\enquote {\bibinfo {title} {Universal
  geometric theory of mesoscopic stochastic pumps and reversible ratchets},}\
  }\href {\doibase 10.1103/PhysRevLett.99.220408} {\bibfield  {journal}
  {\bibinfo  {journal} {Phys. Rev. Lett.}\ }\textbf {\bibinfo {volume} {99}},\
  \bibinfo {pages} {220408} (\bibinfo {year} {2007})}\BibitemShut {NoStop}%
\bibitem [{\citenamefont {Rahav}\ \emph {et~al.}(2008)\citenamefont {Rahav},
  \citenamefont {Horowitz},\ and\ \citenamefont
  {Jarzynski}}]{rahav2008directed}%
  \BibitemOpen
  \bibfield  {author} {\bibinfo {author} {\bibfnamefont {Saar}\ \bibnamefont
  {Rahav}}, \bibinfo {author} {\bibfnamefont {Jordan}\ \bibnamefont
  {Horowitz}}, \ and\ \bibinfo {author} {\bibfnamefont {Christopher}\
  \bibnamefont {Jarzynski}},\ }\bibfield  {title} {\enquote {\bibinfo {title}
  {Directed flow in nonadiabatic stochastic pumps},}\ }\href {\doibase
  10.1103/PhysRevLett.101.140602} {\bibfield  {journal} {\bibinfo  {journal}
  {Phys. Rev. Lett.}\ }\textbf {\bibinfo {volume} {101}},\ \bibinfo {pages}
  {140602} (\bibinfo {year} {2008})}\BibitemShut {NoStop}%
\bibitem [{\citenamefont {Ren}\ \emph {et~al.}(2010)\citenamefont {Ren},
  \citenamefont {H\"anggi},\ and\ \citenamefont {Li}}]{ren2010berry}%
  \BibitemOpen
  \bibfield  {author} {\bibinfo {author} {\bibfnamefont {Jie}\ \bibnamefont
  {Ren}}, \bibinfo {author} {\bibfnamefont {Peter}\ \bibnamefont {H\"anggi}}, \
  and\ \bibinfo {author} {\bibfnamefont {Baowen}\ \bibnamefont {Li}},\
  }\bibfield  {title} {\enquote {\bibinfo {title} {Berry-phase-induced heat
  pumping and its impact on the fluctuation theorem},}\ }\href {\doibase
  10.1103/PhysRevLett.104.170601} {\bibfield  {journal} {\bibinfo  {journal}
  {Phys. Rev. Lett.}\ }\textbf {\bibinfo {volume} {104}},\ \bibinfo {pages}
  {170601} (\bibinfo {year} {2010})}\BibitemShut {NoStop}%
\bibitem [{\citenamefont {Ren}\ \emph {et~al.}(2012)\citenamefont {Ren},
  \citenamefont {Liu},\ and\ \citenamefont {Li}}]{ren2012geometric}%
  \BibitemOpen
  \bibfield  {author} {\bibinfo {author} {\bibfnamefont {Jie}\ \bibnamefont
  {Ren}}, \bibinfo {author} {\bibfnamefont {Sha}\ \bibnamefont {Liu}}, \ and\
  \bibinfo {author} {\bibfnamefont {Baowen}\ \bibnamefont {Li}},\ }\bibfield
  {title} {\enquote {\bibinfo {title} {Geometric heat flux for classical
  thermal transport in interacting open systems},}\ }\href {\doibase
  10.1103/PhysRevLett.108.210603} {\bibfield  {journal} {\bibinfo  {journal}
  {Phys. Rev. Lett.}\ }\textbf {\bibinfo {volume} {108}},\ \bibinfo {pages}
  {210603} (\bibinfo {year} {2012})}\BibitemShut {NoStop}%
\bibitem [{\citenamefont {Yuge}\ \emph {et~al.}(2012)\citenamefont {Yuge},
  \citenamefont {Sagawa}, \citenamefont {Sugita},\ and\ \citenamefont
  {Hayakawa}}]{yuge2012geometrical}%
  \BibitemOpen
  \bibfield  {author} {\bibinfo {author} {\bibfnamefont {Tatsuro}\ \bibnamefont
  {Yuge}}, \bibinfo {author} {\bibfnamefont {Takahiro}\ \bibnamefont {Sagawa}},
  \bibinfo {author} {\bibfnamefont {Ayumu}\ \bibnamefont {Sugita}}, \ and\
  \bibinfo {author} {\bibfnamefont {Hisao}\ \bibnamefont {Hayakawa}},\
  }\bibfield  {title} {\enquote {\bibinfo {title} {Geometrical pumping in
  quantum transport: Quantum master equation approach},}\ }\href {\doibase
  10.1103/PhysRevB.86.235308} {\bibfield  {journal} {\bibinfo  {journal} {Phys.
  Rev. B}\ }\textbf {\bibinfo {volume} {86}},\ \bibinfo {pages} {235308}
  (\bibinfo {year} {2012})}\BibitemShut {NoStop}%
\bibitem [{\citenamefont {Chen}\ \emph {et~al.}(2013)\citenamefont {Chen},
  \citenamefont {Wang},\ and\ \citenamefont {Ren}}]{chen2013dynamic}%
  \BibitemOpen
  \bibfield  {author} {\bibinfo {author} {\bibfnamefont {Tian}\ \bibnamefont
  {Chen}}, \bibinfo {author} {\bibfnamefont {Xiang-Bin}\ \bibnamefont {Wang}},
  \ and\ \bibinfo {author} {\bibfnamefont {Jie}\ \bibnamefont {Ren}},\
  }\bibfield  {title} {\enquote {\bibinfo {title} {Dynamic control of quantum
  geometric heat flux in a nonequilibrium spin-boson model},}\ }\href {\doibase
  10.1103/PhysRevB.87.144303} {\bibfield  {journal} {\bibinfo  {journal} {Phys.
  Rev. B}\ }\textbf {\bibinfo {volume} {87}},\ \bibinfo {pages} {144303}
  (\bibinfo {year} {2013})}\BibitemShut {NoStop}%
\bibitem [{\citenamefont {Wang}\ \emph {et~al.}(2017)\citenamefont {Wang},
  \citenamefont {Ren},\ and\ \citenamefont {Cao}}]{wang2017unifying}%
  \BibitemOpen
  \bibfield  {author} {\bibinfo {author} {\bibfnamefont {Chen}\ \bibnamefont
  {Wang}}, \bibinfo {author} {\bibfnamefont {Jie}\ \bibnamefont {Ren}}, \ and\
  \bibinfo {author} {\bibfnamefont {Jianshu}\ \bibnamefont {Cao}},\ }\bibfield
  {title} {\enquote {\bibinfo {title} {Unifying quantum heat transfer in a
  nonequilibrium spin-boson model with full counting statistics},}\ }\href
  {\doibase 10.1103/PhysRevA.95.023610} {\bibfield  {journal} {\bibinfo
  {journal} {Phys. Rev. A}\ }\textbf {\bibinfo {volume} {95}},\ \bibinfo
  {pages} {023610} (\bibinfo {year} {2017})}\BibitemShut {NoStop}%
\bibitem [{\citenamefont {Watanabe}\ and\ \citenamefont
  {Hayakawa}(2017)}]{watanabe2017geometric}%
  \BibitemOpen
  \bibfield  {author} {\bibinfo {author} {\bibfnamefont {Kota~L.}\ \bibnamefont
  {Watanabe}}\ and\ \bibinfo {author} {\bibfnamefont {Hisao}\ \bibnamefont
  {Hayakawa}},\ }\bibfield  {title} {\enquote {\bibinfo {title} {Geometric
  fluctuation theorem for a spin-boson system},}\ }\href {\doibase
  10.1103/PhysRevE.96.022118} {\bibfield  {journal} {\bibinfo  {journal} {Phys.
  Rev. E}\ }\textbf {\bibinfo {volume} {96}},\ \bibinfo {pages} {022118}
  (\bibinfo {year} {2017})}\BibitemShut {NoStop}%
\bibitem [{\citenamefont {Nie}\ \emph {et~al.}(2020)\citenamefont {Nie},
  \citenamefont {Li}, \citenamefont {Li}, \citenamefont {Chen}, \citenamefont
  {Lan},\ and\ \citenamefont {Zhu}}]{nie2020berry}%
  \BibitemOpen
  \bibfield  {author} {\bibinfo {author} {\bibfnamefont {Wenjie}\ \bibnamefont
  {Nie}}, \bibinfo {author} {\bibfnamefont {Guoyao}\ \bibnamefont {Li}},
  \bibinfo {author} {\bibfnamefont {Xiyun}\ \bibnamefont {Li}}, \bibinfo
  {author} {\bibfnamefont {Aixi}\ \bibnamefont {Chen}}, \bibinfo {author}
  {\bibfnamefont {Yueheng}\ \bibnamefont {Lan}}, \ and\ \bibinfo {author}
  {\bibfnamefont {Shi-Yao}\ \bibnamefont {Zhu}},\ }\bibfield  {title} {\enquote
  {\bibinfo {title} {Berry-phase-like effect of thermo-phonon transport in
  optomechanics},}\ }\href {\doibase 10.1103/PhysRevA.102.043512} {\bibfield
  {journal} {\bibinfo  {journal} {Phys. Rev. A}\ }\textbf {\bibinfo {volume}
  {102}},\ \bibinfo {pages} {043512} (\bibinfo {year} {2020})}\BibitemShut
  {NoStop}%
\bibitem [{\citenamefont {Wang}\ \emph
  {et~al.}(2022{\natexlab{b}})\citenamefont {Wang}, \citenamefont {Wang},
  \citenamefont {Chen}, \citenamefont {Wang},\ and\ \citenamefont
  {Ren}}]{wang2022geometric}%
  \BibitemOpen
  \bibfield  {author} {\bibinfo {author} {\bibfnamefont {Zi}~\bibnamefont
  {Wang}}, \bibinfo {author} {\bibfnamefont {Luqin}\ \bibnamefont {Wang}},
  \bibinfo {author} {\bibfnamefont {Jiangzhi}\ \bibnamefont {Chen}}, \bibinfo
  {author} {\bibfnamefont {Chen}\ \bibnamefont {Wang}}, \ and\ \bibinfo
  {author} {\bibfnamefont {Jie}\ \bibnamefont {Ren}},\ }\bibfield  {title}
  {\enquote {\bibinfo {title} {Geometric heat pump: Controlling thermal
  transport with time-dependent modulations},}\ }\href {\doibase
  10.1007/s11467-021-1095-4} {\bibfield  {journal} {\bibinfo  {journal} {Front.
  Phys.}\ }\textbf {\bibinfo {volume} {17}},\ \bibinfo {pages} {13201}
  (\bibinfo {year} {2022}{\natexlab{b}})}\BibitemShut {NoStop}%
\bibitem [{\citenamefont {Wang}\ \emph
  {et~al.}(2022{\natexlab{c}})\citenamefont {Wang}, \citenamefont {Chen},\ and\
  \citenamefont {Ren}}]{wang2022diffusive}%
  \BibitemOpen
  \bibfield  {author} {\bibinfo {author} {\bibfnamefont {Zi}~\bibnamefont
  {Wang}}, \bibinfo {author} {\bibfnamefont {Jiangzhi}\ \bibnamefont {Chen}}, \
  and\ \bibinfo {author} {\bibfnamefont {Jie}\ \bibnamefont {Ren}},\ }\bibfield
   {title} {\enquote {\bibinfo {title} {Geometric heat pump and no-go
  restrictions of nonreciprocity in modulated thermal diffusion},}\ }\href
  {\doibase 10.1103/PhysRevE.106.L032102} {\bibfield  {journal} {\bibinfo
  {journal} {Phys. Rev. E}\ }\textbf {\bibinfo {volume} {106}},\ \bibinfo
  {pages} {L032102} (\bibinfo {year} {2022}{\natexlab{c}})}\BibitemShut
  {NoStop}%
\bibitem [{\citenamefont {Monsel}\ \emph {et~al.}(2022)\citenamefont {Monsel},
  \citenamefont {Schulenborg}, \citenamefont {Baquet},\ and\ \citenamefont
  {Splettstoesser}}]{monsel2022geometric}%
  \BibitemOpen
  \bibfield  {author} {\bibinfo {author} {\bibfnamefont {Juliette}\
  \bibnamefont {Monsel}}, \bibinfo {author} {\bibfnamefont {Jens}\ \bibnamefont
  {Schulenborg}}, \bibinfo {author} {\bibfnamefont {Thibault}\ \bibnamefont
  {Baquet}}, \ and\ \bibinfo {author} {\bibfnamefont {Janine}\ \bibnamefont
  {Splettstoesser}},\ }\bibfield  {title} {\enquote {\bibinfo {title}
  {Geometric energy transport and refrigeration with driven quantum dots},}\
  }\href {\doibase 10.1103/PhysRevB.106.035405} {\bibfield  {journal} {\bibinfo
   {journal} {Phys. Rev. B}\ }\textbf {\bibinfo {volume} {106}},\ \bibinfo
  {pages} {035405} (\bibinfo {year} {2022})}\BibitemShut {NoStop}%
\bibitem [{\citenamefont {Giri}\ and\ \citenamefont
  {Goswami}(2017)}]{giri2017geometric}%
  \BibitemOpen
  \bibfield  {author} {\bibinfo {author} {\bibfnamefont {Sajal~Kumar}\
  \bibnamefont {Giri}}\ and\ \bibinfo {author} {\bibfnamefont
  {Himangshu~Prabal}\ \bibnamefont {Goswami}},\ }\bibfield  {title} {\enquote
  {\bibinfo {title} {Geometric phaselike effects in a quantum heat engine},}\
  }\href {\doibase 10.1103/PhysRevE.96.052129} {\bibfield  {journal} {\bibinfo
  {journal} {Phys. Rev. E}\ }\textbf {\bibinfo {volume} {96}},\ \bibinfo
  {pages} {052129} (\bibinfo {year} {2017})}\BibitemShut {NoStop}%
\bibitem [{\citenamefont {Bhandari}\ \emph {et~al.}(2020)\citenamefont
  {Bhandari}, \citenamefont {Alonso}, \citenamefont {Taddei}, \citenamefont
  {von Oppen}, \citenamefont {Fazio},\ and\ \citenamefont
  {Arrachea}}]{bhandari2020geometric}%
  \BibitemOpen
  \bibfield  {author} {\bibinfo {author} {\bibfnamefont {Bibek}\ \bibnamefont
  {Bhandari}}, \bibinfo {author} {\bibfnamefont {Pablo~Terr\'en}\ \bibnamefont
  {Alonso}}, \bibinfo {author} {\bibfnamefont {Fabio}\ \bibnamefont {Taddei}},
  \bibinfo {author} {\bibfnamefont {Felix}\ \bibnamefont {von Oppen}}, \bibinfo
  {author} {\bibfnamefont {Rosario}\ \bibnamefont {Fazio}}, \ and\ \bibinfo
  {author} {\bibfnamefont {Liliana}\ \bibnamefont {Arrachea}},\ }\bibfield
  {title} {\enquote {\bibinfo {title} {Geometric properties of adiabatic
  quantum thermal machines},}\ }\href {\doibase 10.1103/PhysRevB.102.155407}
  {\bibfield  {journal} {\bibinfo  {journal} {Phys. Rev. B}\ }\textbf {\bibinfo
  {volume} {102}},\ \bibinfo {pages} {155407} (\bibinfo {year}
  {2020})}\BibitemShut {NoStop}%
\bibitem [{\citenamefont {Hino}\ and\ \citenamefont
  {Hayakawa}(2021)}]{hino2021geometrical}%
  \BibitemOpen
  \bibfield  {author} {\bibinfo {author} {\bibfnamefont {Yuki}\ \bibnamefont
  {Hino}}\ and\ \bibinfo {author} {\bibfnamefont {Hisao}\ \bibnamefont
  {Hayakawa}},\ }\bibfield  {title} {\enquote {\bibinfo {title} {Geometrical
  formulation of adiabatic pumping as a heat engine},}\ }\href {\doibase
  10.1103/PhysRevResearch.3.013187} {\bibfield  {journal} {\bibinfo  {journal}
  {Phys. Rev. Research}\ }\textbf {\bibinfo {volume} {3}},\ \bibinfo {pages}
  {013187} (\bibinfo {year} {2021})}\BibitemShut {NoStop}%
\bibitem [{\citenamefont {Hayakawa}\ \emph {et~al.}(2021)\citenamefont
  {Hayakawa}, \citenamefont {Paasonen},\ and\ \citenamefont
  {Yoshii}}]{hayakawa2021geometrical}%
  \BibitemOpen
  \bibfield  {author} {\bibinfo {author} {\bibfnamefont {Hisao}\ \bibnamefont
  {Hayakawa}}, \bibinfo {author} {\bibfnamefont {Ville~MM}\ \bibnamefont
  {Paasonen}}, \ and\ \bibinfo {author} {\bibfnamefont {Ryosuke}\ \bibnamefont
  {Yoshii}},\ }\bibfield  {title} {\enquote {\bibinfo {title} {Geometrical
  quantum chemical engine},}\ }\href {https://arxiv.org/abs/2112.12370}
  {\bibfield  {journal} {\bibinfo  {journal} {arXiv preprint arXiv:2112.12370}\
  } (\bibinfo {year} {2021})}\BibitemShut {NoStop}%
\bibitem [{\citenamefont {Takahashi}\ \emph {et~al.}(2020)\citenamefont
  {Takahashi}, \citenamefont {Fujii}, \citenamefont {Hino},\ and\ \citenamefont
  {Hayakawa}}]{takahashi2020nonadiabatic}%
  \BibitemOpen
  \bibfield  {author} {\bibinfo {author} {\bibfnamefont {Kazutaka}\
  \bibnamefont {Takahashi}}, \bibinfo {author} {\bibfnamefont {Keisuke}\
  \bibnamefont {Fujii}}, \bibinfo {author} {\bibfnamefont {Yuki}\ \bibnamefont
  {Hino}}, \ and\ \bibinfo {author} {\bibfnamefont {Hisao}\ \bibnamefont
  {Hayakawa}},\ }\bibfield  {title} {\enquote {\bibinfo {title} {Nonadiabatic
  control of geometric pumping},}\ }\href {\doibase
  10.1103/PhysRevLett.124.150602} {\bibfield  {journal} {\bibinfo  {journal}
  {Phys. Rev. Lett.}\ }\textbf {\bibinfo {volume} {124}},\ \bibinfo {pages}
  {150602} (\bibinfo {year} {2020})}\BibitemShut {NoStop}%
\bibitem [{\citenamefont {Funo}\ \emph {et~al.}(2020)\citenamefont {Funo},
  \citenamefont {Lambert}, \citenamefont {Nori},\ and\ \citenamefont
  {Flindt}}]{funo2020shortcuts}%
  \BibitemOpen
  \bibfield  {author} {\bibinfo {author} {\bibfnamefont {Ken}\ \bibnamefont
  {Funo}}, \bibinfo {author} {\bibfnamefont {Neill}\ \bibnamefont {Lambert}},
  \bibinfo {author} {\bibfnamefont {Franco}\ \bibnamefont {Nori}}, \ and\
  \bibinfo {author} {\bibfnamefont {Christian}\ \bibnamefont {Flindt}},\
  }\bibfield  {title} {\enquote {\bibinfo {title} {Shortcuts to adiabatic
  pumping in classical stochastic systems},}\ }\href {\doibase
  10.1103/PhysRevLett.124.150603} {\bibfield  {journal} {\bibinfo  {journal}
  {Phys. Rev. Lett.}\ }\textbf {\bibinfo {volume} {124}},\ \bibinfo {pages}
  {150603} (\bibinfo {year} {2020})}\BibitemShut {NoStop}%
\bibitem [{\citenamefont {Andresen}\ \emph {et~al.}(1984)\citenamefont
  {Andresen}, \citenamefont {Salamon},\ and\ \citenamefont
  {Berry}}]{andresen1984thermodynamics}%
  \BibitemOpen
  \bibfield  {author} {\bibinfo {author} {\bibfnamefont {Bjarne}\ \bibnamefont
  {Andresen}}, \bibinfo {author} {\bibfnamefont {Peter}\ \bibnamefont
  {Salamon}}, \ and\ \bibinfo {author} {\bibfnamefont {R~Stephen}\ \bibnamefont
  {Berry}},\ }\bibfield  {title} {\enquote {\bibinfo {title} {Thermodynamics in
  finite time},}\ }\href {\doibase 10.1063/1.2916405} {\bibfield  {journal}
  {\bibinfo  {journal} {Phys. Today}\ }\textbf {\bibinfo {volume} {37}},\
  \bibinfo {pages} {62} (\bibinfo {year} {1984})}\BibitemShut {NoStop}%
\bibitem [{\citenamefont {Salamon}\ and\ \citenamefont
  {Berry}(1983)}]{salamon1983thermodynamic}%
  \BibitemOpen
  \bibfield  {author} {\bibinfo {author} {\bibfnamefont {Peter}\ \bibnamefont
  {Salamon}}\ and\ \bibinfo {author} {\bibfnamefont {R.~Stephen}\ \bibnamefont
  {Berry}},\ }\bibfield  {title} {\enquote {\bibinfo {title} {Thermodynamic
  length and dissipated availability},}\ }\href {\doibase
  10.1103/PhysRevLett.51.1127} {\bibfield  {journal} {\bibinfo  {journal}
  {Phys. Rev. Lett.}\ }\textbf {\bibinfo {volume} {51}},\ \bibinfo {pages}
  {1127--1130} (\bibinfo {year} {1983})}\BibitemShut {NoStop}%
\bibitem [{\citenamefont {Crooks}(2007)}]{crooks2007measuring}%
  \BibitemOpen
  \bibfield  {author} {\bibinfo {author} {\bibfnamefont {Gavin~E.}\
  \bibnamefont {Crooks}},\ }\bibfield  {title} {\enquote {\bibinfo {title}
  {Measuring thermodynamic length},}\ }\href {\doibase
  10.1103/PhysRevLett.99.100602} {\bibfield  {journal} {\bibinfo  {journal}
  {Phys. Rev. Lett.}\ }\textbf {\bibinfo {volume} {99}},\ \bibinfo {pages}
  {100602} (\bibinfo {year} {2007})}\BibitemShut {NoStop}%
\bibitem [{\citenamefont {Feng}\ and\ \citenamefont
  {Crooks}(2009)}]{feng2009far}%
  \BibitemOpen
  \bibfield  {author} {\bibinfo {author} {\bibfnamefont {Edward~H.}\
  \bibnamefont {Feng}}\ and\ \bibinfo {author} {\bibfnamefont {Gavin~E.}\
  \bibnamefont {Crooks}},\ }\bibfield  {title} {\enquote {\bibinfo {title}
  {Far-from-equilibrium measurements of thermodynamic length},}\ }\href
  {\doibase 10.1103/PhysRevE.79.012104} {\bibfield  {journal} {\bibinfo
  {journal} {Phys. Rev. E}\ }\textbf {\bibinfo {volume} {79}},\ \bibinfo
  {pages} {012104} (\bibinfo {year} {2009})}\BibitemShut {NoStop}%
\bibitem [{\citenamefont {Sivak}\ and\ \citenamefont
  {Crooks}(2012)}]{sivak2012thermodynamic}%
  \BibitemOpen
  \bibfield  {author} {\bibinfo {author} {\bibfnamefont {David~A.}\
  \bibnamefont {Sivak}}\ and\ \bibinfo {author} {\bibfnamefont {Gavin~E.}\
  \bibnamefont {Crooks}},\ }\bibfield  {title} {\enquote {\bibinfo {title}
  {Thermodynamic metrics and optimal paths},}\ }\href {\doibase
  10.1103/PhysRevLett.108.190602} {\bibfield  {journal} {\bibinfo  {journal}
  {Phys. Rev. Lett.}\ }\textbf {\bibinfo {volume} {108}},\ \bibinfo {pages}
  {190602} (\bibinfo {year} {2012})}\BibitemShut {NoStop}%
\bibitem [{\citenamefont {Aurell}\ \emph {et~al.}(2012)\citenamefont {Aurell},
  \citenamefont {Gawedzki}, \citenamefont {Mej{\'\i}a-Monasterio},
  \citenamefont {Mohayaee},\ and\ \citenamefont
  {Muratore-Ginanneschi}}]{aurell2012refined}%
  \BibitemOpen
  \bibfield  {author} {\bibinfo {author} {\bibfnamefont {Erik}\ \bibnamefont
  {Aurell}}, \bibinfo {author} {\bibfnamefont {Krzysztof}\ \bibnamefont
  {Gawedzki}}, \bibinfo {author} {\bibfnamefont {Carlos}\ \bibnamefont
  {Mej{\'\i}a-Monasterio}}, \bibinfo {author} {\bibfnamefont {Roya}\
  \bibnamefont {Mohayaee}}, \ and\ \bibinfo {author} {\bibfnamefont {Paolo}\
  \bibnamefont {Muratore-Ginanneschi}},\ }\bibfield  {title} {\enquote
  {\bibinfo {title} {Refined second law of thermodynamics for fast random
  processes},}\ }\href {\doibase 10.1007/s10955-012-0478-x} {\bibfield
  {journal} {\bibinfo  {journal} {J. Stat. Phys.}\ }\textbf {\bibinfo {volume}
  {147}},\ \bibinfo {pages} {487} (\bibinfo {year} {2012})}\BibitemShut
  {NoStop}%
\bibitem [{\citenamefont {Nakazato}\ and\ \citenamefont
  {Ito}(2021)}]{nakazato2021geometrical}%
  \BibitemOpen
  \bibfield  {author} {\bibinfo {author} {\bibfnamefont {Muka}\ \bibnamefont
  {Nakazato}}\ and\ \bibinfo {author} {\bibfnamefont {Sosuke}\ \bibnamefont
  {Ito}},\ }\bibfield  {title} {\enquote {\bibinfo {title} {Geometrical aspects
  of entropy production in stochastic thermodynamics based on wasserstein
  distance},}\ }\href {\doibase 10.1103/PhysRevResearch.3.043093} {\bibfield
  {journal} {\bibinfo  {journal} {Phys. Rev. Res.}\ }\textbf {\bibinfo {volume}
  {3}},\ \bibinfo {pages} {043093} (\bibinfo {year} {2021})}\BibitemShut
  {NoStop}%
\bibitem [{\citenamefont {Van~Vu}\ and\ \citenamefont
  {Hasegawa}(2021)}]{van2021geometrical}%
  \BibitemOpen
  \bibfield  {author} {\bibinfo {author} {\bibfnamefont {Tan}\ \bibnamefont
  {Van~Vu}}\ and\ \bibinfo {author} {\bibfnamefont {Yoshihiko}\ \bibnamefont
  {Hasegawa}},\ }\bibfield  {title} {\enquote {\bibinfo {title} {Geometrical
  bounds of the irreversibility in markovian systems},}\ }\href {\doibase
  10.1103/PhysRevLett.126.010601} {\bibfield  {journal} {\bibinfo  {journal}
  {Phys. Rev. Lett.}\ }\textbf {\bibinfo {volume} {126}},\ \bibinfo {pages}
  {010601} (\bibinfo {year} {2021})}\BibitemShut {NoStop}%
\bibitem [{\citenamefont {Van~Vu}\ and\ \citenamefont
  {Saito}(2023)}]{van2023thermodynamic}%
  \BibitemOpen
  \bibfield  {author} {\bibinfo {author} {\bibfnamefont {Tan}\ \bibnamefont
  {Van~Vu}}\ and\ \bibinfo {author} {\bibfnamefont {Keiji}\ \bibnamefont
  {Saito}},\ }\bibfield  {title} {\enquote {\bibinfo {title} {Thermodynamic
  unification of optimal transport: Thermodynamic uncertainty relation, minimum
  dissipation, and thermodynamic speed limits},}\ }\href {\doibase
  10.1103/PhysRevX.13.011013} {\bibfield  {journal} {\bibinfo  {journal} {Phys.
  Rev. X}\ }\textbf {\bibinfo {volume} {13}},\ \bibinfo {pages} {011013}
  (\bibinfo {year} {2023})}\BibitemShut {NoStop}%
\bibitem [{\citenamefont {Aurell}\ \emph {et~al.}(2011)\citenamefont {Aurell},
  \citenamefont {Mej\'{\i}a-Monasterio},\ and\ \citenamefont
  {Muratore-Ginanneschi}}]{aurell2011optimal}%
  \BibitemOpen
  \bibfield  {author} {\bibinfo {author} {\bibfnamefont {Erik}\ \bibnamefont
  {Aurell}}, \bibinfo {author} {\bibfnamefont {Carlos}\ \bibnamefont
  {Mej\'{\i}a-Monasterio}}, \ and\ \bibinfo {author} {\bibfnamefont {Paolo}\
  \bibnamefont {Muratore-Ginanneschi}},\ }\bibfield  {title} {\enquote
  {\bibinfo {title} {Optimal protocols and optimal transport in stochastic
  thermodynamics},}\ }\href {\doibase 10.1103/PhysRevLett.106.250601}
  {\bibfield  {journal} {\bibinfo  {journal} {Phys. Rev. Lett.}\ }\textbf
  {\bibinfo {volume} {106}},\ \bibinfo {pages} {250601} (\bibinfo {year}
  {2011})}\BibitemShut {NoStop}%
\bibitem [{\citenamefont {Muratore-Ginanneschi}\ \emph
  {et~al.}(2013)\citenamefont {Muratore-Ginanneschi}, \citenamefont
  {Mej{\'\i}a-Monasterio},\ and\ \citenamefont {Peliti}}]{muratore2013heat}%
  \BibitemOpen
  \bibfield  {author} {\bibinfo {author} {\bibfnamefont {Paolo}\ \bibnamefont
  {Muratore-Ginanneschi}}, \bibinfo {author} {\bibfnamefont {Carlos}\
  \bibnamefont {Mej{\'\i}a-Monasterio}}, \ and\ \bibinfo {author}
  {\bibfnamefont {Luca}\ \bibnamefont {Peliti}},\ }\bibfield  {title} {\enquote
  {\bibinfo {title} {Heat release by controlled continuous-time markov jump
  processes},}\ }\href {\doibase 10.1007/s10955-012-0676-6} {\bibfield
  {journal} {\bibinfo  {journal} {J. Stat. Phys.}\ }\textbf {\bibinfo {volume}
  {150}},\ \bibinfo {pages} {181} (\bibinfo {year} {2013})}\BibitemShut
  {NoStop}%
\bibitem [{\citenamefont {Muratore-Ginanneschi}\ and\ \citenamefont
  {Schwieger}(2014)}]{muratore2014nanomechanical}%
  \BibitemOpen
  \bibfield  {author} {\bibinfo {author} {\bibfnamefont {Paolo}\ \bibnamefont
  {Muratore-Ginanneschi}}\ and\ \bibinfo {author} {\bibfnamefont {Kay}\
  \bibnamefont {Schwieger}},\ }\bibfield  {title} {\enquote {\bibinfo {title}
  {How nanomechanical systems can minimize dissipation},}\ }\href {\doibase
  10.1103/PhysRevE.90.060102} {\bibfield  {journal} {\bibinfo  {journal} {Phys.
  Rev. E}\ }\textbf {\bibinfo {volume} {90}},\ \bibinfo {pages} {060102(R)}
  (\bibinfo {year} {2014})}\BibitemShut {NoStop}%
\bibitem [{\citenamefont {Guarnieri}\ \emph {et~al.}(2019)\citenamefont
  {Guarnieri}, \citenamefont {Landi}, \citenamefont {Clark},\ and\
  \citenamefont {Goold}}]{guarnieri2019thermodynamics}%
  \BibitemOpen
  \bibfield  {author} {\bibinfo {author} {\bibfnamefont {Giacomo}\ \bibnamefont
  {Guarnieri}}, \bibinfo {author} {\bibfnamefont {Gabriel~T.}\ \bibnamefont
  {Landi}}, \bibinfo {author} {\bibfnamefont {Stephen~R.}\ \bibnamefont
  {Clark}}, \ and\ \bibinfo {author} {\bibfnamefont {John}\ \bibnamefont
  {Goold}},\ }\bibfield  {title} {\enquote {\bibinfo {title} {Thermodynamics of
  precision in quantum nonequilibrium steady states},}\ }\href {\doibase
  10.1103/PhysRevResearch.1.033021} {\bibfield  {journal} {\bibinfo  {journal}
  {Phys. Rev. Research}\ }\textbf {\bibinfo {volume} {1}},\ \bibinfo {pages}
  {033021} (\bibinfo {year} {2019})}\BibitemShut {NoStop}%
\bibitem [{\citenamefont {Brandner}\ and\ \citenamefont
  {Saito}(2020)}]{brandner2020thermodynamic}%
  \BibitemOpen
  \bibfield  {author} {\bibinfo {author} {\bibfnamefont {Kay}\ \bibnamefont
  {Brandner}}\ and\ \bibinfo {author} {\bibfnamefont {Keiji}\ \bibnamefont
  {Saito}},\ }\bibfield  {title} {\enquote {\bibinfo {title} {Thermodynamic
  geometry of microscopic heat engines},}\ }\href {\doibase
  10.1103/PhysRevLett.124.040602} {\bibfield  {journal} {\bibinfo  {journal}
  {Phys. Rev. Lett.}\ }\textbf {\bibinfo {volume} {124}},\ \bibinfo {pages}
  {040602} (\bibinfo {year} {2020})}\BibitemShut {NoStop}%
\bibitem [{\citenamefont {Eglinton}\ and\ \citenamefont
  {Brandner}(2022)}]{eglinton2022geometric}%
  \BibitemOpen
  \bibfield  {author} {\bibinfo {author} {\bibfnamefont {Joshua}\ \bibnamefont
  {Eglinton}}\ and\ \bibinfo {author} {\bibfnamefont {Kay}\ \bibnamefont
  {Brandner}},\ }\bibfield  {title} {\enquote {\bibinfo {title} {Geometric
  bounds on the power of adiabatic thermal machines},}\ }\href {\doibase
  10.1103/PhysRevE.105.L052102} {\bibfield  {journal} {\bibinfo  {journal}
  {Phys. Rev. E}\ }\textbf {\bibinfo {volume} {105}},\ \bibinfo {pages}
  {L052102} (\bibinfo {year} {2022})}\BibitemShut {NoStop}%
\bibitem [{\citenamefont {Miller}\ \emph {et~al.}(2019)\citenamefont {Miller},
  \citenamefont {Scandi}, \citenamefont {Anders},\ and\ \citenamefont
  {Perarnau-Llobet}}]{miller2019work}%
  \BibitemOpen
  \bibfield  {author} {\bibinfo {author} {\bibfnamefont {Harry J.~D.}\
  \bibnamefont {Miller}}, \bibinfo {author} {\bibfnamefont {Matteo}\
  \bibnamefont {Scandi}}, \bibinfo {author} {\bibfnamefont {Janet}\
  \bibnamefont {Anders}}, \ and\ \bibinfo {author} {\bibfnamefont {Mart\'{\i}}\
  \bibnamefont {Perarnau-Llobet}},\ }\bibfield  {title} {\enquote {\bibinfo
  {title} {Work fluctuations in slow processes: Quantum signatures and optimal
  control},}\ }\href {\doibase 10.1103/PhysRevLett.123.230603} {\bibfield
  {journal} {\bibinfo  {journal} {Phys. Rev. Lett.}\ }\textbf {\bibinfo
  {volume} {123}},\ \bibinfo {pages} {230603} (\bibinfo {year}
  {2019})}\BibitemShut {NoStop}%
\bibitem [{\citenamefont {Abiuso}\ and\ \citenamefont
  {Perarnau-Llobet}(2020)}]{abiuso2020optimal}%
  \BibitemOpen
  \bibfield  {author} {\bibinfo {author} {\bibfnamefont {Paolo}\ \bibnamefont
  {Abiuso}}\ and\ \bibinfo {author} {\bibfnamefont {Mart\'{\i}}\ \bibnamefont
  {Perarnau-Llobet}},\ }\bibfield  {title} {\enquote {\bibinfo {title} {Optimal
  cycles for low-dissipation heat engines},}\ }\href {\doibase
  10.1103/PhysRevLett.124.110606} {\bibfield  {journal} {\bibinfo  {journal}
  {Phys. Rev. Lett.}\ }\textbf {\bibinfo {volume} {124}},\ \bibinfo {pages}
  {110606} (\bibinfo {year} {2020})}\BibitemShut {NoStop}%
\bibitem [{\citenamefont {Miller}\ and\ \citenamefont
  {Mehboudi}(2020)}]{miller2020geometry}%
  \BibitemOpen
  \bibfield  {author} {\bibinfo {author} {\bibfnamefont {Harry J.~D.}\
  \bibnamefont {Miller}}\ and\ \bibinfo {author} {\bibfnamefont {Mohammad}\
  \bibnamefont {Mehboudi}},\ }\bibfield  {title} {\enquote {\bibinfo {title}
  {Geometry of work fluctuations versus efficiency in microscopic thermal
  machines},}\ }\href {\doibase 10.1103/PhysRevLett.125.260602} {\bibfield
  {journal} {\bibinfo  {journal} {Phys. Rev. Lett.}\ }\textbf {\bibinfo
  {volume} {125}},\ \bibinfo {pages} {260602} (\bibinfo {year}
  {2020})}\BibitemShut {NoStop}%
\bibitem [{\citenamefont {Abiuso}\ \emph {et~al.}(2020)\citenamefont {Abiuso},
  \citenamefont {Miller}, \citenamefont {Perarnau-Llobet},\ and\ \citenamefont
  {Scandi}}]{abiuso2020geometric}%
  \BibitemOpen
  \bibfield  {author} {\bibinfo {author} {\bibfnamefont {Paolo}\ \bibnamefont
  {Abiuso}}, \bibinfo {author} {\bibfnamefont {Harry~JD}\ \bibnamefont
  {Miller}}, \bibinfo {author} {\bibfnamefont {Mart{\'\i}}\ \bibnamefont
  {Perarnau-Llobet}}, \ and\ \bibinfo {author} {\bibfnamefont {Matteo}\
  \bibnamefont {Scandi}},\ }\bibfield  {title} {\enquote {\bibinfo {title}
  {Geometric optimisation of quantum thermodynamic processes},}\ }\href
  {\doibase 10.3390/e22101076} {\bibfield  {journal} {\bibinfo  {journal}
  {Entropy}\ }\textbf {\bibinfo {volume} {22}},\ \bibinfo {pages} {1076}
  (\bibinfo {year} {2020})}\BibitemShut {NoStop}%
\bibitem [{\citenamefont {Proesmans}\ \emph {et~al.}(2020)\citenamefont
  {Proesmans}, \citenamefont {Ehrich},\ and\ \citenamefont
  {Bechhoefer}}]{proesmans2020finite}%
  \BibitemOpen
  \bibfield  {author} {\bibinfo {author} {\bibfnamefont {Karel}\ \bibnamefont
  {Proesmans}}, \bibinfo {author} {\bibfnamefont {Jannik}\ \bibnamefont
  {Ehrich}}, \ and\ \bibinfo {author} {\bibfnamefont {John}\ \bibnamefont
  {Bechhoefer}},\ }\bibfield  {title} {\enquote {\bibinfo {title} {Finite-time
  landauer principle},}\ }\href {\doibase 10.1103/PhysRevLett.125.100602}
  {\bibfield  {journal} {\bibinfo  {journal} {Phys. Rev. Lett.}\ }\textbf
  {\bibinfo {volume} {125}},\ \bibinfo {pages} {100602} (\bibinfo {year}
  {2020})}\BibitemShut {NoStop}%
\bibitem [{\citenamefont {Chen}\ \emph {et~al.}(2021)\citenamefont {Chen},
  \citenamefont {Sun},\ and\ \citenamefont {Dong}}]{chen2021extrapolating}%
  \BibitemOpen
  \bibfield  {author} {\bibinfo {author} {\bibfnamefont {Jin-Fu}\ \bibnamefont
  {Chen}}, \bibinfo {author} {\bibfnamefont {C.~P.}\ \bibnamefont {Sun}}, \
  and\ \bibinfo {author} {\bibfnamefont {Hui}\ \bibnamefont {Dong}},\
  }\bibfield  {title} {\enquote {\bibinfo {title} {Extrapolating the
  thermodynamic length with finite-time measurements},}\ }\href {\doibase
  10.1103/PhysRevE.104.034117} {\bibfield  {journal} {\bibinfo  {journal}
  {Phys. Rev. E}\ }\textbf {\bibinfo {volume} {104}},\ \bibinfo {pages}
  {034117} (\bibinfo {year} {2021})}\BibitemShut {NoStop}%
\bibitem [{\citenamefont {Terr\'en~Alonso}\ \emph {et~al.}(2022)\citenamefont
  {Terr\'en~Alonso}, \citenamefont {Abiuso}, \citenamefont {Perarnau-Llobet},\
  and\ \citenamefont {Arrachea}}]{alonso2021geometric}%
  \BibitemOpen
  \bibfield  {author} {\bibinfo {author} {\bibfnamefont {Pablo}\ \bibnamefont
  {Terr\'en~Alonso}}, \bibinfo {author} {\bibfnamefont {Paolo}\ \bibnamefont
  {Abiuso}}, \bibinfo {author} {\bibfnamefont {Mart\'{\i}}\ \bibnamefont
  {Perarnau-Llobet}}, \ and\ \bibinfo {author} {\bibfnamefont {Liliana}\
  \bibnamefont {Arrachea}},\ }\bibfield  {title} {\enquote {\bibinfo {title}
  {Geometric optimization of nonequilibrium adiabatic thermal machines and
  implementation in a qubit system},}\ }\href {\doibase
  10.1103/PRXQuantum.3.010326} {\bibfield  {journal} {\bibinfo  {journal} {PRX
  Quantum}\ }\textbf {\bibinfo {volume} {3}},\ \bibinfo {pages} {010326}
  (\bibinfo {year} {2022})}\BibitemShut {NoStop}%
\bibitem [{\citenamefont {Frim}\ and\ \citenamefont
  {DeWeese}(2022{\natexlab{a}})}]{frim2022geometric}%
  \BibitemOpen
  \bibfield  {author} {\bibinfo {author} {\bibfnamefont {Adam~G.}\ \bibnamefont
  {Frim}}\ and\ \bibinfo {author} {\bibfnamefont {Michael~R.}\ \bibnamefont
  {DeWeese}},\ }\bibfield  {title} {\enquote {\bibinfo {title} {Geometric bound
  on the efficiency of irreversible thermodynamic cycles},}\ }\href {\doibase
  10.1103/PhysRevLett.128.230601} {\bibfield  {journal} {\bibinfo  {journal}
  {Phys. Rev. Lett.}\ }\textbf {\bibinfo {volume} {128}},\ \bibinfo {pages}
  {230601} (\bibinfo {year} {2022}{\natexlab{a}})}\BibitemShut {NoStop}%
\bibitem [{\citenamefont {Frim}\ and\ \citenamefont
  {DeWeese}(2022{\natexlab{b}})}]{frim2022optimal}%
  \BibitemOpen
  \bibfield  {author} {\bibinfo {author} {\bibfnamefont {Adam~G.}\ \bibnamefont
  {Frim}}\ and\ \bibinfo {author} {\bibfnamefont {Michael~R.}\ \bibnamefont
  {DeWeese}},\ }\bibfield  {title} {\enquote {\bibinfo {title} {Optimal
  finite-time brownian carnot engine},}\ }\href {\doibase
  10.1103/PhysRevE.105.L052103} {\bibfield  {journal} {\bibinfo  {journal}
  {Phys. Rev. E}\ }\textbf {\bibinfo {volume} {105}},\ \bibinfo {pages}
  {L052103} (\bibinfo {year} {2022}{\natexlab{b}})}\BibitemShut {NoStop}%
\bibitem [{\citenamefont {Li}\ \emph {et~al.}(2022)\citenamefont {Li},
  \citenamefont {Chen}, \citenamefont {Sun},\ and\ \citenamefont
  {Dong}}]{li2022geodesic}%
  \BibitemOpen
  \bibfield  {author} {\bibinfo {author} {\bibfnamefont {Geng}\ \bibnamefont
  {Li}}, \bibinfo {author} {\bibfnamefont {Jin-Fu}\ \bibnamefont {Chen}},
  \bibinfo {author} {\bibfnamefont {C.~P.}\ \bibnamefont {Sun}}, \ and\
  \bibinfo {author} {\bibfnamefont {Hui}\ \bibnamefont {Dong}},\ }\bibfield
  {title} {\enquote {\bibinfo {title} {Geodesic path for the minimal energy
  cost in shortcuts to isothermality},}\ }\href {\doibase
  10.1103/PhysRevLett.128.230603} {\bibfield  {journal} {\bibinfo  {journal}
  {Phys. Rev. Lett.}\ }\textbf {\bibinfo {volume} {128}},\ \bibinfo {pages}
  {230603} (\bibinfo {year} {2022})}\BibitemShut {NoStop}%
\bibitem [{\citenamefont {Scandi}\ \emph {et~al.}(2022)\citenamefont {Scandi},
  \citenamefont {Barker}, \citenamefont {Lehmann}, \citenamefont {Dick},
  \citenamefont {Maisi},\ and\ \citenamefont
  {Perarnau-Llobet}}]{scandi2022minimally}%
  \BibitemOpen
  \bibfield  {author} {\bibinfo {author} {\bibfnamefont {Matteo}\ \bibnamefont
  {Scandi}}, \bibinfo {author} {\bibfnamefont {David}\ \bibnamefont {Barker}},
  \bibinfo {author} {\bibfnamefont {Sebastian}\ \bibnamefont {Lehmann}},
  \bibinfo {author} {\bibfnamefont {Kimberly~A.}\ \bibnamefont {Dick}},
  \bibinfo {author} {\bibfnamefont {Ville~F.}\ \bibnamefont {Maisi}}, \ and\
  \bibinfo {author} {\bibfnamefont {Mart\'{\i}}\ \bibnamefont
  {Perarnau-Llobet}},\ }\bibfield  {title} {\enquote {\bibinfo {title}
  {Minimally dissipative information erasure in a quantum dot via thermodynamic
  length},}\ }\href {\doibase 10.1103/PhysRevLett.129.270601} {\bibfield
  {journal} {\bibinfo  {journal} {Phys. Rev. Lett.}\ }\textbf {\bibinfo
  {volume} {129}},\ \bibinfo {pages} {270601} (\bibinfo {year}
  {2022})}\BibitemShut {NoStop}%
\bibitem [{\citenamefont {Abiuso}\ \emph {et~al.}(2022)\citenamefont {Abiuso},
  \citenamefont {Holubec}, \citenamefont {Anders}, \citenamefont {Ye},
  \citenamefont {Cerisola},\ and\ \citenamefont
  {Perarnau-Llobet}}]{abiuso2022thermodynamics}%
  \BibitemOpen
  \bibfield  {author} {\bibinfo {author} {\bibfnamefont {Paolo}\ \bibnamefont
  {Abiuso}}, \bibinfo {author} {\bibfnamefont {Viktor}\ \bibnamefont
  {Holubec}}, \bibinfo {author} {\bibfnamefont {Janet}\ \bibnamefont {Anders}},
  \bibinfo {author} {\bibfnamefont {Zhuolin}\ \bibnamefont {Ye}}, \bibinfo
  {author} {\bibfnamefont {Federico}\ \bibnamefont {Cerisola}}, \ and\ \bibinfo
  {author} {\bibfnamefont {Mart{\'\i}}\ \bibnamefont {Perarnau-Llobet}},\
  }\bibfield  {title} {\enquote {\bibinfo {title} {Thermodynamics and optimal
  protocols of multidimensional quadratic brownian systems},}\ }\href {\doibase
  10.1088/2399-6528/ac72f8} {\bibfield  {journal} {\bibinfo  {journal} {J.
  Phys. Commun.}\ }\textbf {\bibinfo {volume} {6}},\ \bibinfo {pages} {063001}
  (\bibinfo {year} {2022})}\BibitemShut {NoStop}%
\bibitem [{\citenamefont {Dago}\ and\ \citenamefont
  {Bellon}(2022)}]{dago2022dynamics}%
  \BibitemOpen
  \bibfield  {author} {\bibinfo {author} {\bibfnamefont {Salamb\^o}\
  \bibnamefont {Dago}}\ and\ \bibinfo {author} {\bibfnamefont {Ludovic}\
  \bibnamefont {Bellon}},\ }\bibfield  {title} {\enquote {\bibinfo {title}
  {Dynamics of information erasure and extension of landauer's bound to fast
  processes},}\ }\href {\doibase 10.1103/PhysRevLett.128.070604} {\bibfield
  {journal} {\bibinfo  {journal} {Phys. Rev. Lett.}\ }\textbf {\bibinfo
  {volume} {128}},\ \bibinfo {pages} {070604} (\bibinfo {year}
  {2022})}\BibitemShut {NoStop}%
\bibitem [{\citenamefont {Barato}\ and\ \citenamefont
  {Seifert}(2015)}]{barato2015thermodynamic}%
  \BibitemOpen
  \bibfield  {author} {\bibinfo {author} {\bibfnamefont {Andre~C.}\
  \bibnamefont {Barato}}\ and\ \bibinfo {author} {\bibfnamefont {Udo}\
  \bibnamefont {Seifert}},\ }\bibfield  {title} {\enquote {\bibinfo {title}
  {Thermodynamic uncertainty relation for biomolecular processes},}\ }\href
  {\doibase 10.1103/PhysRevLett.114.158101} {\bibfield  {journal} {\bibinfo
  {journal} {Phys. Rev. Lett.}\ }\textbf {\bibinfo {volume} {114}},\ \bibinfo
  {pages} {158101} (\bibinfo {year} {2015})}\BibitemShut {NoStop}%
\bibitem [{\citenamefont {Gingrich}\ \emph {et~al.}(2016)\citenamefont
  {Gingrich}, \citenamefont {Horowitz}, \citenamefont {Perunov},\ and\
  \citenamefont {England}}]{gingrich2016dissipation}%
  \BibitemOpen
  \bibfield  {author} {\bibinfo {author} {\bibfnamefont {Todd~R.}\ \bibnamefont
  {Gingrich}}, \bibinfo {author} {\bibfnamefont {Jordan~M.}\ \bibnamefont
  {Horowitz}}, \bibinfo {author} {\bibfnamefont {Nikolay}\ \bibnamefont
  {Perunov}}, \ and\ \bibinfo {author} {\bibfnamefont {Jeremy~L.}\ \bibnamefont
  {England}},\ }\bibfield  {title} {\enquote {\bibinfo {title} {Dissipation
  bounds all steady-state current fluctuations},}\ }\href {\doibase
  10.1103/PhysRevLett.116.120601} {\bibfield  {journal} {\bibinfo  {journal}
  {Phys. Rev. Lett.}\ }\textbf {\bibinfo {volume} {116}},\ \bibinfo {pages}
  {120601} (\bibinfo {year} {2016})}\BibitemShut {NoStop}%
\bibitem [{\citenamefont {Gingrich}\ \emph {et~al.}(2017)\citenamefont
  {Gingrich}, \citenamefont {Rotskoff},\ and\ \citenamefont
  {Horowitz}}]{gingrich2017inferring}%
  \BibitemOpen
  \bibfield  {author} {\bibinfo {author} {\bibfnamefont {Todd~R}\ \bibnamefont
  {Gingrich}}, \bibinfo {author} {\bibfnamefont {Grant~M}\ \bibnamefont
  {Rotskoff}}, \ and\ \bibinfo {author} {\bibfnamefont {Jordan~M}\ \bibnamefont
  {Horowitz}},\ }\bibfield  {title} {\enquote {\bibinfo {title} {Inferring
  dissipation from current fluctuations},}\ }\href {\doibase
  10.1088/1751-8121/aa672f} {\bibfield  {journal} {\bibinfo  {journal} {J.
  Phys. A}\ }\textbf {\bibinfo {volume} {50}},\ \bibinfo {pages} {184004}
  (\bibinfo {year} {2017})}\BibitemShut {NoStop}%
\bibitem [{\citenamefont {Pal}\ \emph {et~al.}(2020)\citenamefont {Pal},
  \citenamefont {Saryal}, \citenamefont {Segal}, \citenamefont {Mahesh},\ and\
  \citenamefont {Agarwalla}}]{pal2020experimental}%
  \BibitemOpen
  \bibfield  {author} {\bibinfo {author} {\bibfnamefont {Soham}\ \bibnamefont
  {Pal}}, \bibinfo {author} {\bibfnamefont {Sushant}\ \bibnamefont {Saryal}},
  \bibinfo {author} {\bibfnamefont {Dvira}\ \bibnamefont {Segal}}, \bibinfo
  {author} {\bibfnamefont {T.~S.}\ \bibnamefont {Mahesh}}, \ and\ \bibinfo
  {author} {\bibfnamefont {Bijay~Kumar}\ \bibnamefont {Agarwalla}},\ }\bibfield
   {title} {\enquote {\bibinfo {title} {Experimental study of the thermodynamic
  uncertainty relation},}\ }\href {\doibase 10.1103/PhysRevResearch.2.022044}
  {\bibfield  {journal} {\bibinfo  {journal} {Phys. Rev. Research}\ }\textbf
  {\bibinfo {volume} {2}},\ \bibinfo {pages} {022044(R)} (\bibinfo {year}
  {2020})}\BibitemShut {NoStop}%
\bibitem [{\citenamefont {Yang}\ \emph {et~al.}(2020)\citenamefont {Yang},
  \citenamefont {Wei}, \citenamefont {Sheng},\ and\ \citenamefont
  {Wu}}]{yang2020phonon}%
  \BibitemOpen
  \bibfield  {author} {\bibinfo {author} {\bibfnamefont {Cheng}\ \bibnamefont
  {Yang}}, \bibinfo {author} {\bibfnamefont {Xinrui}\ \bibnamefont {Wei}},
  \bibinfo {author} {\bibfnamefont {Jiteng}\ \bibnamefont {Sheng}}, \ and\
  \bibinfo {author} {\bibfnamefont {Haibin}\ \bibnamefont {Wu}},\ }\bibfield
  {title} {\enquote {\bibinfo {title} {Phonon heat transport in cavity-mediated
  optomechanical nanoresonators},}\ }\href {\doibase
  10.1038/s41467-020-18426-4} {\bibfield  {journal} {\bibinfo  {journal} {Nat.
  Commun.}\ }\textbf {\bibinfo {volume} {11}},\ \bibinfo {pages} {4656}
  (\bibinfo {year} {2020})}\BibitemShut {NoStop}%
\bibitem [{\citenamefont {Pietzonka}\ \emph {et~al.}(2017)\citenamefont
  {Pietzonka}, \citenamefont {Ritort},\ and\ \citenamefont
  {Seifert}}]{pietzonka2017finite}%
  \BibitemOpen
  \bibfield  {author} {\bibinfo {author} {\bibfnamefont {Patrick}\ \bibnamefont
  {Pietzonka}}, \bibinfo {author} {\bibfnamefont {Felix}\ \bibnamefont
  {Ritort}}, \ and\ \bibinfo {author} {\bibfnamefont {Udo}\ \bibnamefont
  {Seifert}},\ }\bibfield  {title} {\enquote {\bibinfo {title} {Finite-time
  generalization of the thermodynamic uncertainty relation},}\ }\href {\doibase
  10.1103/PhysRevE.96.012101} {\bibfield  {journal} {\bibinfo  {journal} {Phys.
  Rev. E}\ }\textbf {\bibinfo {volume} {96}},\ \bibinfo {pages} {012101}
  (\bibinfo {year} {2017})}\BibitemShut {NoStop}%
\bibitem [{\citenamefont {Liu}\ \emph {et~al.}(2020)\citenamefont {Liu},
  \citenamefont {Gong},\ and\ \citenamefont {Ueda}}]{liu2020thermodynamic}%
  \BibitemOpen
  \bibfield  {author} {\bibinfo {author} {\bibfnamefont {Kangqiao}\
  \bibnamefont {Liu}}, \bibinfo {author} {\bibfnamefont {Zongping}\
  \bibnamefont {Gong}}, \ and\ \bibinfo {author} {\bibfnamefont {Masahito}\
  \bibnamefont {Ueda}},\ }\bibfield  {title} {\enquote {\bibinfo {title}
  {Thermodynamic uncertainty relation for arbitrary initial states},}\ }\href
  {\doibase 10.1103/PhysRevLett.125.140602} {\bibfield  {journal} {\bibinfo
  {journal} {Phys. Rev. Lett.}\ }\textbf {\bibinfo {volume} {125}},\ \bibinfo
  {pages} {140602} (\bibinfo {year} {2020})}\BibitemShut {NoStop}%
\bibitem [{\citenamefont {Saryal}\ and\ \citenamefont
  {Agarwalla}(2021)}]{saryal2021bounds}%
  \BibitemOpen
  \bibfield  {author} {\bibinfo {author} {\bibfnamefont {Sushant}\ \bibnamefont
  {Saryal}}\ and\ \bibinfo {author} {\bibfnamefont {Bijay~Kumar}\ \bibnamefont
  {Agarwalla}},\ }\bibfield  {title} {\enquote {\bibinfo {title} {Bounds on
  fluctuations for finite-time quantum otto cycle},}\ }\href {\doibase
  10.1103/PhysRevE.103.L060103} {\bibfield  {journal} {\bibinfo  {journal}
  {Phys. Rev. E}\ }\textbf {\bibinfo {volume} {103}},\ \bibinfo {pages}
  {L060103} (\bibinfo {year} {2021})}\BibitemShut {NoStop}%
\bibitem [{\citenamefont {Hasegawa}(2020)}]{hasegawa2020quantum}%
  \BibitemOpen
  \bibfield  {author} {\bibinfo {author} {\bibfnamefont {Yoshihiko}\
  \bibnamefont {Hasegawa}},\ }\bibfield  {title} {\enquote {\bibinfo {title}
  {Quantum thermodynamic uncertainty relation for continuous measurement},}\
  }\href {\doibase 10.1103/PhysRevLett.125.050601} {\bibfield  {journal}
  {\bibinfo  {journal} {Phys. Rev. Lett.}\ }\textbf {\bibinfo {volume} {125}},\
  \bibinfo {pages} {050601} (\bibinfo {year} {2020})}\BibitemShut {NoStop}%
\bibitem [{\citenamefont {Hasegawa}(2021)}]{hasegawa2021thermodynamic}%
  \BibitemOpen
  \bibfield  {author} {\bibinfo {author} {\bibfnamefont {Yoshihiko}\
  \bibnamefont {Hasegawa}},\ }\bibfield  {title} {\enquote {\bibinfo {title}
  {Thermodynamic uncertainty relation for general open quantum systems},}\
  }\href {\doibase 10.1103/PhysRevLett.126.010602} {\bibfield  {journal}
  {\bibinfo  {journal} {Phys. Rev. Lett.}\ }\textbf {\bibinfo {volume} {126}},\
  \bibinfo {pages} {010602} (\bibinfo {year} {2021})}\BibitemShut {NoStop}%
\bibitem [{\citenamefont {Van~Vu}\ and\ \citenamefont
  {Saito}(2022)}]{van2022thermodynamics}%
  \BibitemOpen
  \bibfield  {author} {\bibinfo {author} {\bibfnamefont {Tan}\ \bibnamefont
  {Van~Vu}}\ and\ \bibinfo {author} {\bibfnamefont {Keiji}\ \bibnamefont
  {Saito}},\ }\bibfield  {title} {\enquote {\bibinfo {title} {Thermodynamics of
  precision in markovian open quantum dynamics},}\ }\href {\doibase
  10.1103/PhysRevLett.128.140602} {\bibfield  {journal} {\bibinfo  {journal}
  {Phys. Rev. Lett.}\ }\textbf {\bibinfo {volume} {128}},\ \bibinfo {pages}
  {140602} (\bibinfo {year} {2022})}\BibitemShut {NoStop}%
\bibitem [{\citenamefont {Macieszczak}\ \emph {et~al.}(2018)\citenamefont
  {Macieszczak}, \citenamefont {Brandner},\ and\ \citenamefont
  {Garrahan}}]{macieszczak2018unified}%
  \BibitemOpen
  \bibfield  {author} {\bibinfo {author} {\bibfnamefont {Katarzyna}\
  \bibnamefont {Macieszczak}}, \bibinfo {author} {\bibfnamefont {Kay}\
  \bibnamefont {Brandner}}, \ and\ \bibinfo {author} {\bibfnamefont {Juan~P.}\
  \bibnamefont {Garrahan}},\ }\bibfield  {title} {\enquote {\bibinfo {title}
  {Unified thermodynamic uncertainty relations in linear response},}\ }\href
  {\doibase 10.1103/PhysRevLett.121.130601} {\bibfield  {journal} {\bibinfo
  {journal} {Phys. Rev. Lett.}\ }\textbf {\bibinfo {volume} {121}},\ \bibinfo
  {pages} {130601} (\bibinfo {year} {2018})}\BibitemShut {NoStop}%
\bibitem [{\citenamefont {Proesmans}\ and\ \citenamefont
  {Horowitz}()}]{proesmans2019hysteretic}%
  \BibitemOpen
  \bibfield  {author} {\bibinfo {author} {\bibfnamefont {Karel}\ \bibnamefont
  {Proesmans}}\ and\ \bibinfo {author} {\bibfnamefont {Jordan~M}\ \bibnamefont
  {Horowitz}},\ }\bibfield  {title} {\enquote {\bibinfo {title} {Hysteretic
  thermodynamic uncertainty relation for systems with broken time-reversal
  symmetry},}\ }\href {\doibase 10.1088/1742-5468/ab14da} {\bibfield  {journal}
  {\bibinfo  {journal} {J. Stat. Mech.: Theory Exp}\ }\textbf {\bibinfo
  {volume} {2019}},\ \bibinfo {pages} {054005}}\BibitemShut {NoStop}%
\bibitem [{\citenamefont {Potts}\ and\ \citenamefont
  {Samuelsson}(2019)}]{potts2019thermodynamic}%
  \BibitemOpen
  \bibfield  {author} {\bibinfo {author} {\bibfnamefont {Patrick~P.}\
  \bibnamefont {Potts}}\ and\ \bibinfo {author} {\bibfnamefont {Peter}\
  \bibnamefont {Samuelsson}},\ }\bibfield  {title} {\enquote {\bibinfo {title}
  {Thermodynamic uncertainty relations including measurement and feedback},}\
  }\href {\doibase 10.1103/PhysRevE.100.052137} {\bibfield  {journal} {\bibinfo
   {journal} {Phys. Rev. E}\ }\textbf {\bibinfo {volume} {100}},\ \bibinfo
  {pages} {052137} (\bibinfo {year} {2019})}\BibitemShut {NoStop}%
\bibitem [{\citenamefont {Esposito}\ \emph {et~al.}(2009)\citenamefont
  {Esposito}, \citenamefont {Harbola},\ and\ \citenamefont
  {Mukamel}}]{esposito2009nonequilibrium}%
  \BibitemOpen
  \bibfield  {author} {\bibinfo {author} {\bibfnamefont {Massimiliano}\
  \bibnamefont {Esposito}}, \bibinfo {author} {\bibfnamefont {Upendra}\
  \bibnamefont {Harbola}}, \ and\ \bibinfo {author} {\bibfnamefont {Shaul}\
  \bibnamefont {Mukamel}},\ }\bibfield  {title} {\enquote {\bibinfo {title}
  {Nonequilibrium fluctuations, fluctuation theorems, and counting statistics
  in quantum systems},}\ }\href {\doibase 10.1103/RevModPhys.81.1665}
  {\bibfield  {journal} {\bibinfo  {journal} {Rev. Mod. Phys.}\ }\textbf
  {\bibinfo {volume} {81}},\ \bibinfo {pages} {1665--1702} (\bibinfo {year}
  {2009})}\BibitemShut {NoStop}%
\bibitem [{\citenamefont {Campisi}\ \emph {et~al.}(2011)\citenamefont
  {Campisi}, \citenamefont {H\"anggi},\ and\ \citenamefont
  {Talkner}}]{campisi2011colloquium}%
  \BibitemOpen
  \bibfield  {author} {\bibinfo {author} {\bibfnamefont {Michele}\ \bibnamefont
  {Campisi}}, \bibinfo {author} {\bibfnamefont {Peter}\ \bibnamefont
  {H\"anggi}}, \ and\ \bibinfo {author} {\bibfnamefont {Peter}\ \bibnamefont
  {Talkner}},\ }\bibfield  {title} {\enquote {\bibinfo {title} {Colloquium:
  Quantum fluctuation relations: Foundations and applications},}\ }\href
  {\doibase 10.1103/RevModPhys.83.771} {\bibfield  {journal} {\bibinfo
  {journal} {Rev. Mod. Phys.}\ }\textbf {\bibinfo {volume} {83}},\ \bibinfo
  {pages} {771--791} (\bibinfo {year} {2011})}\BibitemShut {NoStop}%
\bibitem [{\citenamefont {Hasegawa}\ and\ \citenamefont
  {Van~Vu}(2019)}]{hasegawa2019fluctuation}%
  \BibitemOpen
  \bibfield  {author} {\bibinfo {author} {\bibfnamefont {Yoshihiko}\
  \bibnamefont {Hasegawa}}\ and\ \bibinfo {author} {\bibfnamefont {Tan}\
  \bibnamefont {Van~Vu}},\ }\bibfield  {title} {\enquote {\bibinfo {title}
  {Fluctuation theorem uncertainty relation},}\ }\href {\doibase
  10.1103/PhysRevLett.123.110602} {\bibfield  {journal} {\bibinfo  {journal}
  {Phys. Rev. Lett.}\ }\textbf {\bibinfo {volume} {123}},\ \bibinfo {pages}
  {110602} (\bibinfo {year} {2019})}\BibitemShut {NoStop}%
\bibitem [{\citenamefont {Horowitz}\ and\ \citenamefont
  {Gingrich}(2020)}]{horowitz2020thermodynamic}%
  \BibitemOpen
  \bibfield  {author} {\bibinfo {author} {\bibfnamefont {Jordan~M}\
  \bibnamefont {Horowitz}}\ and\ \bibinfo {author} {\bibfnamefont {Todd~R}\
  \bibnamefont {Gingrich}},\ }\bibfield  {title} {\enquote {\bibinfo {title}
  {Thermodynamic uncertainty relations constrain non-equilibrium
  fluctuations},}\ }\href {\doibase 10.1038/s41567-019-0702-6} {\bibfield
  {journal} {\bibinfo  {journal} {Nat. Phys.}\ }\textbf {\bibinfo {volume}
  {16}},\ \bibinfo {pages} {15--20} (\bibinfo {year} {2020})}\BibitemShut
  {NoStop}%
\bibitem [{\citenamefont {Seifert}(2019)}]{seifert2019stochastic}%
  \BibitemOpen
  \bibfield  {author} {\bibinfo {author} {\bibfnamefont {Udo}\ \bibnamefont
  {Seifert}},\ }\bibfield  {title} {\enquote {\bibinfo {title} {From stochastic
  thermodynamics to thermodynamic inference},}\ }\href {\doibase
  10.1146/annurev-conmatphys-031218-013554} {\bibfield  {journal} {\bibinfo
  {journal} {Annu. Rev. Condens. Matter Phys.}\ }\textbf {\bibinfo {volume}
  {10}},\ \bibinfo {pages} {171--192} (\bibinfo {year} {2019})}\BibitemShut
  {NoStop}%
\bibitem [{\citenamefont {Cao}\ \emph {et~al.}(2022)\citenamefont {Cao},
  \citenamefont {Su}, \citenamefont {Jiang},\ and\ \citenamefont
  {Hou}}]{cao2022effective}%
  \BibitemOpen
  \bibfield  {author} {\bibinfo {author} {\bibfnamefont {Zhiyu}\ \bibnamefont
  {Cao}}, \bibinfo {author} {\bibfnamefont {Jie}\ \bibnamefont {Su}}, \bibinfo
  {author} {\bibfnamefont {Huijun}\ \bibnamefont {Jiang}}, \ and\ \bibinfo
  {author} {\bibfnamefont {Zhonghuai}\ \bibnamefont {Hou}},\ }\bibfield
  {title} {\enquote {\bibinfo {title} {Effective entropy production and
  thermodynamic uncertainty relation of active brownian particles},}\ }\href
  {\doibase 10.1063/5.0094211} {\bibfield  {journal} {\bibinfo  {journal}
  {Phys. Fluids}\ }\textbf {\bibinfo {volume} {34}},\ \bibinfo {pages} {053310}
  (\bibinfo {year} {2022})}\BibitemShut {NoStop}%
\bibitem [{\citenamefont {Cao}\ and\ \citenamefont
  {Hou}(2022)}]{cao2022improved}%
  \BibitemOpen
  \bibfield  {author} {\bibinfo {author} {\bibfnamefont {Zhiyu}\ \bibnamefont
  {Cao}}\ and\ \bibinfo {author} {\bibfnamefont {Zhonghuai}\ \bibnamefont
  {Hou}},\ }\bibfield  {title} {\enquote {\bibinfo {title} {Improved estimation
  for energy dissipation in biochemical oscillations},}\ }\href {\doibase
  10.1063/5.0092126} {\bibfield  {journal} {\bibinfo  {journal} {J. Chem.
  Phys.}\ }\textbf {\bibinfo {volume} {157}},\ \bibinfo {pages} {025102}
  (\bibinfo {year} {2022})}\BibitemShut {NoStop}%
\bibitem [{\citenamefont {Koyuk}\ \emph {et~al.}(2018)\citenamefont {Koyuk},
  \citenamefont {Seifert},\ and\ \citenamefont
  {Pietzonka}}]{koyuk2018generalization}%
  \BibitemOpen
  \bibfield  {author} {\bibinfo {author} {\bibfnamefont {Timur}\ \bibnamefont
  {Koyuk}}, \bibinfo {author} {\bibfnamefont {Udo}\ \bibnamefont {Seifert}}, \
  and\ \bibinfo {author} {\bibfnamefont {Patrick}\ \bibnamefont {Pietzonka}},\
  }\bibfield  {title} {\enquote {\bibinfo {title} {A generalization of the
  thermodynamic uncertainty relation to periodically driven systems},}\ }\href
  {\doibase 10.1088/1751-8121/aaeec4} {\bibfield  {journal} {\bibinfo
  {journal} {J. Phys. A: Math. Theor}\ }\textbf {\bibinfo {volume} {52}},\
  \bibinfo {pages} {02LT02} (\bibinfo {year} {2018})}\BibitemShut {NoStop}%
\bibitem [{\citenamefont {Koyuk}\ and\ \citenamefont
  {Seifert}(2019)}]{koyuk2019operationally}%
  \BibitemOpen
  \bibfield  {author} {\bibinfo {author} {\bibfnamefont {Timur}\ \bibnamefont
  {Koyuk}}\ and\ \bibinfo {author} {\bibfnamefont {Udo}\ \bibnamefont
  {Seifert}},\ }\bibfield  {title} {\enquote {\bibinfo {title} {Operationally
  accessible bounds on fluctuations and entropy production in periodically
  driven systems},}\ }\href {\doibase 10.1103/PhysRevLett.122.230601}
  {\bibfield  {journal} {\bibinfo  {journal} {Phys. Rev. Lett.}\ }\textbf
  {\bibinfo {volume} {122}},\ \bibinfo {pages} {230601} (\bibinfo {year}
  {2019})}\BibitemShut {NoStop}%
\bibitem [{\citenamefont {Koyuk}\ and\ \citenamefont
  {Seifert}(2020)}]{koyuk2020thermodynamic}%
  \BibitemOpen
  \bibfield  {author} {\bibinfo {author} {\bibfnamefont {Timur}\ \bibnamefont
  {Koyuk}}\ and\ \bibinfo {author} {\bibfnamefont {Udo}\ \bibnamefont
  {Seifert}},\ }\bibfield  {title} {\enquote {\bibinfo {title} {Thermodynamic
  uncertainty relation for time-dependent driving},}\ }\href {\doibase
  10.1103/PhysRevLett.125.260604} {\bibfield  {journal} {\bibinfo  {journal}
  {Phys. Rev. Lett.}\ }\textbf {\bibinfo {volume} {125}},\ \bibinfo {pages}
  {260604} (\bibinfo {year} {2020})}\BibitemShut {NoStop}%
\bibitem [{\citenamefont {Maes}(2020)}]{maes2020frenesy}%
  \BibitemOpen
  \bibfield  {author} {\bibinfo {author} {\bibfnamefont {Christian}\
  \bibnamefont {Maes}},\ }\bibfield  {title} {\enquote {\bibinfo {title}
  {Frenesy: Time-symmetric dynamical activity in nonequilibria},}\ }\href
  {\doibase 10.1016/j.physrep.2020.01.002} {\bibfield  {journal} {\bibinfo
  {journal} {Phys. Rep.}\ }\textbf {\bibinfo {volume} {850}},\ \bibinfo {pages}
  {1--33} (\bibinfo {year} {2020})}\BibitemShut {NoStop}%
\bibitem [{sup()}]{supp}%
  \BibitemOpen
  \href@noop {} {\bibinfo  {journal} {For details, see the supplement, which
  includes Ref. [19, 36, 47, 64, 68, 85, 89]}\ }\BibitemShut {NoStop}%
\bibitem [{\citenamefont {Aharonov}\ and\ \citenamefont
  {Anandan}(1987)}]{aharonov1987phase}%
  \BibitemOpen
\bibfield  {journal} {  }\bibfield  {author} {\bibinfo {author} {\bibfnamefont
  {Y.}~\bibnamefont {Aharonov}}\ and\ \bibinfo {author} {\bibfnamefont
  {J.}~\bibnamefont {Anandan}},\ }\bibfield  {title} {\enquote {\bibinfo
  {title} {Phase change during a cyclic quantum evolution},}\ }\href {\doibase
  10.1103/PhysRevLett.58.1593} {\bibfield  {journal} {\bibinfo  {journal}
  {Phys. Rev. Lett.}\ }\textbf {\bibinfo {volume} {58}},\ \bibinfo {pages}
  {1593--1596} (\bibinfo {year} {1987})}\BibitemShut {NoStop}%
\bibitem [{\citenamefont {Dechant}\ and\ \citenamefont
  {Sasa}(2020)}]{dechant2020fluctuation}%
  \BibitemOpen
  \bibfield  {author} {\bibinfo {author} {\bibfnamefont {Andreas}\ \bibnamefont
  {Dechant}}\ and\ \bibinfo {author} {\bibfnamefont {Shin-ichi}\ \bibnamefont
  {Sasa}},\ }\bibfield  {title} {\enquote {\bibinfo {title}
  {Fluctuation--response inequality out of equilibrium},}\ }\href {\doibase
  10.1073/pnas.1918386117} {\bibfield  {journal} {\bibinfo  {journal} {Proc.
  Natl. Acad. Sci. U.S.A.}\ }\textbf {\bibinfo {volume} {117}},\ \bibinfo
  {pages} {6430} (\bibinfo {year} {2020})}\BibitemShut {NoStop}%
\bibitem [{\citenamefont {Lu}\ \emph {et~al.}(2022)\citenamefont {Lu},
  \citenamefont {Wang}, \citenamefont {Peng}, \citenamefont {Wang},
  \citenamefont {Jiang},\ and\ \citenamefont {Ren}}]{lu2022geometric}%
  \BibitemOpen
  \bibfield  {author} {\bibinfo {author} {\bibfnamefont {Jincheng}\
  \bibnamefont {Lu}}, \bibinfo {author} {\bibfnamefont {Zi}~\bibnamefont
  {Wang}}, \bibinfo {author} {\bibfnamefont {Jiebin}\ \bibnamefont {Peng}},
  \bibinfo {author} {\bibfnamefont {Chen}\ \bibnamefont {Wang}}, \bibinfo
  {author} {\bibfnamefont {Jian-Hua}\ \bibnamefont {Jiang}}, \ and\ \bibinfo
  {author} {\bibfnamefont {Jie}\ \bibnamefont {Ren}},\ }\bibfield  {title}
  {\enquote {\bibinfo {title} {Geometric thermodynamic uncertainty relation in
  a periodically driven thermoelectric heat engine},}\ }\href {\doibase
  10.1103/PhysRevB.105.115428} {\bibfield  {journal} {\bibinfo  {journal}
  {Phys. Rev. B}\ }\textbf {\bibinfo {volume} {105}},\ \bibinfo {pages}
  {115428} (\bibinfo {year} {2022})}\BibitemShut {NoStop}%
\bibitem [{\citenamefont {Astumian}\ and\ \citenamefont
  {H{\"a}nggi}(2002)}]{astumian2002brownian}%
  \BibitemOpen
  \bibfield  {author} {\bibinfo {author} {\bibfnamefont {R~Dean}\ \bibnamefont
  {Astumian}}\ and\ \bibinfo {author} {\bibfnamefont {Peter}\ \bibnamefont
  {H{\"a}nggi}},\ }\bibfield  {title} {\enquote {\bibinfo {title} {Brownian
  motors},}\ }\href {\doibase 10.1063/1.1535005} {\bibfield  {journal}
  {\bibinfo  {journal} {Phys. Today}\ }\textbf {\bibinfo {volume} {55}},\
  \bibinfo {pages} {33--39} (\bibinfo {year} {2002})}\BibitemShut {NoStop}%
\bibitem [{\citenamefont {Kosloff}\ and\ \citenamefont
  {Levy}(2014)}]{kosloff2014quantum}%
  \BibitemOpen
  \bibfield  {author} {\bibinfo {author} {\bibfnamefont {Ronnie}\ \bibnamefont
  {Kosloff}}\ and\ \bibinfo {author} {\bibfnamefont {Amikam}\ \bibnamefont
  {Levy}},\ }\bibfield  {title} {\enquote {\bibinfo {title} {Quantum heat
  engines and refrigerators: Continuous devices},}\ }\href {\doibase
  10.1146/annurev-physchem-040513-103724} {\bibfield  {journal} {\bibinfo
  {journal} {Annu. Rev. Phys. Chem.}\ }\textbf {\bibinfo {volume} {65}},\
  \bibinfo {pages} {365--393} (\bibinfo {year} {2014})}\BibitemShut {NoStop}%
\bibitem [{\citenamefont {Chernyak}\ and\ \citenamefont
  {Sinitsyn}(2008)}]{chernyak2008pumping}%
  \BibitemOpen
  \bibfield  {author} {\bibinfo {author} {\bibfnamefont {V.~Y.}\ \bibnamefont
  {Chernyak}}\ and\ \bibinfo {author} {\bibfnamefont {N.~A.}\ \bibnamefont
  {Sinitsyn}},\ }\bibfield  {title} {\enquote {\bibinfo {title} {Pumping
  restriction theorem for stochastic networks},}\ }\href {\doibase
  10.1103/PhysRevLett.101.160601} {\bibfield  {journal} {\bibinfo  {journal}
  {Phys. Rev. Lett.}\ }\textbf {\bibinfo {volume} {101}},\ \bibinfo {pages}
  {160601} (\bibinfo {year} {2008})}\BibitemShut {NoStop}%
\bibitem [{\citenamefont {Brandner}\ \emph {et~al.}(2017)\citenamefont
  {Brandner}, \citenamefont {Bauer},\ and\ \citenamefont
  {Seifert}}]{brandner2017universal}%
  \BibitemOpen
  \bibfield  {author} {\bibinfo {author} {\bibfnamefont {Kay}\ \bibnamefont
  {Brandner}}, \bibinfo {author} {\bibfnamefont {Michael}\ \bibnamefont
  {Bauer}}, \ and\ \bibinfo {author} {\bibfnamefont {Udo}\ \bibnamefont
  {Seifert}},\ }\bibfield  {title} {\enquote {\bibinfo {title} {Universal
  coherence-induced power losses of quantum heat engines in linear response},}\
  }\href {\doibase 10.1103/PhysRevLett.119.170602} {\bibfield  {journal}
  {\bibinfo  {journal} {Phys. Rev. Lett.}\ }\textbf {\bibinfo {volume} {119}},\
  \bibinfo {pages} {170602} (\bibinfo {year} {2017})}\BibitemShut {NoStop}%
\bibitem [{\citenamefont {Ptaszy\ifmmode~\acute{n}\else
  \'{n}\fi{}ski}(2018)}]{ptaszynski2018coherence}%
  \BibitemOpen
  \bibfield  {author} {\bibinfo {author} {\bibfnamefont {Krzysztof}\
  \bibnamefont {Ptaszy\ifmmode~\acute{n}\else \'{n}\fi{}ski}},\ }\bibfield
  {title} {\enquote {\bibinfo {title} {Coherence-enhanced constancy of a
  quantum thermoelectric generator},}\ }\href {\doibase
  10.1103/PhysRevB.98.085425} {\bibfield  {journal} {\bibinfo  {journal} {Phys.
  Rev. B}\ }\textbf {\bibinfo {volume} {98}},\ \bibinfo {pages} {085425}
  (\bibinfo {year} {2018})}\BibitemShut {NoStop}%
\bibitem [{\citenamefont {Camati}\ \emph {et~al.}(2019)\citenamefont {Camati},
  \citenamefont {Santos},\ and\ \citenamefont {Serra}}]{camati2019coherence}%
  \BibitemOpen
  \bibfield  {author} {\bibinfo {author} {\bibfnamefont {Patrice~A.}\
  \bibnamefont {Camati}}, \bibinfo {author} {\bibfnamefont {Jonas F.~G.}\
  \bibnamefont {Santos}}, \ and\ \bibinfo {author} {\bibfnamefont {Roberto~M.}\
  \bibnamefont {Serra}},\ }\bibfield  {title} {\enquote {\bibinfo {title}
  {Coherence effects in the performance of the quantum otto heat engine},}\
  }\href {\doibase 10.1103/PhysRevA.99.062103} {\bibfield  {journal} {\bibinfo
  {journal} {Phys. Rev. A}\ }\textbf {\bibinfo {volume} {99}},\ \bibinfo
  {pages} {062103} (\bibinfo {year} {2019})}\BibitemShut {NoStop}%
\bibitem [{\citenamefont {Schaller}(2014)}]{schaller2014open}%
  \BibitemOpen
  \bibfield  {author} {\bibinfo {author} {\bibfnamefont {Gernot}\ \bibnamefont
  {Schaller}},\ }\href@noop {} {\emph {\bibinfo {title} {Open quantum systems
  far from equilibrium}}},\ Vol.\ \bibinfo {volume} {881}\ (\bibinfo
  {publisher} {Springer},\ \bibinfo {year} {2014})\BibitemShut {NoStop}%
\bibitem [{\citenamefont {Wang}\ \emph {et~al.}(2021)\citenamefont {Wang},
  \citenamefont {Chen},\ and\ \citenamefont {Liao}}]{wang2021nonequilibrium}%
  \BibitemOpen
  \bibfield  {author} {\bibinfo {author} {\bibfnamefont {Chen}\ \bibnamefont
  {Wang}}, \bibinfo {author} {\bibfnamefont {Hua}\ \bibnamefont {Chen}}, \ and\
  \bibinfo {author} {\bibfnamefont {Jie-Qiao}\ \bibnamefont {Liao}},\
  }\bibfield  {title} {\enquote {\bibinfo {title} {Nonequilibrium thermal
  transport and photon squeezing in a quadratic qubit-resonator system},}\
  }\href {\doibase 10.1103/PhysRevA.104.033701} {\bibfield  {journal} {\bibinfo
   {journal} {Phys. Rev. A}\ }\textbf {\bibinfo {volume} {104}},\ \bibinfo
  {pages} {033701} (\bibinfo {year} {2021})}\BibitemShut {NoStop}%
\end{thebibliography}%


\begin{thebibliography}{7}%
\makeatletter
\providecommand \@ifxundefined [1]{%
 \@ifx{#1\undefined}
}%
\providecommand \@ifnum [1]{%
 \ifnum #1\expandafter \@firstoftwo
 \else \expandafter \@secondoftwo
 \fi
}%
\providecommand \@ifx [1]{%
 \ifx #1\expandafter \@firstoftwo
 \else \expandafter \@secondoftwo
 \fi
}%
\providecommand \natexlab [1]{#1}%
\providecommand \enquote  [1]{``#1''}%
\providecommand \bibnamefont  [1]{#1}%
\providecommand \bibfnamefont [1]{#1}%
\providecommand \citenamefont [1]{#1}%
\providecommand \href@noop [0]{\@secondoftwo}%
\providecommand \href [0]{\begingroup \@sanitize@url \@href}%
\providecommand \@href[1]{\@@startlink{#1}\@@href}%
\providecommand \@@href[1]{\endgroup#1\@@endlink}%
\providecommand \@sanitize@url [0]{\catcode `\\12\catcode `\$12\catcode
  `\&12\catcode `\#12\catcode `\^12\catcode `\_12\catcode `\%12\relax}%
\providecommand \@@startlink[1]{}%
\providecommand \@@endlink[0]{}%
\providecommand \url  [0]{\begingroup\@sanitize@url \@url }%
\providecommand \@url [1]{\endgroup\@href {#1}{\urlprefix }}%
\providecommand \urlprefix  [0]{URL }%
\providecommand \Eprint [0]{\href }%
\providecommand \doibase [0]{https://doi.org/}%
\providecommand \selectlanguage [0]{\@gobble}%
\providecommand \bibinfo  [0]{\@secondoftwo}%
\providecommand \bibfield  [0]{\@secondoftwo}%
\providecommand \translation [1]{[#1]}%
\providecommand \BibitemOpen [0]{}%
\providecommand \bibitemStop [0]{}%
\providecommand \bibitemNoStop [0]{.\EOS\space}%
\providecommand \EOS [0]{\spacefactor3000\relax}%
\providecommand \BibitemShut  [1]{\csname bibitem#1\endcsname}%
\let\auto@bib@innerbib\@empty
%</preamble>
\bibitem [{\citenamefont {Gingrich}\ \emph {et~al.}(2017)\citenamefont
  {Gingrich}, \citenamefont {Rotskoff},\ and\ \citenamefont
  {Horowitz}}]{gingrich2017inferring}%
  \BibitemOpen
  \bibfield  {author} {\bibinfo {author} {\bibfnamefont {T.~R.}\ \bibnamefont
  {Gingrich}}, \bibinfo {author} {\bibfnamefont {G.~M.}\ \bibnamefont
  {Rotskoff}},\ and\ \bibinfo {author} {\bibfnamefont {J.~M.}\ \bibnamefont
  {Horowitz}},\ }\bibfield  {title} {\bibinfo {title} {Inferring dissipation
  from current fluctuations},\ }\href
  {https://doi.org/10.1088/1751-8121/aa672f} {\bibfield  {journal} {\bibinfo
  {journal} {J. Phys. A}\ }\textbf {\bibinfo {volume} {50}},\ \bibinfo {pages}
  {184004} (\bibinfo {year} {2017})}\BibitemShut {NoStop}%
\bibitem [{\citenamefont {Dechant}\ and\ \citenamefont
  {Sasa}(2020)}]{dechant2020fluctuation}%
  \BibitemOpen
  \bibfield  {author} {\bibinfo {author} {\bibfnamefont {A.}~\bibnamefont
  {Dechant}}\ and\ \bibinfo {author} {\bibfnamefont {S.-i.}\ \bibnamefont
  {Sasa}},\ }\bibfield  {title} {\bibinfo {title} {Fluctuation--response
  inequality out of equilibrium},\ }\href
  {https://doi.org/10.1073/pnas.1918386117} {\bibfield  {journal} {\bibinfo
  {journal} {Proc. Natl. Acad. Sci. U.S.A.}\ }\textbf {\bibinfo {volume}
  {117}},\ \bibinfo {pages} {6430} (\bibinfo {year} {2020})}\BibitemShut
  {NoStop}%
\bibitem [{\citenamefont {Liu}\ \emph {et~al.}(2020)\citenamefont {Liu},
  \citenamefont {Gong},\ and\ \citenamefont {Ueda}}]{liu2020thermodynamic}%
  \BibitemOpen
  \bibfield  {author} {\bibinfo {author} {\bibfnamefont {K.}~\bibnamefont
  {Liu}}, \bibinfo {author} {\bibfnamefont {Z.}~\bibnamefont {Gong}},\ and\
  \bibinfo {author} {\bibfnamefont {M.}~\bibnamefont {Ueda}},\ }\bibfield
  {title} {\bibinfo {title} {Thermodynamic uncertainty relation for arbitrary
  initial states},\ }\href {https://doi.org/10.1103/PhysRevLett.125.140602}
  {\bibfield  {journal} {\bibinfo  {journal} {Phys. Rev. Lett.}\ }\textbf
  {\bibinfo {volume} {125}},\ \bibinfo {pages} {140602} (\bibinfo {year}
  {2020})}\BibitemShut {NoStop}%
\bibitem [{\citenamefont {Koyuk}\ and\ \citenamefont
  {Seifert}(2020)}]{koyuk2020thermodynamic}%
  \BibitemOpen
  \bibfield  {author} {\bibinfo {author} {\bibfnamefont {T.}~\bibnamefont
  {Koyuk}}\ and\ \bibinfo {author} {\bibfnamefont {U.}~\bibnamefont
  {Seifert}},\ }\bibfield  {title} {\bibinfo {title} {Thermodynamic uncertainty
  relation for time-dependent driving},\ }\href
  {https://doi.org/10.1103/PhysRevLett.125.260604} {\bibfield  {journal}
  {\bibinfo  {journal} {Phys. Rev. Lett.}\ }\textbf {\bibinfo {volume} {125}},\
  \bibinfo {pages} {260604} (\bibinfo {year} {2020})}\BibitemShut {NoStop}%
\bibitem [{\citenamefont {Ren}\ \emph {et~al.}(2012)\citenamefont {Ren},
  \citenamefont {Liu},\ and\ \citenamefont {Li}}]{ren2012geometric}%
  \BibitemOpen
  \bibfield  {author} {\bibinfo {author} {\bibfnamefont {J.}~\bibnamefont
  {Ren}}, \bibinfo {author} {\bibfnamefont {S.}~\bibnamefont {Liu}},\ and\
  \bibinfo {author} {\bibfnamefont {B.}~\bibnamefont {Li}},\ }\bibfield
  {title} {\bibinfo {title} {Geometric heat flux for classical thermal
  transport in interacting open systems},\ }\href
  {https://doi.org/10.1103/PhysRevLett.108.210603} {\bibfield  {journal}
  {\bibinfo  {journal} {Phys. Rev. Lett.}\ }\textbf {\bibinfo {volume} {108}},\
  \bibinfo {pages} {210603} (\bibinfo {year} {2012})}\BibitemShut {NoStop}%
\bibitem [{\citenamefont {Crooks}(2007)}]{crooks2007measuring}%
  \BibitemOpen
  \bibfield  {author} {\bibinfo {author} {\bibfnamefont {G.~E.}\ \bibnamefont
  {Crooks}},\ }\bibfield  {title} {\bibinfo {title} {Measuring thermodynamic
  length},\ }\href {https://doi.org/10.1103/PhysRevLett.99.100602} {\bibfield
  {journal} {\bibinfo  {journal} {Phys. Rev. Lett.}\ }\textbf {\bibinfo
  {volume} {99}},\ \bibinfo {pages} {100602} (\bibinfo {year}
  {2007})}\BibitemShut {NoStop}%
\bibitem [{\citenamefont {Brandner}\ and\ \citenamefont
  {Saito}(2020)}]{brandner2020thermodynamic}%
  \BibitemOpen
  \bibfield  {author} {\bibinfo {author} {\bibfnamefont {K.}~\bibnamefont
  {Brandner}}\ and\ \bibinfo {author} {\bibfnamefont {K.}~\bibnamefont
  {Saito}},\ }\bibfield  {title} {\bibinfo {title} {Thermodynamic geometry of
  microscopic heat engines},\ }\href
  {https://doi.org/10.1103/PhysRevLett.124.040602} {\bibfield  {journal}
  {\bibinfo  {journal} {Phys. Rev. Lett.}\ }\textbf {\bibinfo {volume} {124}},\
  \bibinfo {pages} {040602} (\bibinfo {year} {2020})}\BibitemShut {NoStop}%
\end{thebibliography}%

\onecolumngrid
\end{document}

% \linenumbers
\title{Supplement:\\Thermodynamic Geometry of Nonequilibrium Fluctuations in Cyclically Driven Transport}

\author{Zi Wang}
\affiliation{\tongji}

\author{Jie Ren}
\email{Corresponding Email: Xonics@tongji.edu.cn}
\affiliation{\tongji}

\date{\today}

\begin{abstract}
In Sec.~\ref{supp:sec1}, we provide the details of the full counting statistics (FCS) in the Markovian dynamics. In Sec.~\ref{supp:sec2}, we present the derivation of our geometric separation of FCS in a periodically driven thermal device and show its manifestation as a metric term. In Sec.~\ref{supp:sec3}, the formulae for average currents are derived based on our geometric theory on FCS. In Sec.~\ref{supp:sec_nonad_control}, we illustrate our geometric non-adiabatic control principle. In Sec.~\ref{supp:sec4}, we derive our geometric thermodynamic uncertainty relation (TUR) and its implication on the fluctuation of entropy production. Finally, the details for the two models exemplified in the main text are provided in Sec.~\ref{supp:sec5}. 
\end{abstract}
%\pacs{}
%\keywords{}
\maketitle{}
%\tableofcontents

\section{General Markovian Dynamics}
\label{supp:sec1}
{\bf In this section, we discuss the basic theory for the generating function of a fluctuating current. }

We consider a periodically driven discrete Markovian system coupled to several reservoirs. Its probability distribution $p_i$ is labeled by $i$ ($1 \leq i \leq N$), with $N$ being the overall number of states. The rate along the transition $j \to i$ induced by the $\nu$-th reservoir is $k_{ij}^\nu$ ($i \neq j$), whose time-dependence is kept implicit for clarity. As demanded by the thermodynamic consistency, we assume the local detailed balance condition $k_{ij}^\nu/k_{ji}^\nu=e^{\beta_\nu(E_j-E_i)}$, with $\beta_\nu :=1/T_\nu$ being the inverse temperature of the $\nu$-th reservoir (Boltzmann constant is set to $1$) and $E_i$ the energy level of the system. The master equation of $p_i$ is $\partial_t \ket{p(t)}=\hat{L} \ket{p(t)}$, with $L_{ij} := \sum_\nu (k_{ij}^\nu-\delta_{ij} \sum_{j \neq i}k_{ji}^\nu)$ preserving probability by $\sum_i L_{ij}=0$. 

The transition $k_{ij}^\nu$ is associated with an accumulation of exchange current $\Delta Q = d_{ij}^\nu$ ($d_{ij}^\nu = -d_{ji}^\nu$). The accumulated current $Q$ can be particle number, energy, entropy, work, and so on, which in the special case of $d_{ij}^\nu = \ln(k_{ij}^\nu/k_{ji}^\nu)$ corresponds to the reservoir entropy production $\Sigma$. To consider the fluctuation of this current, we take advantage of the full counting statistics formalism. We twist the operator $\hat{L}$ to $\hat{L}_\chi$, with $L_{\chi, ij} = \sum_\nu k_{ij}^\nu e^{\chi d_{ij}^\nu}$ for $i \neq j$ and $L_{\chi, ii} = L_{ii}$. Here, $\chi$ is the counting field of $Q$. Taking the derivative of the cumulant generating function (CGF)
\begin{equation}
{\mathcal G} = \ln {\mathcal Z} = \ln \braket{1|p_\chi(t)}
\end{equation}
with respect to $\chi$ yields the $n$-th cumulant of the accumulated current $Q$ during $[0, t]$ 
\begin{equation}
\braket{Q^n}_c = \frac{\partial^n {\mathcal G}}{\partial \chi^n}|_{\chi = 0}. 
\end{equation}
Here, $\ket{p_\chi(t)}$ is generated by the equation of motion $\partial_t \ket{p_\chi(t)} = \hat{L}_\chi \ket{p_\chi(t)}$ and $\bra{1}$ is a vector with all elements being $1$. We note that the characteristic function ${\mathcal Z}$ satisfies ${\mathcal Z}|_{\chi=0} = 1$ due to the probability normalization condition $\braket{1|p_\chi}|_{\chi=0} = 1$. As a result, the CGF has the property ${\mathcal G}|_{\chi = 0} = 0$. 

This is equivalent to the path integral formalism. By assigning the weight to a trajectory $\omega = x(t)$ ($0 \leq t \leq \tau$)
\begin{equation}
P[\omega] = p(x_0) e^{-\int_0^\tau d t \sum_{ i \neq j; \nu}[\delta_{x(t), i} k_{ji}^\nu(t) - \dot{m}_{ji}^\nu \ln k_{ji}^\nu(t)]}, 
\end{equation}
with $p(x_0)$ being the initial distribution at $t = 0$ and $m_{ij}^\nu$ counting the accumulated number of transition $k_{ij}^\nu$, the CGF is also expressed as
\begin{equation}
{\mathcal G} = \ln \int d\omega p(x_0) e^{-\int_0^\tau d t \sum_{i \neq j; \nu}[\delta_{x(t), i} k_{ji}^\nu (t) - \dot{m}_{ji}^\nu (\chi d_{ji}^\nu(t)+ \ln k_{ji}^\nu (t))]}. 
\end{equation}

The overdamped Brownian dynamics can be taken as a specific limit case of this discrete transition dynamics~\cite{gingrich2017inferring}.

\section{Dynamic and Geometric Components of the CGF}
\label{supp:sec2}
{\bf Here, we derive the general geometric components of the current fluctuation and its metric component. }

We start from the equation of motion $\partial_t \ket{p_\chi(t)}=\hat{L}_\chi({\boldsymbol \Lambda}(t)) \ket{p_\chi(t)}$, where ${\boldsymbol \Lambda}(t)$ is the periodically driven parameter vector with period $\tau_p$. Suppose that $\hat{L}_\chi$ can be diagonalized $\hat{L}_\chi = \sum_n E_n \ket{r_n}\bra{l_n}$, with $E_n$, $\ket{r_n}$ and $\bra{l_n}$ being respectively the eigenvalue, right- and left-eigenvectors satisfying $\hat{L}_\chi \ket{r_n} = E_n \ket{r_n}$ and $\bra{l_n} \hat{L}_\chi = E_n \bra{l_n}$ and the orthonormal condition $\braket{l_m|r_n} = \delta_{m,n}$. The eigenvectors $\ket{r_n}$ and $\bra{l_n}$ and the eigenvalues $E_n$ are all functions of the counting parameter $\chi$. $n=0$ corresponds to the steady state. We also note that by taking $\chi=0$, the twisted operator $\hat{L}_\chi$ reduces to $\hat{L}$, $\ket{r_0}$ to the steady state distribution $\ket{\pi}$ and $\bra{l_0}$ to the one-vector $\bra{1}$. After a large number of periods, the system enters its cyclic state. According to the Floquet theorem, the system state is of the form
\begin{equation}
\label{supp: Floquet theorem eq}
\ket{p_\chi(t)} = e^{{\mathcal G}(t)} \ket{\phi(t)},  
\end{equation}
where $\ket{\phi(t+\tau_p)} = \ket{\phi(t)}$ is a cyclic state and $\ket{p_\chi(t)}$ only accumulates an CGF phase ${\mathcal G}(\tau_p)$ during one driving period. Inserting this ansatz into the equation of motion, the dynamics of $\ket{\phi(t)}$ is governed by
\begin{equation}
\label{supp:floquet eq 1}
(\hat{L}_\chi - \partial_t) \ket{\phi(t)} = (\partial_t {\mathcal G}(t)) \ket{\phi(t)}. 
\end{equation}

The operator $\partial_t$ can be seen as a perturbation to the steady state. In the absence of $\partial_t$, Eq.~(\ref{supp:floquet eq 1}) is solved by the state $\ket{\phi(t)}=\ket{r_0}$ and the equation for CGF 
\begin{equation}
\partial_t {\mathcal G}_{\rm dyn}(t) = E_0(t), 
\end{equation}
which corresponds to the dynamic-phase-like contribution
\begin{equation}
\label{supp: dynamic eq}
{\mathcal G}_{\rm dyn}(\tau_{\rm p}) = \int_0^{\tau_{\rm p}} dt E_0(t). 
\end{equation}

Denoting the deviation of ${\mathcal G}(t)$ from ${\mathcal G}_{\rm dyn}(t)$ as ${\mathcal G}_{\rm geo}(t) = {\mathcal G}(t) - {\mathcal G}_{\rm dyn}(t)$, which is later shown to be a general geometric phase like contribution applicable to arbitrary driving frequency, we can write Eq.~(\ref{supp:floquet eq 1}) as 
\begin{equation}
\label{supp:floquet eq 1'}
[\hat{L}_\chi - E_0(t)]\ket{\phi(t)} = [\partial_t + \partial_t {\mathcal G}_{\rm geo}(t)] \ket{\phi(t)}. 
\end{equation}
Taking the inner product of both sides with $\bra{l_0}$, the left hand side is zero. Thus, the right hand side is orthogonal to $\ket{r_0}$ and we derive the equation for ${\mathcal G}_{\rm geo}$ as $(\partial_t {\mathcal G}_{\rm geo}) \braket{l_0|\phi} = - \braket{l_0|\partial_t \phi}$. This leaves a gauge degree of freedom for $\ket{\phi}$. If we make the gauge transform $\ket{\phi} \to e^{\alpha(t)} \ket{\phi}$, $\partial_t {\mathcal G}_{\rm geo}$ is changed to $\partial_t {\mathcal G}_{\rm geo} + \partial_t \alpha(t)$, ensuring the gauge-invariance of ${\mathcal G}_{\rm geo}(\tau_{\rm p})$ since $\int_0^{\tau_{\rm p}} dt \partial_t \alpha = 0$.

We can fix the gauge of $\ket{\phi(t)}$ by requiring its form 
\begin{equation}
\label{supp: phi eq}
\ket{\phi} = \ket{r_0} + \ket{\phi_\bot}, 
\end{equation}
where $\braket{l_0|\phi_\bot} = 0$. Substituting this expression into Eq.~(\ref{supp:floquet eq 1'}), we obtain 
\begin{equation}
\label{supp:floquet eq 2}
(\hat{L}_\chi - E_0) \ket{\phi_\bot} = \ket{\partial_t r_0} + \ket{\partial_t \phi_\bot} + (\partial_t {\mathcal G}_{\rm geo}) (\ket{r_0}+\ket{\phi_\bot}), 
\end{equation}
the inner product of which and $\bra{l_0}$ provides an equation for ${\mathcal G}_{\rm geo}$:
\begin{equation}
\begin{split}
\partial_t {\mathcal G}_{\rm geo} &= -\braket{l_0|\partial_t r_0} - \braket{l_0|\partial_t \phi_\bot} := \partial_t {\mathcal G}_{\rm curv} + \partial_t {\mathcal G}_{\rm metr}, 
\end{split}
\end{equation}
where we identify the curvature term and the metric term. They describe respectively the adiabatic and nonadiabatic effect of the driving. Through a whole period, the curvature term is 
\begin{equation}
\label{supp: G_curv connection eq}
{\mathcal G}_{\rm curv}(\tau_{\rm p}) = \oint_{\partial \Omega} d \Lambda_\mu A_\mu := - \oint_{\partial \Omega} d \Lambda_\mu \braket{l_0|\partial_\mu r_0}, 
\end{equation}
where we define the geometric connection $A_\mu = - \braket{l_0|\partial_\mu r_0}$. Using the Stokes formula, this is equivalent to 
\begin{equation}
\label{supp: G_curv curvature eq}
{\mathcal G}_{\rm curv}(\tau_{\rm p}) = \int_{\Omega} dS_{\mu \nu} F_{\mu \nu}, 
\end{equation}
in terms of the geometric curvature $F_{\mu \nu} := \partial_\mu A_\nu - \partial_\nu A_\mu = -\braket{\partial_\mu l_0|\partial_\nu r_0} + \braket{\partial_\nu l_0|\partial_\mu r_0}$.

Now we focus on the metric term ${\mathcal G}_{\rm metr}$. We first note
\begin{equation}
\label{supp: G_metr eq}
\begin{split}
{\mathcal G}_{\rm metr}(\tau_{\rm p}) &= -\int_0^{\tau_{\rm p}} dt \braket{l_0|\partial_t \phi_\bot}\\
&= \int_0^{\tau_{\rm p}} dt \braket{\partial_t l_0| \phi_\bot}. 
\end{split}
\end{equation}
Since Eq.~(\ref{supp:floquet eq 2}) can be formally solved as 
\begin{equation}
\label{supp: self-consistent eq}
\begin{split}
\ket{\phi_\bot} &= (\hat{L}_\chi - E_0)^{+} \ket{\partial_t \phi} + (\partial_t {\mathcal G}_{\rm geo}) (\hat{L}_\chi - E_0)^{+} \ket{\phi_\bot} \\
&= (\hat{L}_\chi - E_0)^{+} \ket{\partial_t \phi} - (\hat{L}_\chi - E_0)^{+} \ket{\phi} \braket{l_0|\partial_t \phi}, 
\end{split}
\end{equation}
with the pseudo-inverse being $(\hat{L}_\chi - E_0)^+ := \sum_{n \neq 0} \frac{1}{E_n - E_0} \ket{r_n} \bra{l_n}$, we derive the metric expression 
\begin{equation}
\label{supp: metr eq1}
\begin{split}
{\mathcal G}_{\rm metr} (\tau_{\rm p}) &= \int_0^{\tau_{\rm p}} dt g_{tt} := \int_0^{\tau_{\rm p}} dt \braket{\partial_t l_0|\hat{G}|\partial_t \phi}. 
\end{split}
\end{equation}
Here, we define the metric tensor as 
\begin{equation}
\label{supp: metric tensor eq}
\hat{G} = (\hat{L}_\chi - E_0)^+ - (\hat{L}_\chi - E_0)^+ \ket{\phi} \bra{l_0}. 
\end{equation}
This metric structure also endows a pseudo-Riemannian manifold in the driven parameter space. In terms of this, ${\mathcal G}_{\rm metr}$ can be written as 
\begin{eqnarray}
\label{supp: metr eq2}
{\mathcal G}_{\rm metr} &=& \int_0^{\tau_{\rm p}} g_{\mu \nu} \dot{\Lambda}_\mu \dot{\Lambda}_\nu dt, 
\\
\text{with} \;\; g_{\mu \nu} &:=& \frac{1}{2} \left [ \braket{\partial_\mu l_0|\hat{G}|\partial_\nu \phi} + \braket{\partial_\nu l_0|\hat{G}|\partial_\mu \phi} \right ]. 
\nonumber
\end{eqnarray}
In contrast to the Riemannian manifold, the metric $[g_{\mu \nu}]$ is not necessarily positive-definite. This sacrifice makes our formalism applicable to the whole fluctuation properties of an arbitrary current.

When the near-adiabatic regime is concerned, the Floquet state $\ket{\phi}$ can be approximated, to the leading order, as $\ket{r_0}$. This provides the nonadiabatic metric structure
\begin{equation}
\label{supp: metric near-ad eq}
\begin{split}
\mathfrak{g}_{\mu \nu} &= \frac{1}{2} \left [ \braket{\partial_\mu l_0|\hat{G}|\partial_\nu r_0} + \braket{\partial_\nu l_0|\hat{G}|\partial_\mu r_0} \right ] \\
&= \sum_{n \neq 0} \frac{\braket{\partial_\mu l_0|r_n} \braket{l_n|\partial_\nu r_0}+\braket{\partial_\nu l_0|r_n} \braket{l_n|\partial_\mu r_0}
}{2(E_n - E_0)}, 
\end{split}
\end{equation}
which describes the leading finite-time effect in the near-adiabatic regime. Away from this regime, the exact $\ket{\phi}$ can be obtained by using the Dyson-series-like
self-consistent equation
Eq.~(\ref{supp: self-consistent eq}) with iteration.

Therefore, as a summary the CGF has the decomposition:
\begin{equation}
{\mathcal G}={\mathcal G}_{\rm dyn}+{\mathcal G}_{\rm geo}={\mathcal G}_{\rm dyn}+{\mathcal G}_{\rm curv}+{\mathcal G}_{\rm metr}.
\end{equation}
We note that the above formalism is applicable to arbitrary time antisymmetric current in general nonequilibrium conditions. This provides the basis for analyzing the fluctuation properties of a finite-time cyclically driven thermal device.

\section{Average geometric currents}
\label{supp:sec3}
{\bf Here, we derive the average current based on our FCS theory. }

In this section, $\ket{p}$ and $\hat{L}$ without the subscript $\chi$ means $\ket{p_\chi}|_{\chi=0}$ and $\hat{L}_\chi|_{\chi=0}$. 

First, we consider the total accumulated current and show that the mean value $\braket{Q} = \partial_\chi {\mathcal G}(\tau_p)|_{\chi = 0}$ is also expressed as
\begin{equation}
\label{supp: avg total eq}
\braket{Q} = \int_0^{\tau_p} dt \braket{1|\hat{J}(t)|p(t)}, 
\end{equation}
where the current density operator is given by $\hat{J} = \partial_\chi \hat{L}_\chi |_{\chi = 0}$. Eq.~(\ref{supp: avg total eq}) is important in studying both the average current in stochastic systems and the current in deterministic systems. Defining the propagator for the twisted dynamics $\partial_t \ket{p_\chi(t)} = \hat{L}_\chi \ket{p_\chi(t)}$ as 
\begin{equation}
\hat{U}_\chi(t_2, t_1) = {\mathcal T}e^{\int_{t_1}^{t_2} dt \hat{L}_\chi(t)}, 
\end{equation}
in terms of the time-ordering operator ${\mathcal T}$, we note that the Floquet state Eq.~(\ref{supp: Floquet theorem eq}) is the eigenstate of $\hat{U}_\chi(\tau_p, 0)$
\begin{equation}
\hat{U}_\chi(\tau_p, 0) \ket{p_\chi(0)} = e^{{\mathcal G}(\tau_p)} \ket{p_\chi(0)}. 
\end{equation}
We differentiate both side with respect to $\chi$, set $\chi=0$ and take the inner product with $\bra{1}$. The left hand side is
\begin{equation}
\begin{split}
& \braket{1|[\partial_\chi \hat{U}_\chi(\tau_p, 0)]|p_\chi(0)} |_{\chi =0} + \braket{1|\hat{U}_\chi(\tau_p, 0)|\partial_\chi p_\chi(0)}|_{\chi=0} \\
&= \int_0^{\tau_p} dt \braket{1|\hat{J}|p(t)} + \braket{1|\partial_\chi p_\chi(0)}|_{\chi = 0}, 
\end{split}
\end{equation}
using $\partial_\chi \hat{U}_\chi(\tau_p, 0) = \int_0^{\tau_p} dt \hat{U}_\chi(\tau_p,t) [\partial_\chi \hat{L}_\chi(t)] \hat{U}_\chi(t,0)$ and $\bra{1} \hat{U}_\chi |_{\chi =0}=\bra{1}$. The right hand side is
\begin{equation}
\begin{split}
& [\partial_\chi {\mathcal G}(\tau_p)] e^{{\mathcal G}(\tau_p)}\braket{1|p_\chi(0)}|_{\chi=0} + e^{{\mathcal G}(\tau_p)} \braket{1|\partial_\chi p_\chi(0)}|_{\chi=0} \\
&= \braket{Q} + \braket{1|\partial_\chi p_\chi(0)}|_{\chi =0}, 
\end{split}
\end{equation}
using ${\mathcal G}(\tau_p)|_{\chi = 0} = 0$, the definition $\braket{Q}:= \partial_\chi {\mathcal G}(\tau_p)|_{\chi = 0}$ and the normalization condition $\braket{1|p_\chi(0)}|_{\chi=0}=\braket{1|p(0)} = 1$. Combining these results, we arrive at the identity for average total accumulated current Eq.~(\ref{supp: avg total eq}). 

Second, we consider the dynamic current component. To prove 
\begin{equation}
\label{supp: avg dyn eq}
\braket{Q_{\rm dyn}} := \partial_\chi {\mathcal G}_{\rm dyn}(\tau_p)|_{\chi=0} = \int_0^{\tau_p} dt \braket{1|\hat{J}|\pi(t)}, 
\end{equation}
with $\ket{\pi} = \ket{r_0}|_{\chi=0}$ being the instantaneous steady state, we start from the eigenvalue equation
\begin{equation}
\hat{L}_\chi \ket{r_0} = E_0 \ket{r_0}. 
\end{equation}
Calculating the derivative at $\chi=0$ and taking inner product with $\bra{1}$, the left hand side is $\braket{1|\hat{J}|\pi}+\braket{1|\hat{L}_\chi|\partial_\chi r_0}|_{\chi=0} = \braket{1|\hat{J}|\pi}$ using the property $\bra{1}\hat{L}=0$, while the right hand side is $\partial_\chi E_0|_{\chi=0}$ since $\ket{\pi}$ is normalized and $E_0|_{\chi=0}=0$. Therefore, we have the result $\partial_\chi E_0|_{\chi=0} = \braket{1|\hat{J}(t)|\pi(t)}$ and consequently Eq.~(\ref{supp: avg dyn eq}). 

Thirdly, we consider the general geometric current. For the cyclic state $\ket{\phi(t)}$, we note that the corresponding average current density $-\partial_\chi \braket{l_0|\partial_t \phi}|_{\chi=0}$ is identical to $-\braket{\partial_\chi l_0|\partial_t \phi}|_{\chi=0}$ up to a complete differential term $-\braket{l_0|\partial_\chi \partial_t \phi}|_{\chi=0} = -\partial_t \braket{1|\partial_\chi \phi}|_{\chi=0}$, for any cyclic $\ket{\phi}$. This complete differential term makes no contribution when integrating over a whole period. By further showing the identity 
\begin{equation}
\bra{\partial_\chi l_0} \hat{L}|_{\chi =0} + \bra{1}\hat{J} = \bra{1} (\partial_\chi E_0) |_{\chi = 0}, 
\end{equation}
and therefore
\begin{equation}
\begin{split}
\bra{\partial_\chi l_0} |_{\chi = 0} &= \braket{\partial_\chi l_0|r_0}|_{\chi =0} \bra{1} - \bra{1} \hat{J} \hat{L}^+, 
\end{split}
\end{equation}
where $\hat{L}^+$ is the pseudo-inverse of $\hat{L}$, we prove that our formulation in terms of CGF provides the expression for average accumulated geometric current
\begin{equation}
\begin{split}
\braket{Q_{\rm geo}} &:= \partial_\chi {\mathcal G}_{\rm geo}|_{\chi=0} = -\int_0^{\tau_{\rm p}} dt \partial_\chi \braket{l_0|\partial_t \phi} |_{\chi = 0}  \\
&= - \int_0^{\tau_{\rm p}} dt \braket{\partial_\chi l_0|\partial_t \phi}|_{\chi = 0}\\
&= -\int_0^{\tau_{\rm p}} dt \left [ \braket{\partial_\chi l_0|r_0} \braket{1|\partial_t \phi} - \braket{1|\hat{J} \hat{L}^+|\partial_t \phi} \right ]|_{\chi = 0}\\
&= \int_0^{\tau_{\rm p}} dt \braket{1|\hat{J}\hat{L}^+|\partial_t p(t)}. 
\end{split}
\end{equation}
The fourth equality is due to the constant normalization condition $\partial_t \braket{1|\phi}|_{\chi=0} = 0$. This expression of $\braket{Q_{\rm geo}}$ simplifies to the adiabatic geometric current 
\begin{equation}
\begin{split}
\braket{Q_{\rm curv}} &= \oint_{\partial \Omega} d\Lambda_\mu \braket{1|\hat{J}\hat{L}^+|\partial_\mu \pi(t)} \\
&= \int_\Omega d \Lambda_\mu d\Lambda_\nu [\braket{1|\partial_\mu(\hat{J}\hat{L}^+)|\partial_\nu \pi}-\braket{1|\partial_\nu(\hat{J}\hat{L}^+)|\partial_\mu \pi}], 
\end{split}
\end{equation}
and the nonadiabatic geometric current
\begin{equation}
\label{supp:avg metric eq}
\begin{split}
\braket{Q_{\rm metr}} &= \int_0^{\tau_{\rm p}} dt \braket{1|\hat{J}\hat{L}^+|\partial_t \phi_{\bot}}|_{\chi = 0} \\
&= \int_0^{\tau_{\rm p}} dt \braket{1|\hat{J}\hat{L}^+ \partial_t [(\hat{L}_\chi - E_0)^+ |\partial_t \phi} ]_{\chi =0} \\ 
& - \int_0^{\tau_{\rm p}}dt \braket{1|\hat{J}\hat{L}^+ \partial_t [(\hat{L}_\chi-E_0)^+|\phi}\braket{l_0|\partial_t \phi}]_{\chi = 0} \\
&=\int_0^{\tau_{\rm p}} dt \braket{1|\hat{J}\hat{L}^+ \partial_t [\hat{L}^+|\partial_t p}] \\
&= \int_0^{\tau_{\rm p}} dt \dot{\Lambda}_\mu \dot{\Lambda}_\nu \frac{1}{2} \left [\bra{1}\hat{J} \hat{L}^+ \partial_\mu (\hat{L}^+ \ket{\partial_\nu p})+(\mu \leftrightarrow \nu) \right]\\
&:= \int_0^{\tau_{\rm p}} dt \dot{\Lambda}_\mu \dot{\Lambda}_\nu g_{\mu \nu}^Q. 
\end{split}
\end{equation}
Here, $g_{\mu \nu}^Q$ is the metric related to the average current $Q$.

Correspondingly, the near-adiabatic metric for average $Q$ is obtained by replacing $\ket{p}$ in Eq.~(\ref{supp:avg metric eq}) by $\ket{\pi}$: 
\begin{equation}
\label{supp:near_ad avg metric eq}
\mathfrak{g}_{\mu \nu}^Q = \frac{1}{2} \left [\bra{1}\hat{J} \hat{L}^+ \partial_\mu (\hat{L}^+ \ket{\partial_\nu \pi})+(\mu \leftrightarrow \nu) \right]. 
\end{equation}

Therefore, we have shown the decomposition for the average currents
\begin{equation}
\braket{Q} = \braket{Q_{\rm dyn}}+\braket{Q_{\rm geo}} = \braket{Q_{\rm dyn}}+\braket{Q_{\rm curv}}+\braket{Q_{\rm metr}}. 
\end{equation}

We note here that the above derived average current components are part of our general geometric formulation. Our CGF theory is indispensable in analyzing the fluctuations of finite-time periodically driven systems.

\section{Non-adiabatic control over current pump}
\label{supp:sec_nonad_control}
{\bf Here, we discuss the non-adiabatic geometric control over current pump effect. }

The metric $\mathfrak{g}_{\mu \nu}^Q$ for an arbitrary current $Q$ is not guaranteed to be positive definite. It can have both positive and negative eigenvalues. If we pass through the positive (negative) eigenvector directions, the non-adiabatic geometric current is along its positive (negative) direction. Between the negative and positive eigenvectors, there must exist a direction where $\mathfrak{g}_{\mu \nu} \dot{\Lambda}_\mu \dot{\Lambda}_\nu = 0$ is satisfied. If we design our driving protocol to be along these zero-vector directions, we can eliminate the non-adiabatic pump effect. This constructs a non-adiabatic geometric control over the current pump effect.

\begin{figure}[tp]
\centering
\includegraphics[width=\linewidth]{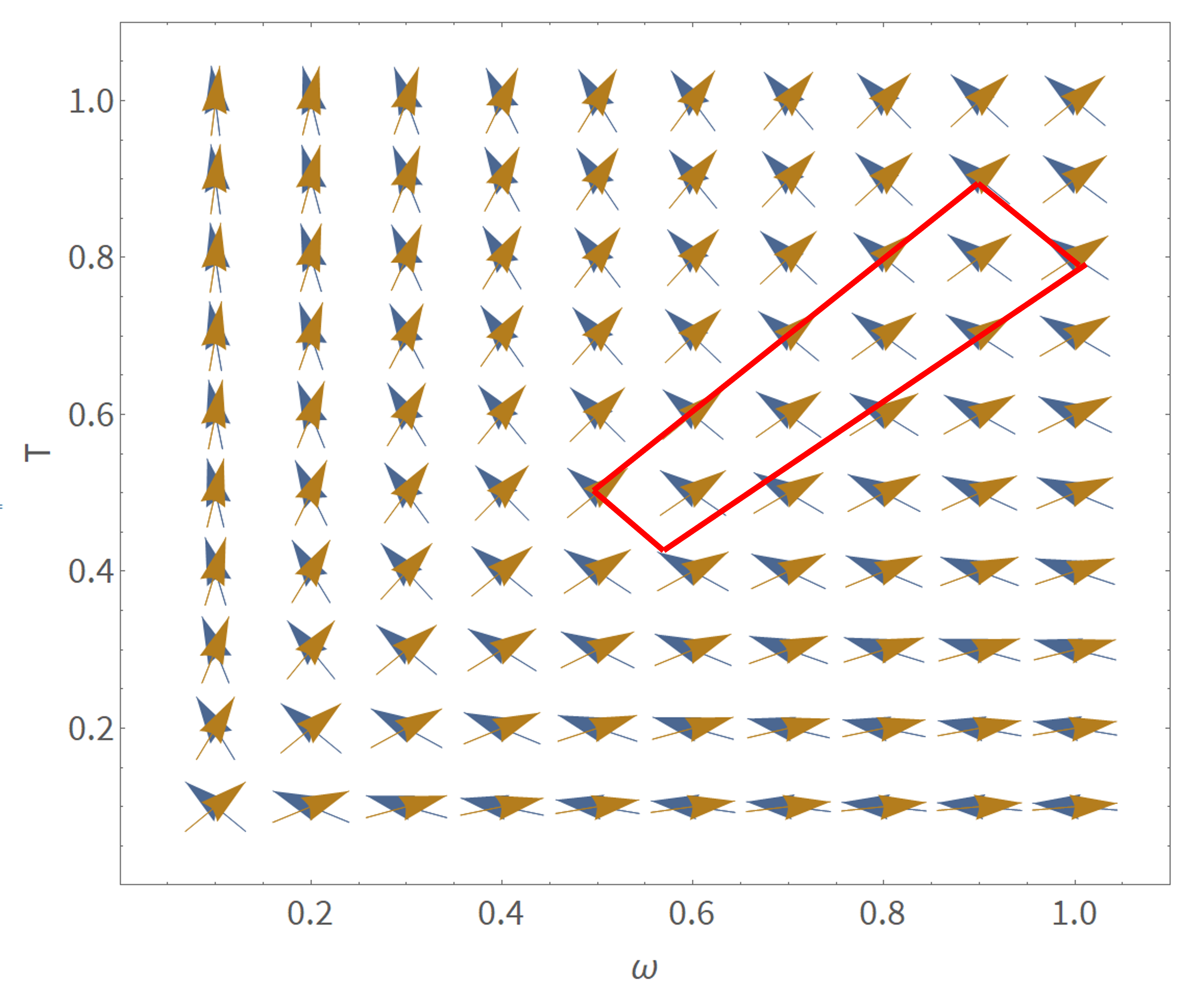}
\caption{
{\bf The non-adiabatic geometric control over current pump.} In the parameter space composed of system energy level difference $\omega$ and the bath temperature $T$, the arrows illustrates the direction of zero non-adiabatic directions, along which $\mathfrak{g}_{\mu \nu}^Q \dot{\Lambda}_\mu \dot{\Lambda}_\nu = 0$ is satisfied. By designing contours along this zero direction field, as shown in the red curve, we can eliminate the non-adiabatic effect of current pump effect. 
}
\label{supp_current_opt}
\end{figure}

Here, we demonstrate this control principle with a two-level-system (TLS) model in Fig.~\ref{supp_current_opt}. It is composed of a TLS of energy difference $\omega$ and is coupled to two reservoirs of the same temperature $T$. We drive the parameter $\boldsymbol{\Lambda} := (\omega, T)^{\rm T}$. By counting the heat current $Q$ from the left reservoir into the TLS, we obtain the twisted master equation operator 
\begin{equation}
\hat{L}_{\chi}=
\begin{pmatrix}
 -2 n & (1+n)(e^{-\chi \omega}+1) \\
 n(e^{\chi \omega} + 1) & -2 (1+n)
\end{pmatrix},
\end{equation}
where $\chi$ is the counting parameter and $n=1/(e^{\omega/T}-1)$ is the Bose-Einstein distribution. Using Eq.~(\ref{supp: metric near-ad eq}), we obtain the metric in parameter space as
\begin{equation}
\mathfrak{g}^Q = \partial_\chi \mathfrak{g}|_{\chi =0} = 
\frac{1}{4T^2} {\rm csch}^3(\frac{\omega}{T}) \sinh^4(\frac{\omega}{2T})
\begin{pmatrix}
 -2T & \omega\\
 \omega & 0
\end{pmatrix}. 
\end{equation}
It indeed has both positive and negative eigenvalues and we illustrate its corresponding zero-vector direction field in Fig.~\ref{supp_current_opt}. This non-adiabatic control principle can also be used to regulate other types of generic currents and their cumulants.

%%%%%%%%%%%%%%%%%%%%%%%%%%%%%%%%%%%%%%%%%%%%%
%%%%%%%%%%%%%%%%%%%%%%%%%%%%%%%%%%%%%%%%%%%%%%%%%

\section{Derivation of Geometric TURs}
\label{supp:sec4}
{\bf The geometric TUR for cyclically driven thermal device is derived in this section. }

We set the counting parameter $\chi=0$ in this proof. Starting from $\partial_t \ket{p(t)}=\hat{L}(t) \ket{p(t)}$, we obtain a self-consistent equation for $\ket{\delta(t)} = \ket{p(t)} - \ket{\pi(t)}$: 
\begin{equation}
\ket{\delta(t)} = \hat{L}^+ \ket{\partial_t \pi(t)} +\hat{L}^+ \ket{\partial_t \delta}, 
\end{equation}
where $\ket{\pi}$ is the instantaneous steady state satisfying $\hat{L}(t) \ket{\pi(t)}=0$ and $\hat{L}^+$ is the pseudo-inverse of $\hat{L}$. $\hat{L}^+$ is defined by $\hat{L}^{+} \hat{L} = \hat{L} \hat{L}^+ = \hat{I} - \ket{\pi}\bra{1}$, with $\hat{I}$ being the identity operator and $\bra{1}$ the one-vector. By recurrently substituting, it is obvious to obtain the adiabatic and nonadiabatic corrections to the distribution $\ket{p} = \ket{\pi}+\ket{p_{\rm curv}} + \ket{p_{\rm metr}}$ respectively, where $\ket{p_{\rm curv}} = \hat{L}^+ \ket{\partial_t \pi(t)}$ and $\ket{p_{\rm metr}} = \hat{L}^+ \partial_t (\hat{L}^+ \ket{\partial_t \pi(t)})$. 

The current operator $\hat{J}$ is defined by $J_{ij} = \sum_\mu k_{ij}^\mu d_{ij}^\mu$ in terms of the transition rates $k_{ij}^\mu$ and the antisymmetric tensor $d_{ij}^\mu = -d_{ji}^\mu$. The component $d_{ij}^\mu$ is the stochastic current accompanying the transition $j \to i$ induced by the reservoir $\mu$. In these notations, the adiabatic and nonadiabatic geometric currents are 
\begin{equation}
\begin{split}
\braket{Q_{\rm curv}} &= \int_{\Omega} dS_{\mu \nu} F_{\mu \nu}^Q, \\
\braket{Q_{\rm metr}} &= \int_0^{\tau_p} d t \dot{\Lambda}_\mu \dot{\Lambda}_\nu \mathfrak{g}_{\mu \nu}^Q,
\end{split}
\end{equation}
where $F_{\mu \nu}^Q = \braket{1|\partial_\mu(\hat{J} \hat{L}^+)|\partial_\nu \pi} - \braket{1|\partial_\nu(\hat{J} \hat{L}^+)|\partial_\mu \pi}$, $\mathfrak{g}_{\mu \nu}^Q = \frac{1}{2} \left [\bra{1}\hat{J} \hat{L}^+ \partial_\mu (\hat{L}^+ \ket{\partial_\nu \pi})+(\mu \leftrightarrow \nu) \right]$. We note here that these expression for the average current can also be derived from Eq.~(\ref{supp: G_curv curvature eq}) and Eq.~(\ref{supp: metric near-ad eq}) by taking derivative to $\chi$ and then setting $\chi \to 0$, as shown in Sec.~\ref{supp:sec3}. 

Now, let us start our proof of Geometric TURs from the fluctuation-response inequality~\cite{dechant2020fluctuation}, which is equivalent to the classical Cramer-Rao bound in the linear response regime. Similar method has been adopted by~\cite{liu2020thermodynamic, koyuk2020thermodynamic} in deriving their TURs. Our method here is similar to those used by Ref.~\cite{dechant2020fluctuation} in proving their generalized TURs. We consider a virtual perturbation on the transition rates, with $\theta$ being its perturbation strength. Considering the Kullback-Leibler (KL) divergence of the perturbed and unperturbed trajectory weight to the order of $\theta^2$, one can write 
\begin{equation}
\begin{split}
D_{\rm KL} &= \int d\omega P_\theta (\omega) \ln(\frac{P_\theta(\omega)}{P_0 (\omega)}) \\
&= \frac{\theta^2}{2} \int d \omega P_0(\omega) [\partial_\theta \ln P_\theta(\omega)|_{\theta \to 0}]^2 + {\mathcal O}(\theta^3) \\
&= -\frac{\theta^2}{2} \int d\omega P_0(\omega) [\partial_\theta^2 \ln P_\theta(\omega)|_{\theta \to 0}] + {\mathcal O}(\theta^3) \\
&= - \frac{\theta^2}{2} \braket{\partial_\theta ^2 \ln P_\theta|_{\theta \to 0}}_0 + {\mathcal O}(\theta^3). 
\end{split}
\end{equation}

For the discrete Markovian systems, the weight for the trajectory $\omega = {x(t)}$ is given by~\cite{liu2020thermodynamic} 
\begin{equation}
P_\theta(\omega) = p(x_0) e^{-\int_0^{\tau_p} dt \sum_{i \neq j; \mu}[\delta_{x(t), i} k_{ji}^\mu (\theta, t) - \dot{m}_{ji}^\mu \ln k_{ji}^\mu(\theta, t)]},
\end{equation}
where $p(x_0)$ is the initial condition and $m_{ij}^\mu$ is the accumulated number of transitions. The basic average quantities are given by $\braket{\delta_{x(t), i}}_\theta = p_i(\theta, t)$ and $\braket{\dot{m}_{ij}^\mu}_\theta = p_j(\theta, t) k_{ij}^\mu (\theta, t)$. We choose the form of the virtual perturbation to be $k_{ij}^\mu (\theta,t)=k_{ij}^\mu e^{\theta \alpha_{ij}^\mu(t)}$. By showing $\partial_\theta^2 \ln P_\theta |_{\theta \to 0}=-\sum_{i \neq j;\mu} \int_0^{\tau_p} dt \delta_{x(t), i} k_{ji}^\mu (\alpha_{ji}^\mu)^2$, we express the KL divergence as 
\begin{equation}
\begin{split}
D_{\rm KL} & \approx -\frac{\theta^2}{2} \braket{\partial_\theta^2 \ln P_\theta|_{\theta \to 0}}_0  \\
&= \frac{\theta^2}{2} \sum_{i \neq j;\mu} \int_0^{\tau_p} dt \braket{\delta_{x(t), i}}_0 k_{ji}^\mu(t) [\alpha_{ji}^\mu(t)]^2 \\
&= \frac{\theta^2}{2} \sum_{i \neq j; \mu} \int_0^{\tau_p} dt p_i(t) k_{ji}^\mu (t) [\alpha_{ji}^\mu(t)]^2 \\
& = \frac{\theta^2}{2} \sum_{i>j; \mu} \int_0^{\tau_p} dt \frac{[j_{ji}^\mu(t)]^2}{t_{ji}^\mu (t)}. 
\end{split}
\end{equation}
Here we select $\alpha_{ij}^\mu = j_{ij}^\mu / t_{ij}^\mu$ in the last line, where we define the detailed current $j_{ij}^\mu = p_j k_{ij}^\mu - p_i k_{ji}^\mu$ and the activity $t_{ij}^\mu = p_j k_{ij}^\mu + p_i k_{ji}^\mu$. By using the log-sum inequality $\frac{2(b-a)^2}{a+b} \leq (b-a)(\ln b - \ln a)$ valid for an arbitrary pair of $a$ and $b$, $D_{\rm KL}$ provides a lower bound on the entropy production $\braket{\Sigma}_0 = \int_0^{\tau_p} dt \sum_{i \neq j;\mu} p_j k_{ij}^\mu \ln[(p_j k_{ij}^\mu)/(p_i k_{ji}^\mu)]$, as given by Ref.~\cite{dechant2020fluctuation}, 
\begin{equation}
D_{\rm KL} \leq \frac{\theta^2}{4} \braket{\Sigma}_0. 
\end{equation}
Also, since $(b-a)^2/(a+b) \leq a+b$ for positive $a$ and $b$, the KL divergence is also bounded by 
\begin{equation}
D_{\rm KL} \leq \frac{\theta^2}{2} \braket{A}_{0}, 
\end{equation}
with the average activity being $\braket{A}_0 = \int_0^{\tau_p} dt \sum_{i \neq j; \mu}(p_j k_{ij}^\mu + p_i k_{ji}^\mu)$. 

Using the form of $\alpha_{ij}^\mu$, it is simple to show that~\cite{dechant2020fluctuation}, to the first order of $\theta$, 
\begin{equation}
\begin{split}
\hat{L}(\theta, t) \ket{p(0, t)} &= (1+\theta)\hat{L}(0, t)\ket{p(0, t)}+\mathcal{ O} (\theta^2) \\
&= (1+\theta)\partial_t \ket{p(0, t)}+\mathcal{ O} (\theta^2). 
\end{split}
\end{equation}
By applying $\hat{L}^+ (\theta, t)$ on both sides, we have 
\begin{equation}
\ket{p(0, t)}=\ket{\pi(\theta, t)} + (1+\theta) \hat{L}^+(\theta, t) \ket{\partial_t p(0, t)}. 
\end{equation}
By applying the adiabatic perturbation method, we derive $\ket{p(0, t)} = \ket{\pi(\theta, t)} +(1+\theta) \hat{L}^+ (\theta, t) \ket{\pi(\theta, t)} + (1+\theta)^2 \hat{L}^+(\theta, t) \partial_t (\hat{L}^+(\theta, t) \ket{\partial_t \pi(\theta, t)}) + {\mathcal O}(\theta^2) = \ket{p(\theta, t)} + \theta \ket{p_{\rm curv} (\theta, t)} + 2\theta \ket{p_{\rm metr}(\theta, t)} +{\mathcal O}(\theta^2)$. Therefore, we have
\begin{equation}
\begin{split}
\partial_\theta \ket{p(\theta, t)} |_{\theta \to 0} &= \lim_{\theta \to 0} \frac{\ket{p(\theta, t)} - \ket{p(0, t)}}{\theta} \\
&= -\ket{p_{\rm curv}(0, t)} -2 \ket{p_{\rm metr}(0, t)}. 
\end{split}
\end{equation}
Also, we can obtain $\braket{1|\partial_\theta \hat{J}(\theta)|_{\theta \to 0}|p(0, t)} = \braket{1|\hat{J}(0, t)|p(0, t)}$ by using the form of the perturbed transition rates $k_{ij}^\mu(\theta, t)$. 

In this way, we show that the linear response of $\braket{Q}$ due to the perturbation is given by
\begin{equation}
\begin{split}
\lim_{\theta \to 0} \frac{\braket{Q}_\theta - \braket{Q}_0}{\theta} &= \int_0^{\tau_p} dt \partial_\theta \braket{1|\hat{J}(\theta, t)|p(\theta, t)}|_{\theta \to 0} \\
&= \braket{Q_{\rm dyn}}_0 - \braket{Q_{\rm metr}}_0.  
\end{split}
\end{equation}

According to the linear fluctuation-response inequality $(\braket{Q}_\theta - \braket{Q}_0)^2 \leq 2D_{\rm KL} \braket{Q^2}_c$, first propose by Ref.~\cite{dechant2020fluctuation}, we reach the final result 
\begin{equation}
\label{supp: tur eq}
\braket{\Sigma} \braket{Q^2}_c \geq 2(\braket{Q_{\rm dyn}} - \braket{Q_{\rm metr}})^2, 
\end{equation}
where we omit the subscript $0$ since all averages are with regard to the unperturbed dynamics. The average entropy production is thus bounded by $\Sigma_g$ 
\begin{equation}
\label{supp: GTUR eq}
\braket{\Sigma} \geq \frac{2(\braket{Q_{\rm dyn}} - \braket{Q_{\rm metr}})^2}{\braket{Q^2}_c}:= \Sigma_{\rm g}. 
\end{equation}
This is the so called Geometric TURs, one of our main results. 

Similarly, the average activity is bounded by 
\begin{equation}
\braket{A} \geq \frac{(\braket{Q_{\rm dyn}}-\braket{Q_{\rm metr}})^2}{\braket{Q^2}_c}:= A_{\rm g}. 
\end{equation}

We further discuss some consequences of Eq.~(\ref{supp: tur eq}). If the instantaneous steady state $\ket{\pi(t)}$ is actually the equilibrium state, the dynamic part $\braket{\Sigma_{\rm dyn}}$ and $\braket{Q_{\rm dyn}}$ and the adiabatic curvature part $\braket{\Sigma_{\rm curv}}$ vanish. The TUR simplifies to
\begin{equation}
\braket{\Sigma_{\rm metr}} \braket{Q^2}_c \geq 2\braket{Q_{\rm metr}}^2. 
\end{equation}
If we consider the current $Q$ to be $\Sigma$, this inequality further simplifies to 
\begin{equation}
\label{supp:entropy geometric bound}
\braket{\Sigma^2}_c \geq 2 \braket{\Sigma_{\rm metr}} \geq \frac{2{\mathcal L}^2}{\tau_p}, 
\end{equation}
bounding the fluctuation of entropy production with the thermodynamic length given by the metric of $\braket{\Sigma_{\rm metr}}$, i.e., for $\braket{\Sigma_{\rm metr}} = \int_0^{\tau_p} dt g_{\mu \nu}^{\Sigma} \dot{\Lambda}_\mu \dot{\Lambda}_\nu$, ${\mathcal L} := \oint_{\partial \Omega} dt \sqrt{g_{\mu \nu}^{\Sigma}\dot{\Lambda}_\mu \dot{\Lambda}_\nu}$. Here, the second inequality can actually saturated by optimizing the parametrization of protocol. We note ${\mathcal L}$ is independent of the detailed protocol but only dependent on the path in the parameter space. According to this inequality, the underlying thermodynamic length is bounded from above by an arbitrary parametrization of the path in the parameter space
\begin{equation}
{\mathcal L} \leq \sqrt{\tau_p \braket{\Sigma^2}_c/2}. 
\end{equation}
This form a basis for future inference of the thermodynamic length.

\section{Models}
\label{supp:sec5}
{\bf We treat the theoretical details of our two models with continuous and discrete degrees of freedom in this section. }

%%%%%%%%%%%%%%%%%%%%%%%%%%%%%%%%%%%%%%%%
\subsection{Chiral current in the nonequilibrium tricycle}

We consider a system composed of three quantum dots with tunable energy levels. Electrons can tunnel between one of the dots and an electron thermal reservoir. We also indirectly couple these dots with three photonic/phononic baths. The bosonic baths compensate the energy difference between dots and renders the system a nonequilibrium quantum tricycle. To be specific, the system Hamiltonian $\hat{H} = \hat{H}_{S} + \hat{H}_{R} + \hat{H}_{SR}+\hat{H}_{B}+\hat{H}_{SB}$ is composed of the three quantum dot levels $\hat{H}_S = \sum_{n = 1}^3 \epsilon_n \hat{c}_n^\dagger \hat{c}_n$, the electron reservoirs $\hat{H}_R = \sum_{k} \epsilon_{k} \hat{d}_{k}^\dagger \hat{d}_{k}$, the tunneling term $\hat{H}_{SR} = \sum_{k} t_{k} (\hat{d}_{k}^\dagger \hat{c}_3 + \hat{c}_3^\dagger \hat{d}_{k})$, the Bosonic thermal bath $\hat{H}_{B} = \sum_{\nu=1;k}^{\nu=3} \epsilon_{\nu, k} \hat{a}_{\nu,k}^\dagger \hat{a}_{\nu,k}$, and the system-bath coupling term $\hat{H}_{SB} = \sum_{\nu=1;k}^{\nu=3} r_{\nu,k} (\hat{a}_{\nu,k} +\hat{a}_{\nu,k}^\dagger)(\hat{c}_{\nu}^\dagger \hat{c}_{\nu+1} + \hat{c}_{\nu+1}^\dagger \hat{c}_{\nu})$. Here, $\nu = 4$ denotes the same site as $\nu=1$. We restrict ourselves to the weak coupling regime. Under the standard Born-Markov approximation, we obtain the transition rates between different states. 

By concentrating on the strong Coulomb blockade regime, the master equation is $\partial_t \ket{p_{\boldsymbol \chi}} = \hat{L}_{\boldsymbol \chi} \ket{p_{\boldsymbol \chi}}$, with $\ket{p} = (p_0, p_1, p_2, p_3)^{\rm T}$ and  
\begin{widetext}
\begin{equation}
\hat{L}_{\boldsymbol \chi}=
\begin{pmatrix}
-k_+ & 0 & 0 & k_- e^{\chi_2(\epsilon_3-\mu)/T_R} \\
 0 & -(k_{21}+k_{31}) & k_{12}e^{\chi_2(\epsilon_2-\epsilon_1)/T_{1}} & k_{13}e^{-\chi_1+\chi_2(\epsilon_3-\epsilon_1)/T_{3}} \\
 0 & k_{21}e^{-\chi_2(\epsilon_2-\epsilon_1)/T_{1}} & -(k_{12}+k_{32}) & k_{23}e^{\chi_2(\epsilon_3-\epsilon_2)/T_{2}} \\
 k_+ e^{-\chi_2(\epsilon_3-\mu)/T_R} & k_{31}e^{\chi_1-\chi_2(\epsilon_3-\epsilon_1)/T_{3}} & k_{32}e^{-\chi_2(\epsilon_3-\epsilon_2)/T_{2}} & -(k_- + k_{13}+k_{23})
\end{pmatrix},
\end{equation}
\end{widetext}
where $k_{ij}$ represents the transition from $j$ to $i$. To be concrete, we suppose $\epsilon_3 > \epsilon_2 > \epsilon_1$. Here, the rates $\Gamma_R = 2\pi \sum_{k} t_{k}^2 \delta(\epsilon - \epsilon_{k})$ and $\Gamma_{\nu} = 2\pi \sum_{k} r_{\nu,k}^2 \delta(\epsilon - \epsilon_{\nu, k})$ are taken as constant under the flat-band limit. $f(\epsilon) = 1/
[e^{(\epsilon-\mu)/T_R} + 1]$ and $n_\nu(\epsilon) = 1/(e^{\epsilon/T_{\nu}} -1)$ are respectively the Fermi and Bose distribution. 

\begin{equation}
\begin{split}
k_{+} &= \Gamma_R f(\epsilon_3), \\
k_{-} &= \Gamma_R[1- f(\epsilon_3)],
\end{split}
\end{equation}

\begin{equation}
\begin{split}
k_{21} &= \Gamma_{1} n_1(\epsilon_2-\epsilon_1), \\
k_{12} &= \Gamma_{1} [1+n_1(\epsilon_2-\epsilon_1)], 
\end{split}
\end{equation}

\begin{equation}
\begin{split}
k_{32} &= \Gamma_{2} n_2(\epsilon_3-\epsilon_2), \\
k_{23} &= \Gamma_{2} [1+n_2(\epsilon_3-\epsilon_2)],
\end{split}
\end{equation}

\begin{equation}
\begin{split}
k_{31} &= \Gamma_{3} n_3(\epsilon_3-\epsilon_1), \\
k_{13} &= \Gamma_{3} [1+n_3(\epsilon_3-\epsilon_1)]. 
\end{split}
\end{equation}
We use $\chi_1$ to count the chiral current and $\chi_2$ to count the entropy production. 

Similarly, with regard to the dynamic activity of the whole system, which is the total number of bidirectional transitions between each pair of states, we can also count it with $\chi_3$, with the corresponding 
\begin{widetext}
\begin{equation}
\hat{L}_{\boldsymbol \chi}=
\begin{pmatrix}
-k_+ & 0 & 0 & k_-e^{\chi_3}  \\
 0 & -(k_{21}+k_{31}) & k_{12}e^{\chi_3} & k_{13}e^{\chi_3} \\
 0 & k_{21}e^{\chi_3} & -(k_{12}+k_{32}) & k_{23}e^{\chi_3} \\
 k_+e^{\chi_3}  & k_{31}e^{\chi_3} & k_{32}e^{\chi_3} & -(k_- + k_{13}+k_{23})
\end{pmatrix},
\end{equation}
\end{widetext}

\begin{figure}[ht]
\centering
\includegraphics[width=0.8\linewidth]{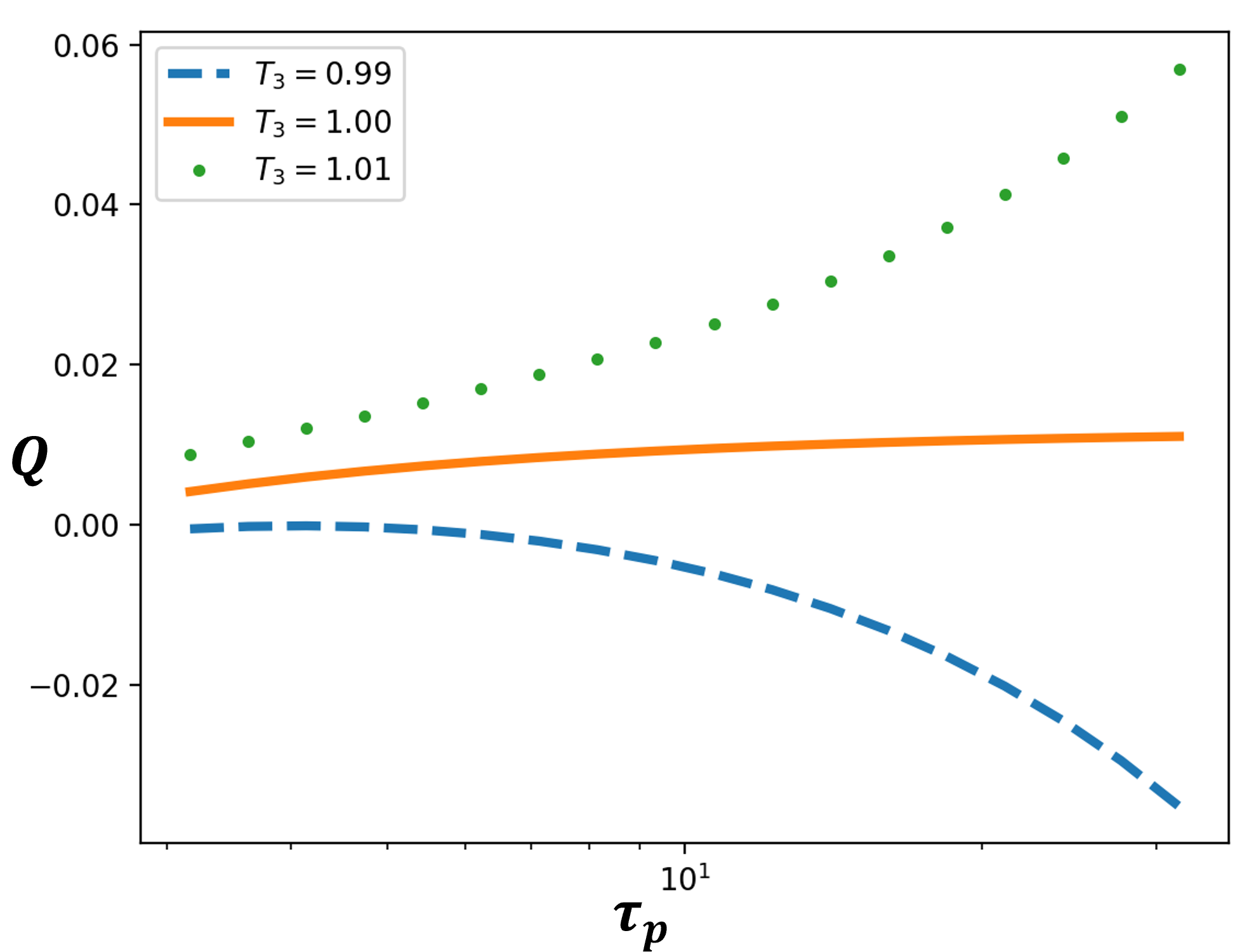}
\caption{{\bf The nonequilibrium chiral current versus the driving period.} $T_3=1.00$ has no dynamic component and other two systems have nonzero dynamic chiral current. Here, the parameters are given by $\phi = 2\pi/3$, $\delta=0.2$, $T_1=T_2=T_R=1.00$, and $\mu=5.0$. }
\label{supp_fig2}
\end{figure}

We drive the system according to the protocol $\epsilon_3(t) = 1+\delta \sin(2\pi t/\tau_p)$, $\epsilon_2(t) = 0.5+\delta \sin(2\pi t/\tau_p+\phi)$, $\epsilon_1 (t) = \delta \sin(2\pi t/\tau_p+2\phi)$ ($0 \leq \delta < 0.25$), and $\Gamma_{\nu} = \Gamma_R = \Gamma$. The parameters $\delta$, $\phi$, $\tau_p$, $\Gamma_B$, $T_{\nu}$ ($\nu=1,2,3$), $T_R$ and $\mu$ determine our model. The characteristic time scale of this system is $\tau_c \approx 1/\Gamma$. The pumped chiral current is shown in Fig.~\ref{supp_fig2}. $T_3=1.00$ corresponds to the case with no dynamic components. 

In the Fig.~2(c) of the main text, we select $T_1=T_2=T_R=1, \mu =5$, the period $10^{1/2} \leq \tau_p \leq 100$, the driving amplitude $\delta = 0.2$, the temperature $0.97 \leq T_3 \leq 1.03$ and the phase difference $0\leq \phi \leq 2\pi$.

In the Fig.~2(d) of the main text, we select the temperatures all being $T_1=T_2=T_3=T_R=1$, $\mu =5$, the driving amplitude $\delta \in \{ 0.05, 0.10, 0.15, 0.20 \}$ and the phase difference $0 \leq \phi \leq 2\pi$. 

%%%%%%%%%%%%%%%%%%%%%%%%%%%%%%%%%%%%%%%%%%%%%%%%%%%%%%%%%%%

%%%%%%%%%%%%%%%%%%%%%%%%%%%%%%%%%%%%%%%%%%%%%%%%%%%%%%%%%%%%%%%%%%%%%

%%%%%%%%%%%%%%%%%%%%%%%%%%%%%%%%%%%%%%%%%%%%%%%%%%%%%%%%%%%%%%%%%%%%%

\subsection{Brownian heat pump}

%%%%%%%%%%%%%%%%%%%%%%%%%%%%%%%%%%%%%%%%%%%%%
\begin{figure}[tp]
\centering
\includegraphics[width=\linewidth]{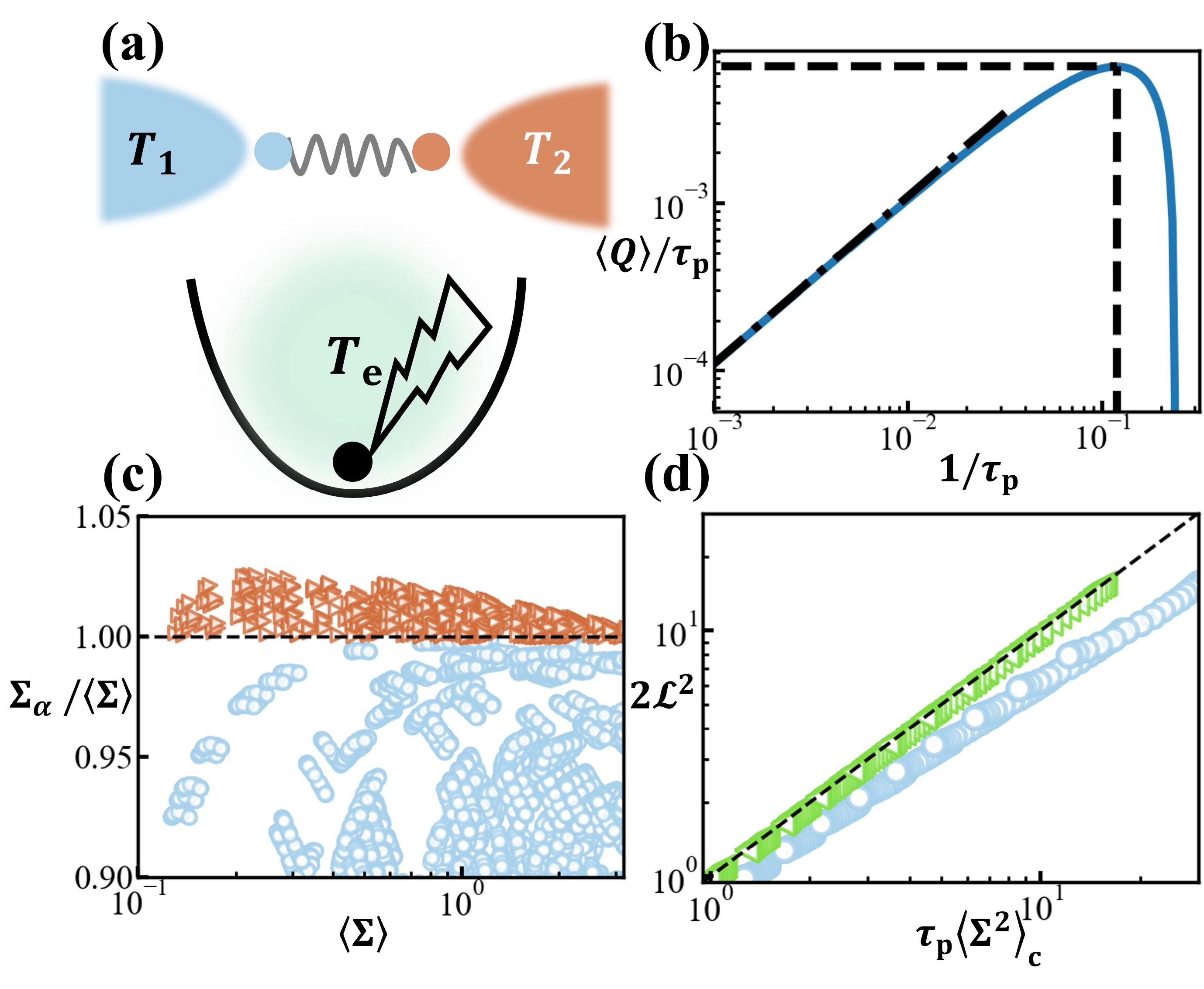}
\caption{
{\bf The Brownian heat pump model by cyclically driving the stiffness of the oscillator $k$ and the coupling to the left reservoir $\gamma_1$.} (a) The Brownian model and its effective system. (b) The geometrically pumped heat density $Q/\tau_{\rm}$ versus the inverse period $1/\tau_{\rm p}$ (the solid line). Here, we choose the protocol $k(t)/k_0 = 1+\sin(2\pi t/\tau_{\rm p})/2, \gamma_1(t)/\gamma_2 = 1+\sin(2\pi t/\tau_{\rm p}+\pi/2)/2$ and $T_1=T_2$, with $k_0$ being a stiffness constant. The dot-dash line is for the adiabatic component $\braket{Q_{\rm curv}}/\tau_{\rm p}$ and the dash line denotes the optimal driving period. (c) The thermodynamic uncertainty relations ($\Sigma_\alpha = \Sigma_{\rm b}$ or $\Sigma_{\rm g}$). Our geometric bound $\braket{\Sigma} \geq \Sigma_{\rm g}$ is satisfied (blue circles), while the steady state bound $\braket{\Sigma} \geq \Sigma_{\rm b}$ can be broken (orange triangles); we only show the breaking situations of the latter bound for clarity. (d) The geometric bound of the fluctuation of the entropy production (Eq.~(\ref{supp:entropy geometric bound})). Blue circles are for constant speed driving protocols and green triangles (on the dashed line) are for the optimal protocols parametrized in terms of the length ${\mathcal L}$. 
}
\label{supp_bh}
\end{figure}
%%%%%%%%%%%%%%%%%%%%%%%%%%%%%%%%%%%%%%%%%%%%%%%%%

Consider a harmonic oscillator composed of two particles coupling two reservoirs, which is illustrated in Fig.~\ref{supp_bh}(a). The Langevin equation of motion is given by
\begin{equation}
\begin{split}
\gamma_1 \dot{x}_1 &= k(x_2 - x_1) + \xi_1, \\
\gamma_2 \dot{x}_2 &= k(x_1 - x_2) + \xi_2, 
\end{split}
\end{equation}
where $k$ is the stiffness of the oscillator and $\xi_i$ is the Gaussian white noise satisfying $\braket{\xi_i(t_1) \xi_j(t_2)} = 2 \gamma_i T_i \delta_{ij} \delta(t_1 - t_2)$. Due to the translational invariance of the above equations, it is equivalent to the dynamics of a single overdamped oscillator $\gamma \dot{y} = -k y + \xi(t)$, with $y = x_1-x_2$ being the effective degree of freedom, $\gamma = \gamma_1 \gamma_2/(\gamma_1 + \gamma_2)$, and $\xi = (\gamma_2 \xi_1 - \gamma_1 \xi_2)/(\gamma_1 + \gamma_2)$. We note the effective noise $\xi$ has zero mean and the variance $\braket{\xi(t_1) \xi(t_2)} = 2\gamma T_{\rm e} \delta(t_1 - t_2)$, with the effective temperature given by $T_{\rm e} = (\gamma_2 T_1 + \gamma_1 T_2)/(\gamma_1 + \gamma_2)$. 

Here, the stochastic heat flowing into the system from the first reservoir is $\dot{Q}_1 = (-\gamma_1 \dot{x}_1 + \xi_1) \dot{x}_1 = -\frac{k^2}{\gamma_1} y^2 + \frac{k}{\gamma_1} \xi_1 y$, and the stochastic entropy production rate $\dot{\Sigma} = -\dot{Q}_1/T_1 - \dot{Q}_2/T_2$ is given by
\begin{equation}
\dot{\Sigma} = (\frac{1}{\gamma_1 T_1} + \frac{1}{\gamma_2 T_2}) k^2 y^2 - (\frac{\xi_1}{\gamma_1 T_1} - \frac{\xi_2}{\gamma_2 T_2})ky. 
\end{equation}
Defining the joint distribution function $\ket{p(y, Q, \Sigma, t)}$, its dynamics is governed by the stochastic Liouville equation $\partial_t \ket{p(y, Q, \Sigma, t)} = \hat{\Omega} \ket{p(y, Q, \Sigma, t)}$ following the continuity relation $\hat{\Omega} = - \partial_y \dot{y} - \partial_{Q_1} \dot{Q}_1 - \partial_{\Sigma} \dot{\Sigma}$. By taking an ensemble average over the reservoir noise, we derive the generalized Fokker-Planck operator as $\hat{L} = \lim_{\Delta t \to 0}\frac{1}{\Delta t}[\int_t^{t+\Delta t}dt_1 \braket{\hat{\Omega}(t_1)} + \int_t^{t+\Delta t} dt_1 \int_t^{t_1} dt_2 \braket{\hat{\Omega}(t_1)\hat{\Omega}(t_2)}]$. The first term is the ballistic term $\hat{L}_1 = \frac{k}{\gamma} \partial_y y + \frac{k^2}{\gamma_1}y^2 \partial_{Q_1} - (\frac{1}{\gamma_1 T_1} + \frac{1}{\gamma_2 T_2})k^2 y^2 \partial_\Sigma$, while the second term is the diffusion term $\hat{L}_2 = \frac{T_{\rm e}}{\gamma}\partial_y^2 + \frac{k T_1}{\gamma_1}(\partial_y y + y \partial_y) \partial_{Q_1} - \frac{k}{\gamma}(\partial_y y + y\partial_y) \partial_\Sigma - \frac{2 k^2 }{\gamma_1}y^2 \partial_{Q_1} \partial_\Sigma + \frac{k^2 T_1}{\gamma_1} y^2 \partial_{Q_1}^2 + (\frac{1}{\gamma_1 T_1}+\frac{1}{\gamma_2 T_2})k^2 y^2\partial_\Sigma^2$.

We use the counting field ${\boldsymbol \chi} = (\chi_1, \chi_2)^{\rm T}$ with two components generating the statistics of $Q_1$ and $\Sigma$ respectively. By making a Fourier-Laplace transformation $\ket{p(y, {\boldsymbol \chi}, t)} = \int dQ d\Sigma \ket{p(y, Q, \Sigma, t)}e^{\chi_1 Q+\chi_2 \Sigma}$, we derive the twisted Fokker-Planck operator as a generator for the dynamics of $\ket{p(y, {\boldsymbol \chi}, t)}$ 
\begin{equation}
\begin{split}
\hat{L}_{\boldsymbol \chi} = \frac{T_{\rm e}}{\gamma} \partial_y^2 + a_1 y \partial_y + a_2 y^2 +a_3,
\end{split}
\end{equation}
with the coefficients $a_1 = \frac{k}{\gamma}[1+2(\chi_2 - \frac{\gamma}{\gamma_1}T_1 \chi_1)]$, $a_2 = k^2 [\frac{\chi_1}{\gamma_1}(T_1 \chi_1 - 1)+(\frac{1}{\gamma_1 T_1} + \frac{1}{\gamma_2 T_2})\chi_2 (\chi_2 + 1) - \frac{2}{\gamma_1} \chi_1 \chi_2]$ and $a_3 = \frac{k}{\gamma}(1+\chi_2 -\frac{\gamma}{\gamma_1}T_1 \chi_1)$. A similarity transformation  $\hat{U} = e^{\beta k y^2 /2}$ would bring $\hat{L}_{\boldsymbol \chi}$ into a Hermitian operator $\tilde{L}_{\boldsymbol \chi} = \tilde{L}_{\boldsymbol \chi}^{\dagger} = \hat{U} \hat{L}_{\boldsymbol \chi} \hat{U}^{-1} = \frac{T_{\rm e}}{\gamma} \partial_y^2 + (a_2 - \frac{a_1^2 \gamma}{4 T_{\rm e}}) y^2 + (a_3 - \frac{a_1}{2})$, where $\beta = \frac{a_1 \gamma}{2 T_{\rm e} k}$. 

Analogous to the solution of quantum oscillators, in terms of the lowering $\hat{b} = \sqrt{\epsilon} \partial_y + \frac{1}{2\sqrt{\epsilon}}y$ and raising operator $\hat{b}^\dagger = -\sqrt{\epsilon} \partial_y + \frac{1}{2\sqrt{\epsilon}}y$ ($[\hat{b}, \hat{b}^\dagger] = 1$), where the factor $\epsilon = \frac{T_{\rm e}}{\sqrt{a_1^2 \gamma^2 - 4 a_2 T_{\rm e} \gamma}}$, we can write $\tilde{L}_{\boldsymbol \chi} = - \frac{T_{\rm e}}{\gamma \epsilon} \hat{b}^\dagger \hat{b} + (a_3 - \frac{a_1}{2} - \frac{T_{\rm e}}{2 \gamma \epsilon})$. Therefore, we derive the eigenvalues of $\hat{L}_{\boldsymbol \chi}$ to be $E_n = -n\frac{T_{\rm e}}{\gamma \epsilon} + (a_3 - \frac{a_1}{2} - \frac{T_{\rm e}}{2 \gamma \epsilon})$ ($E_0$ corresponding to the steady state) and the eigenvectors to be
\begin{equation}
\begin{split}
\ket{r_n} &= \hat{U}^{-1} \ket{\psi_n} = \frac{1}{\sqrt{n!}} \hat{U}^{-1} (\hat{b}^\dagger)^n \ket{\psi_0}, \\
\bra{l_n} &= \bra{\psi_n} \hat{U} = \frac{1}{\sqrt{n!}} \bra{\psi_0} \hat{b}^n \hat{U}, 
\end{split}
\end{equation}
where $\ket{\psi_0} = (\bra{\psi_0})^\dagger$ is the ground eigenstate of $\tilde{L}_{\boldsymbol \chi}$ satisfying $\hat{b} \ket{\psi_0} = 0$. Here, $\{\ket{\psi_n}\}$ forms an orthonormal basis $\braket{\psi_m|\psi_n} = \delta_{mn}$.

The ground state satisfies $\hat{b} \ket{\psi_0}=0$ ($(\sqrt{\epsilon}\partial_y + \frac{1}{2\sqrt{\epsilon}}y)\ket{\psi_0} = 0$) and is thus given by a Gaussian distribution~\cite{ren2012geometric}
\begin{equation}
\ket{\psi_0} = (\frac{1}{2\pi \epsilon})^{1/4} e^{-\frac{y^2}{4\epsilon}},
\end{equation}
with an expression similar to the ground state wavefunction of the quantum harmonic oscillator. To calculate the geometric CGF, we note the relations $y = \sqrt{\epsilon}(\hat{b}+\hat{b}^\dagger)$, $\ket{\partial_\mu \psi_0} = \frac{(y^2-\epsilon)\partial_\mu \epsilon}{4\epsilon^2} \ket{\psi_0}$, $\partial_\mu \hat{U} = (\partial_\mu \beta k) \frac{y^2}{2} \hat{U}$ and $\partial_\mu \hat{U}^{-1} = -(\partial_\mu \beta k)\frac{y^2}{2} \hat{U}^{-1}$. We show
\begin{equation}
\begin{split}
\ket{\partial_\mu r_0} &= -\frac{\epsilon \partial_\mu(\beta k)}{2} \ket{r_0} + \sqrt{2}[\frac{\partial_\mu \epsilon}{4\epsilon}-\frac{\epsilon \partial_\mu (\beta k)}{2}]\ket{r_2}, \\
\bra{\partial_\mu l_0} &= \frac{\epsilon \partial_\mu (\beta k)}{2}\bra{l_0} + \sqrt{2}[\frac{\partial_\mu \epsilon}{4\epsilon}+\frac{\epsilon \partial_\mu (\beta k)}{2}]\bra{l_2}. 
\end{split}
\end{equation}

The relation $\braket{l_0|\partial_\mu r_0} +\braket{\partial_\mu l_0|r_0}=0$ is satisfied due to the normalization condition $\braket{l_0|r_0}=1$. For arbitrary driven parameters, the geometric connection is $A_\mu = -\braket{l_0|\partial_\mu r_0} = \epsilon \partial_\mu(\beta k)/2$. The corresponding geometric curvature is $F_{\mu \nu} = \partial_\mu A_\nu - \partial_\nu A_\mu = [(\partial_\mu \epsilon) (\partial_\nu \beta k)-(\partial_\nu \epsilon)(\partial_\mu \beta k)]/2$. The metric tensor is given by
\begin{equation}
g_{\mu \nu} = \frac{\gamma}{16 T_{\rm e}\epsilon}[4\epsilon^4(\partial_\mu \beta k)(\partial_\nu \beta k) - (\partial_\mu \epsilon)(\partial_\nu \epsilon)]. 
\end{equation}

To illustrate our theory, we consider a heat engine powered by driving $k$ and $\gamma_1$, i.e. ${\boldsymbol \Lambda} = (k, \gamma_1)^{\rm T}$. We first consider the driving protocol $k = k_0 [1+a\sin(2 \pi t/\tau_p)]$, $\gamma_1 = \gamma_0[1+a\sin(2\pi t/\tau_p+\phi)]$, and $\gamma_2=\gamma_0$. The parameters $\tau_p$, $T_1$, $T_2$ and $\phi$ can be taken as configurations of the driving protocol. We further define a time scale of this system as $\tau_c = \gamma_0/k_0$, which is of the same order of the characteristic time scale $\gamma/k$. 

The instantaneous dynamic components for $\braket{Q}$, $\braket{Q^2}_c$, $\braket{\Sigma}$ and $\braket{\Sigma^2}_c$ are given by
\begin{equation}
\begin{split}
E^{Q} &:= \partial_{\chi_1} E_0|_{{\boldsymbol \chi}=0} = \frac{1}{\tau_c} \frac{\tilde{k} (T_1 - T_2)}{\tilde{\gamma}_1 + 1}, 
\end{split}
\end{equation}
\begin{equation}
\begin{split}
E^{Q^2} &:= \partial_{\chi_1}^2 E_0|_{{\boldsymbol \chi}=0} = \frac{1}{\tau_c} \frac{2 \tilde{k}(T_2 \tilde{\gamma}_1 + T_1)(T_1 \tilde{\gamma}_1 +T_2)}{(\tilde{\gamma}_1 + 1)^3}, 
\end{split}
\end{equation}
\begin{equation}
\label{supp:Sigma_dyn}
\begin{split}
E^{\Sigma} &:= \partial_{\chi_2} E_0|_{{\boldsymbol \chi}=0} = \frac{1}{\tau_c} \frac{\tilde{k}(T_1 - T_2)^2}{T_1 T_2 (\tilde{\gamma}_1 +1)}, 
\end{split}
\end{equation}
\begin{equation}
\label{supp:Sigma_2_dyn}
\begin{split}
E^{\Sigma^2} &:= \partial_{\chi_2}^2 E_0|_{{\boldsymbol \chi}=0} \\
&= \frac{1}{\tau_c}\frac{2\tilde{k}(T_1-T_2)^2 [T_1^2 \tilde{\gamma}_1 +T_2^2 \tilde{\gamma}_1 + T_1 T_2(\tilde{\gamma}_1^2+1)]}{T_1^2 T_2^2 (\tilde{\gamma}_1 + 1)^3}, 
\end{split}
\end{equation}
where we define the dimensionless factors $\tilde{k}:=k/k_0$ and $\tilde{\gamma}_1 := \gamma_1/\gamma_0$.

The geometric connection vectors for $\braket{Q_{\rm curv}}$, $\braket{Q_{\rm curv}^2}_c$ and $\braket{\Sigma_{\rm curv}}$ are given by
\begin{equation}
\begin{split}
A^Q_{\mu} d\Lambda_\mu &= -\frac{(T_2 \tilde{\gamma}_1+T_1)}{2\tilde{k}(\tilde{\gamma}_1+1)^2}d\tilde{k}+\frac{T_1 \tilde{\gamma}_1 + T_2}{2(\tilde{\gamma}_1+1)^3}d\tilde{\gamma}_1,
\end{split}
\end{equation}

\begin{equation}
\begin{split}
A^{Q^2}_\mu d\Lambda_\mu &=\frac{\tilde{\gamma}_1 (T_2 \tilde{\gamma}_1 +T_1)[T_1(\tilde{\gamma}_1-2)+3 T_2]}{\tilde{k}(\tilde{\gamma}_1+1)^4}d\tilde{k} \\
&+ \frac{3\tilde{\gamma}_1(T_1-T_2)(T_1 \tilde{\gamma}_1 + T_2)}{(\tilde{\gamma}_1 + 1)^5} d\tilde{\gamma_1},
\end{split}
\end{equation}

\begin{equation}
\label{supp:Sigma_connection}
\begin{split}
A^\Sigma_\mu d\Lambda_\mu &= \frac{(T_2 \tilde{\gamma}_1+T_1)(T_1 \tilde{\gamma}_1 + T_2)}{2\tilde{k}T_1 T_2 (\tilde{\gamma}_1+1)^2}d\tilde{k} \\
&+ \frac{(T_1-T_2)(T_1 \tilde{\gamma}_1 +T_2)}{2T_1 T_2(\tilde{\gamma}_1+1)^3} d\tilde{\gamma}_1,
\end{split}
\end{equation}

\begin{equation}
\label{supp:Sigma_2_connection}
\begin{split}
A^{\Sigma^2}_\mu d\Lambda_\mu &= \frac{3(T_1-T_2)^2(T_1 \tilde{\gamma}_1 + T_2)}{T_1^2 T_2^2 (\tilde{\gamma}_1 + 1)^4}[\frac{\tilde{\gamma}_1 (T_2 \tilde{\gamma}_1 + T_1)}{\tilde{k}} d\tilde{k} \\
&+ \frac{(T_1-T_2)\tilde{\gamma}_1}{\tilde{\gamma}_1 + 1} d\tilde{\gamma}_1]. 
\end{split}
\end{equation}
Obviously, from Eq.~(\ref{supp:Sigma_dyn}), Eq.~(\ref{supp:Sigma_2_dyn}), Eq.~(\ref{supp:Sigma_connection}) and Eq.~(\ref{supp:Sigma_2_connection}), we can easily show that both the dynamic and adiabatic geometric components of the entropy production (and its variance) vanish in the isothermal case where $T_1(t) = T_2(t)$. Particularly, although $A^{\Sigma}_{\mu \nu} d \Lambda_\mu \neq 0$ in this situation, it simplifies to $\frac{1}{2} d \ln \tilde{k}$ as a total derivative of $\tilde{k}$ and therefore the accumulation during one cycle $\int_{\partial \Omega} A^{\Sigma}_{\mu \nu} d \Lambda_\mu$ is zero. 

\begin{widetext}
The geometric metrics are given by
\begin{equation}
\begin{split}
\mathfrak{g}^{Q}_{\mu \nu} d\Lambda_\mu d\Lambda_\nu &= \frac{\tau_c \tilde{\gamma}_1}{4\tilde{k}^3(\tilde{\gamma}_1+1)^3}[-(T_2 \tilde{\gamma}_1 + T_1) (d\tilde{k})^2+\tilde{k} T_2 d\tilde{k} d\tilde{\gamma}_1 + \frac{\tilde{k}^2(T_1-T_2)}{(\tilde{\gamma}_1 + 1)^2}(d \tilde{\gamma}_1)^2],
\end{split}
\end{equation}

\begin{equation}
\begin{split}
\mathfrak{g}^{Q^2}_{\mu \nu} d\Lambda_\mu d\Lambda_\nu &= \tau_c [\frac{\tilde{\gamma}_1 (T_2 \tilde{\gamma}_1 + T_1)[6T_2 \tilde{\gamma}_1 +T_1(\tilde{\gamma}_1^2-4\tilde{\gamma}_1 + 1)]}{2 \tilde{k}^3 (\tilde{\gamma}_1 + 1)^5} (d\tilde{k})^2 - \frac{T_2 \tilde{\gamma}_1 [6T_2 \tilde{\gamma}_1 + T_1 (\tilde{\gamma}_1^2 - 4\tilde{\gamma}_1 +1)]}{2 \tilde{k}^2 (\tilde{\gamma}_1 + 1)^5} d\tilde{k} d\tilde{\gamma}_1 \\
&- \frac{\tilde{\gamma}_1 [T_1 T_2 (-3 \tilde{\gamma}_1^2 + 14 \tilde{\gamma}_1 -3)+T_1^2 (\tilde{\gamma}_1^2-8\tilde{\gamma}_1 +1)+T_2^2 (\tilde{\gamma}_1^2 - 8\tilde{\gamma}_1 + 1)]}{2 \tilde{k}(\tilde{\gamma}_1+1)^7} (d \tilde{\gamma}_1)^2], 
\end{split}
\end{equation}

\begin{equation}
\mathfrak{g}^\Sigma_{\mu \nu} d\Lambda_\mu d\Lambda_\nu = \tau_c [\frac{\tilde{\gamma}_1 (T_2 \tilde{\gamma}_1 + T_1)(T_1 \tilde{\gamma}_1 + T_2)}{4\tilde{k}^3 T_1 T_2(\tilde{\gamma}_1+1)^3}(d \tilde{k})^2 + \frac{(T_1^2-T_2^2) \tilde{\gamma}_1}{4\tilde{k}^2 T_1 T_2 (\tilde{\gamma}_1 + 1)^3} d\tilde{k} d\tilde{\gamma}_1 + \frac{(T_1-T_2)^2 \tilde{\gamma}_1}{4 \tilde{k} T_1 T_2 (\tilde{\gamma}_1 + 1)^5}(d \tilde{\gamma}_1)^2],
\end{equation}

\begin{equation}
\label{supp: metric CGF bh}
\begin{split}
\mathfrak{g}^{\Sigma^2}_{\mu \nu} d\Lambda_\mu d\Lambda_\nu &= \tau_c [\frac{\tilde{\gamma}_1 (T_2 \tilde{\gamma}_1 + T_1)[6T_1^3 \tilde{\gamma}_1^2 + 6 T_2^3 \tilde{\gamma}_1 + T_1 T_2^2 (7 \tilde{\gamma}_1^2-10\tilde{\gamma}_1 +1)+T_1^2 T_2 \tilde{\gamma}_1 (\tilde{\gamma}_1^2-10\tilde{\gamma}_1 + 7)]}{2 \tilde{k}^3 T_1^2 T_2^2 (\tilde{\gamma}_1 + 1)^5} (d\tilde{k})^2 \\
&+ \frac{\tilde{\gamma}_1(T_1^2 - T_2^2) [6T_1^2 \tilde{\gamma}_1 + 6T_2^2 \tilde{\gamma}_1 +T_1 T_2(\tilde{\gamma}_1^2 - 10 \tilde{\gamma}_1 + 1)]}{2 \tilde{k}^2 T_1^2 T_2^2 (\tilde{\gamma}_1 + 1)^5}d\tilde{k} d\tilde{\gamma}_1 \\
&+ \frac{\tilde{\gamma}_1 (T_1-T_2)^2 [T_1 T_2 (3 \tilde{\gamma}_1^2 - 14 \tilde{\gamma}_1 + 3) - T_1^2(\tilde{\gamma}_1^2 - 8 \tilde{\gamma}_1 + 1)-T_2^2(\tilde{\gamma}_1^2 - 8\tilde{\gamma}_1 +1)]}{2 \tilde{k} T_1^2 T_2^2 (\tilde{\gamma}_1 + 1)^7} (d \tilde{\gamma}_1)^2]. 
\end{split}
\end{equation}
\end{widetext}

In the non-biased situation. $T_1 = T_2$ and the metric structure for $\braket{\Sigma}$ and $\braket{\Sigma^2}_c$ simplifies to be
\begin{equation}
\begin{split}
\mathfrak{g}^\Sigma_{\mu \nu} d\Lambda_\mu d\Lambda_\nu &= \tau_c \frac{\tilde{\gamma}_1}{4 \tilde{k}^3 (\tilde{\gamma}_1 + 1)} (d\tilde{k})^2, \\
\mathfrak{g}^{\Sigma^2}_{\mu \nu} d\Lambda_\mu d\Lambda_\nu &= 2 (\mathfrak{g}_\Sigma)_{\mu \nu} d\Lambda_\mu d\Lambda_\nu. 
\end{split}
\end{equation}
In this Brownian model, as an illustration, the metric structure of the average and variance of entropy production is of the form that our bound 
Eq.~(\ref{supp:entropy geometric bound}) can actually be saturated. In numerical simulation,s we randomly choose $10^{1/2} \leq \tau_p \leq 100, T_1 = 1, 0.1 \leq T_2 \leq 1.9$, the driving amplitude $a=0.5$ and the phase difference $0 \leq \phi \leq 2\pi$. The results validate the Geometric TURs in Fig~\ref{supp_bh}(c). 

To consider the geometric bound on the variance of entropy production in Fig.~\ref{supp_bh}(d), we fix $T_1 = T_2 =1$ and select the driving amplitude $a \in [0.4, 0.9]$ and the phase difference $0 \leq \phi \leq 2\pi$.

As shown in Fig.~\ref{supp_bh}(b), the nonadiabatic heat pump effect is bounded from above by $\braket{Q}_{\rm curv}/\tau_{\rm p}$. We can easily show that the average heat flux $\braket{Q}/\tau_{\rm p}$ reaches its maximum $-\braket{Q_{\rm curv}}^2/(4\tau_{\rm p} \braket{Q_{\rm metr}})$ at the optimal period $-2 \tau_{\rm p} \braket{Q_{\rm metr}}/\braket{Q_{\rm curv}}$. In Fig.~\ref{supp_bh}(c), independent of the driving protocols, the entropy production $\braket{\Sigma}$ is bounded from below by $\Sigma_{\rm g}$ (blue circles) as derived in Eq.~(\ref{supp: GTUR eq}), but breaks the corresponding steady state bound $\braket{\Sigma} \geq \Sigma_{\rm b} := 2\braket{Q}^2/\braket{Q^2}_c$ (orange triangles). Furthermore, in this system, the metric expression for the average of entropy production $\mathfrak{g}_{\mu \nu}^\Sigma$ and entropy variance $\mathfrak{g}_{\mu \nu}^{\Sigma^2}$ for $T_1=T_2$, i.e. Eq.~(\ref{supp: metric CGF bh}), satisfies
\begin{equation}
\mathfrak{g}_{\mu \nu}^{\Sigma^2} = 2\mathfrak{g}_{\mu \nu}^{\Sigma} = \frac{\gamma_1 \gamma_2}{2k^3 (\gamma_1+\gamma_2)} \begin{pmatrix}
 1 & 0 \\
 0 & 0
\end{pmatrix}, 
\end{equation}
implying that our geometric bound Eq.~(\ref{supp:entropy geometric bound}) is actually saturable by reparametrizing the protocol in terms of the thermodynamic length, i.e. the time spend around a parameter point $dt = (\tau_{\rm p}/{\mathcal L}) \sqrt{\mathfrak{g}_{\mu \nu}^{\Sigma} d\Lambda_\nu d\Lambda_\nu}$~\cite{crooks2007measuring,brandner2020thermodynamic}, as shown by the green triangles in Fig.~\ref{supp_bh}(d). Here, $\mathfrak{g}_{\mu \nu}^{\Sigma} := \partial_{\chi_2} \mathfrak{g}_{\mu \nu}|_{{\boldsymbol \chi}=0}$ and $\mathfrak{g}_{\mu \nu}^{\Sigma^2} := \partial_{\chi_2}^2 \mathfrak{g}_{\mu \nu}|_{{\boldsymbol \chi}=0}$. The blue circles in Fig.~\ref{supp_bh}(d) shows the validity of Eq.~(\ref{supp:entropy geometric bound}) in the constant speed protocols.

%%%%%%%%%%%%%%%%%%%%%%%%%%%%%%%%%%%%%%%%%%

\bibliography{ref.bib}

\onecolumngrid